\documentclass[12pt]{elsarticle}
\usepackage{graphicx} 
\usepackage{amsfonts}
\usepackage{mystyle}

\usepackage{titlesec}
\titleformat{\paragraph}
{\normalfont\normalsize\bfseries}{\theparagraph}{1em}{}
\titlespacing*{\paragraph}
{0pt}{3.25ex plus 1ex minus .2ex}{1.5ex plus .2ex}

\usepackage{mathrsfs}
\usepackage{amssymb}
\usepackage{amsmath}
\usepackage{lipsum}
\usepackage{multirow}
\usepackage[colorlinks=True]{hyperref}
\usepackage{xcolor}
\usepackage{comment}
\usepackage[normalem]{ulem}
\usepackage[modulo]{lineno}
\newcommand{\p}{\partial}
\usepackage{subfigure}
\usepackage{epstopdf}

\newcommand\figref[1]{Figure \ref{fig:#1}} 
\newcommand\tabref[1]{Table \ref{tab:#1}} 
\newcommand\eref[1]{Eq. (\ref{eq:#1})} 

\begin{document}

\begin{frontmatter}

\title{Structure-preserving variational neural fields: Uncertainty-quantified reduced-order modeling of nonlinear conservation laws}
\author[LANL]{Aviral Prakash\corref{cor1}}
\ead{aviralp@lanl.gov}\cortext[cor1]{Corresponding author}
\author[LANL]{Marc L. Klasky}

\address[LANL]{Theoretical Division, Los Alamos National Laboratory, Los Alamos, NM 87545, USA}


\begin{abstract}
Reduced-order models, such as latent dynamics models, are becoming mainstream for accelerating simulations for parameterized physical systems governed by nonlinear conservation laws. However, most existing latent dynamics frameworks suffer from two important limitations: they do not provide uncertainty estimates for model predictions, and they do not guarantee adherence to the underlying conservation laws. While these challenges have been addressed separately in prior work, a unified framework that simultaneously provides uncertainty quantification and exact conservation-law preservation remains largely unexplored. In this work, we develop a variational latent neural field framework that integrates Gaussian process-inspired surrogates, enabling estimation of predictive confidence for both in-distribution and out-of-distribution parameter regimes. Three variants of the framework are considered:
IRS-UQ, PI-IRS-UQ, and ECLEIRS-UQ, corresponding to unconstrained,
physics-informed, and conservation-structure-preserving formulations,
respectively. Exact conservation-structure preservation is achieved by embedding the solution dynamics within a conservation-law manifold through a space-time divergence-free representation of the solution-flux field. We demonstrate the applicability of the framework through three numerical experiments: 1) 1-D advection, 2) 2-D Euler and 3) 2-D shallow water equations in parameterized settings. Numerical experiments demonstrate that the proposed approach provides accurate predictions together with uncertainty estimates, while remaining robust to sparse and noisy training data. Comparisons between the proposed three approaches show that conservation-structure preserving latent representations improve robustness to degraded training data while maintaining competitive predictive accuracy and uncertainty quantification capability. 
 
\end{abstract}


\end{frontmatter}

\section{Introduction}

Over the past decade, there have been several new developments under the umbrella of reduced-order models (ROMs) for accelerating computational physics simulations for multi-query applications. Traditional ROMs, which are commonly known as projection-based ROMs \cite{Sirovich1987, Aubry1988, Benner2015}, rely on vast amounts of data and governing physical equations. While these methods are typically intrusive, meaning they require changes in the source code, nonintrusive alternatives are available \cite{Peherstorfer2016, Kramer2024, Gkimisis2024, Prakash2024b, Prakash2025} but their applicability is constrained to scenarios where the governing physical equations are fully described. Advances in modern deep-learning architecture have paved the way for pure data-driven ROMs, commonly categorized under the umbrella term of neural operators \cite{Li2021, Lu2021} or latent/reduced dynamics models \cite{Fries2022, Lee2021a, Bonneville2024, Prakash2026}, that require an immense amount of data, but overcome the need for governing equations. These methods are ideal for scenarios where governing physics is not completely known or complex to implement, such as weather forecasting \cite{Pathak2022} and fusion energy systems \cite{Gopakumar2024}. In this article, we discuss the key shortcomings of existing latent dynamics models and propose a new method to overcome these issues.

\subsection{Survey of latent dynamics models}

While neural operators and latent dynamics models are designed to achieve the same goal, that is accelerating physical simulations, these methods differ in the philosophy of modeling. For spatio-temporal dynamical problems, neural operators \cite{Li2021, Lu2021} are an operator map from solution at one time to the solution at another time. On the other hand, latent-dynamics models follow an alternate approach where a low-dimensional manifold on which the solution lies is identified and the dynamics is evolved in this low-dimensional manifold. Therefore, latent dynamics models are similar to projection-based ROMs, and are, in some cases, inspired by them. Although we focus on latent-dynamics models in this work, the proposed framework can also be applied to nonlinear projection-based ROMs in settings where the governing equations are known and manifold projection techniques are available. \cite{Lee2020, Kim2022, Chen2023, Puri2025}. Latent-dynamics models have been shown to provide robust reduced-order representations of parameterized spatio-temporal dynamics, even for systems governed by highly nonlinear and complex equations \cite{Chung2026, Prakash2025b}. 

There are several variants of these latent dynamics models that are proposed in the literature. The literature survey in \cite{Bonneville2024} summarizes the major developments in this area especially in the context of Latent Space Dynamics Identification (LaSDI) class of models. These models rely on a grid-based autoencoder-decoder method for identifying a low-dimensional manifold and sparse identification of nonlinear dynamics (SINDy) class of methods to identify the dynamics on the low-dimensional manifold. More recent work in this area uses auto-decoder-based neural network architectures \cite{Wen2023, Prakash2026} to extend these model to scenarios where the data does not lie on a fixed space-time grid. 

Although recent advances have broadened the applicability of these models to high-dimensional and complex physical systems, two important limitations remain. First, latent dynamics models are solely data-driven and lack the required information of the underlying physical structure of the problem \cite{Park2024b, Prakash2026}. Recent work in \cite{Prakash2026} highlighted that the absence of conservation structure-preservation reduces the accuracy of these models when used for scenarios with lower data quality due to either lack of resolution or added noise in measurements. Second, common latent dynamics models only give point estimates of the prediction, thereby providing overconfident predictions for scenarios outside the training dataset \cite{Bonneville2024b}. Since these models are frequently deployed to predict system responses in parameter regimes beyond those observed during training, overconfident predictions can have serious consequences for downstream applications, particularly in out-of-distribution and extrapolatory settings. Therefore, it is important to develop latent dynamics models that resolve these two key issues. While  several approaches exist for incorporating physical constraints in neural operators and latent dynamics models, there is limited work on identifying model uncertainties. A detailed literature review on incorporating conservation constraints in neural operators and latent dynamics models can be found in \cite{Prakash2026}.

\subsection{Enabling uncertainty quantification for parameterized solution modeling}

In the past few years, a growing body of work has focused on uncertainty quantification (UQ) for neural operators and, to a lesser extent, latent dynamics models. In the context of neural operators, several approaches integrate Bayesian neural-network methodology into operator-learning architectures \cite{Magnani2025, Lin2023, Garg2025, Zou2024}. These approximate Bayesian neural operators do not, by themselves, provide a reliable mechanism for detecting distribution shift, and recent studies have shown that several neural-operator UQ methods can fail on even moderately out-of-domain regimes \cite{Mouli2024}. Deep ensemble-based approaches have also been proposed for neural-operator UQ \cite{Yang2022, Garg2025, Mouli2024}. These methods can produce uncertainty estimates that correlate with prediction error in some out-of-distribution settings, but ensemble disagreement is not an intrinsic measure of distance from the training distribution. Conformal prediction has also been integrated with neural operators \cite{Ma2024, Moya2025}, providing finite-sample coverage guarantees under exchangeability between calibration and test data. However, standard conformal prediction does not provide an intrinsic out-of-distribution-detection mechanism, and its coverage guarantees may not hold when deployment scenarios differ substantially from the calibration distribution. Another line of work treats neural operators themselves as probabilistic models, learning distributions over output functions using proper scoring rules or generative objectives \cite{Bulte2025}. Despite their methodological differences, all these approaches for estimating uncertainty share a common limitation: their uncertainty estimates primarily characterize variability in the predicted solution conditional on the supplied input scenario. Bayesian methods encode posterior variability over model parameters or functions, ensembles use inter-model disagreement, conformal methods calibrate prediction sets using residuals from an exchangeable calibration distribution, and probabilistic neural operators learn conditional distributions over output functions. However, these quantities are not explicit measures of the distance between a new input scenario and the training or calibration distribution. Thus, they do not directly define the applicability domain of the learned operator. For reliable deployment under scenario-level distribution shift, predictive uncertainty should be complemented by a separate physically meaningful distributional-distance metric over the input function or scenario space.

This issue can be tackled by Gaussian process-based approaches, as they introduce an explicit covariance structure over the input \cite{Kumar2024} or latent \cite{Kumar2025, Bonneville2024b} space. This covariance defines a similarity measure between a new scenario and the observed training scenarios, thereby providing a more direct mechanism for assessing applicability-domain uncertainty. When the new input is strongly correlated with the training inputs, the posterior variance is reduced. When it is weakly correlated, the posterior variance remains close to the prior variance, which is typically higher than the posterior variance near observed data. This makes GP-based operator-learning methods attractive for distinguishing inputs that are close to the training distribution from those that are far away. In addition to the work in \cite{Kumar2024, Kumar2025}, a recent work in this area learns operators with Gaussian processes to approximate the real-valued bilinear form \cite{Mora2025}. Compared with neural operators, there is relatively less work on integrating uncertainty-estimation strategies with latent dynamics models. Bonneville et al. \cite{Bonneville2024b} extend latent dynamics models by incorporating Gaussian processes to model the coefficients governing latent-vector dynamics. This approach propagates uncertainty through the latent state evolution in a computationally efficient reduced-order space. However, the offline stage can be expensive because it involves simultaneous identification of latent dynamics and dimensionality reduction. 

From this literature survey, we conclude that Gaussian process-based UQ approaches are ideal as they have an intrinsic mechanism to handle out-of-distribution
uncertainties. However, covariance estimation can scale poorly with training-set size, limiting applicability to massive datasets. Therefore, there is a need to integrate Gaussian process \cite{Rasmussen2005} in a scalable way, while ensuring that the predictive operator structure
is not overly restrictive for problems requiring asymmetric, or multimodal uncertainty estimates. 


\subsection{Proposed method and contributions}
The performance of reduced-order models, including projection-based methods, neural operators, and latent-dynamics models, is fundamentally limited by the quality and coverage of the data used to characterize the parameter space of interest.  Therefore, it is important not only to develop methods that maintain robust predictive performance when training data are limited, but also to quantify the uncertainty introduced by this parametric data scarcity. To date, most neural-operator and latent-dynamics modeling approaches have addressed either robustness in data-limited regimes through physics-informed constraints or UQ, but rarely both simultaneously. In this work, we address both challenges by proposing a structure-preserving framework that not only preserves the underlying physical constraints but also quantifies the uncertainty resulting from limited training data across the parameter space. Our proposed approach, which we refer to as variational latent neural fields, integrates ideas from latent neural fields \cite{Du2024, Chen2023, Puri2025, Prakash2025} for grid-free representation of spatio-temporal data with approximate data-distribution sampling approaches such as variational
autoencoders \cite{Kingma2019}, to model solutions of parameterized physical systems while providing uncertainty estimates. The parameterized latent space is approximated as Gaussian process-inspired surrogates, which allow uncertainty estimation based on the covariance structure of new input to the training inputs while also enabling asymmetric, multi-modal and heavy tailed solution distribution modeling via the neural decoder. Unlike other Gaussian process-integrated neural operators, this framework is made scalable to large datasets through learning surrogates for Gaussian process posterior mean and variance in the latent space. Finally, the framework exactly preserves the nonlinear conservation law by parameterizing the high-dimensional vector potential of space-time divergence-free solution-flux vector using variational neural field representation. We demonstrate the applicability of the proposed approach through three numerical examples: 1) 1-D advection problem, 2) 2-D Euler problem and 3) 2-D shallow water equations. The results also assess the impact of conservation-structure preservation by comparing the proposed framework against both unconstrained models and models that impose physical constraints through soft regularization, particularly when data quality is degraded by spatio-temporal sparsity and noise. Although we demonstrate this framework in the context of neural field-based latent dynamics models, the framework is directly applicable to DeepONet-based neural operators, such as latent neural operators \cite{Kontolati2023}. 

\subsection{Outline}

The outline of this manuscript is as follows. In Section \ref{sec:MathBack}, we provide the mathematical background for parametric nonlinear conservation-law systems, introduce the notation used throughout the paper, and present the latent-dynamics framework employed for accelerated surrogate modeling of such systems. In Section \ref{sec:Section3}, we discuss the probabilistic approach for modeling latent dynamics solutions using variational latent neural fields. In Section \ref{sec:Section4}, we discuss the weak inclusion of nonlinear conservation law constraint through a penalty-based approach for training the model. In Section \ref{sec:Section5}, we exactly enforce the nonlinear conservation law constraint by learning the high-dimensional vector potential of space-time divergence-free solution-flux vector using variational latent neural fields. In Section \ref{sec:Section6}, we discuss the computational implementation of different latent dynamics modeling approaches proposed in this work. In Section \ref{sec:Section7}, we demonstrate the applicability of the proposed approach in accurately modeling parameterized systems with uncertainty estimates through examination of three numerical experiments. Lastly, Section \ref{sec:Section8} concludes this article by summarizing the main contributions and results, while also listing a few key directions of future research.

\section{Mathematical background}
\label{sec:MathBack}

\subsection{System description}

We consider parameterized systems of partial differential equations of the form
\begin{equation}
    \frac{\p \pmb{q}}{\p t} + \mathcal{N}(\pmb{q}; \pmb{\nu}) = 0,
\end{equation}
where $\pmb{q}(\pmb{x}, t, \pmb{\nu}): \mathcal{X} \times \mathcal{T} \times \mathcal{V} \to \mathbb{R}^m$ is the pointwise solution variable, $\pmb{x} \in \mathcal{X} \subset \mathbb{R}^{d}$ is the spatial coordinate, $t \in \mathcal{T} \subset  \mathbb{R}^+$ is the temporal coordinate, $ \pmb{\nu} \in \mathcal{V} \subset \mathbb{R}^{d_{\nu}}$ is the system parametric coordinate, and $\mathcal{N}(\cdot; \pmb{\nu})$ is the nonlinear differential operator. The parameter vector $\pmb{\nu}$ may include quantities such as initial conditions, boundary conditions, forcing terms, and other PDE coefficients. For each fixed time $t$ and parameter instance $\pmb{\nu}$, the solution field is $\pmb{q}(\cdot, t, \pmb{\nu}) \in \mathcal{Q}$, where the Hilbert space $\mathcal{Q} = L^2(\mathcal{X}; \mathbb{R}^m)$, equipped with the standard $L^2$ inner product.

Often the nonlinear solution operator has a nonlinear conservation law form
\begin{equation}
    \mathcal{R}(\pmb{q}; \pmb{\nu}) = \frac{\p \pmb{q}}{\p t} + \nabla_x \cdot \pmb{f} (\pmb{q}; \pmb{\nu}) = 0,
    \label{eq:consv_law}
\end{equation}
where $\pmb{f} (\pmb{u}; \pmb{\nu})$ defines the flux of the solution and $\nabla_x\cdot\pmb{f}  =
    \sum_{j=1}^{d}
    \frac{\partial \pmb{f}_j}{\partial x_j}$ is the spatial divergence operator. When this conservation form is invoked, we regard the full trajectory $\pmb q(\cdot,\cdot,\pmb\nu)$ as belonging to an admissible space-time solution class $\mathcal{Y}$. Assuming sufficient regularity and appropriate boundary conditions, integration of Eq.~\eqref{eq:consv_law} over the spatial domain and application of the divergence theorem gives

\begin{equation}
    \frac{d}{dt}
    \int_{\mathcal X}
    \pmb q(\pmb{x},t,\pmb{\nu})\,d\pmb{x}
    =
    -
    \int_{\partial\mathcal X}
    \pmb f(\pmb q;\pmb\nu)\cdot\pmb n\,dS
\end{equation}
where \(\pmb n\) is the outward unit normal on \(\partial\mathcal X\). Under periodic, zero-flux, or other boundary conditions that eliminate the boundary integral, the total conserved quantity $\int_{\mathcal X}\pmb q(\pmb{x},t,\pmb{\nu})\,d\pmb{x}$ remains invariant in time. Conservation is fundamental to the accurate modeling of many physical systems, especially those without source or sink terms or with no net flux through the boundary.

For general nonlinear solution operators, there is no established theory that characterizes the regions of phase space in which the solutions evolve. However, for nonlinear conservation-law systems, the solutions are constrained by the governing conservation laws. To characterize this admissible set, we define the solution manifold associated with parameter instance $\pmb{\nu}$ as
\begin{equation}
    \mathcal{M}_{\nu} = \{\pmb{q} (\cdot, \cdot, \pmb{\nu}) \in \mathcal{Y} \vert \;\mathcal{R}(\pmb{q}; \pmb{\nu}) = 0\},
    \label{eq:sol_manifold_def}
\end{equation}
where $\mathcal{M}_{\nu}$ contains admissible solution trajectories satisfying the conservation law constraint for the parameter instance $\pmb{\nu}$. Following the common terminology in reduced-order modeling, we refer to this admissible set as the solution manifold.

\subsection{Modeling PDE solutions  using latent dynamics models}

Latent-dynamics models provide computationally efficient surrogates for high-fidelity simulations through a two-step process: (1) identifying a low-dimensional latent representation of the high-dimensional state and (2) learning the temporal evolution of the latent variables.  Implicit neural representation (INR)-based latent dynamics models offer a space-time mesh-free approach for reduced-order modeling of parameterized PDE systems. 

The dimensionality reduction stage approximates the PDE solution as \begin{equation}
    \pmb{q}(\pmb{x}, t, \pmb{\nu}) \approx \pmb{q}^m (\tilde{\pmb{q}} (t, \pmb{\nu}), \pmb{x}),
    \label{eq:sol_approx_deterministic}
\end{equation}
where $\pmb{q}^m$ is the modeled approximation of the PDE solution and $\tilde{\pmb{q}}: \mathcal{T} \times \mathcal{V} \to \mathbb{R}^r$ is the $r$-dimensional latent vector. The latent vector $\tilde{\pmb{q}}(t, \pmb{\nu})$ is typically obtained through linear approaches, such as proper orthogonal decomposition \cite{Aubry1988}, or nonlinear approaches, such as auto-encoder-decoder \cite{Fries2022} or auto-decoder neural network representations \cite{Wen2023, Puri2025, Prakash2026}. In this article, we consider the auto-decoder neural network representation, where the deterministic decoder is taken as $\pmb q^m:\mathbb{R}^r\times\mathcal X\to\mathbb{R}^m$, such that \(\pmb q^m(\tilde{\pmb q}(t,\pmb\nu),\cdot)\in\mathcal Q\). The image of this map defines the latent approximation manifold
\begin{equation}
    \mathcal M_r
    =
    \left\{
    \pmb q^m(\pmb z,\cdot):\pmb z\in\mathbb{R}^r
    \right\}\subset\mathcal Q.
\end{equation}
The representation is trained so that the state-approximation error, for example \(\|\pmb q(\cdot,t,\pmb\nu)-\pmb q^m(\tilde{\pmb q}(t,\pmb\nu),\cdot)\|_{\mathcal Q}\), is small over the sampled trajectories.

The low dimensionality of the latent vector enables inexpensive propagation of dynamics. For a fixed parameter instance $\pmb{\nu}$, the equation governing these dynamics of the latent vector is written as an autonomous ordinary differential equation (ODE) of the form
\begin{equation}
    \frac{d \tilde{\pmb{q}}}{d t} = \pmb{g} (\tilde{\pmb{q}}; \pmb{\nu}),
    \label{eq:NODE_ls}
\end{equation}
where $\pmb{g}(\cdot; \pmb{\nu})$ is a nonlinear function of the latent vector. We assume that $\pmb{g}$ is sufficiently regular for the latent initial-value problem to be well defined.

Recent work has explored physics-informed variants of latent dynamics approaches that allow for fine-tuning predictions during forecasting \cite{Wen2023} and improved robustness in the presence of sparse and noisy measurements \cite{Prakash2026}. Although latent-dynamics models can accurately reproduce solution trajectories, neither the latent evolution model nor the decoder is generally constrained to satisfy the conservation law in Eq.~\eqref{eq:consv_law}. Consequently, the reconstructed solution is not guaranteed to lie on the solution manifold defined in Eq.~\eqref{eq:sol_manifold_def}, and the desired conservation properties may not be preserved. In conservation-preserving constructions, one instead seeks a decoder or constrained representation whose reconstructed trajectories satisfy \(\mathcal R(\pmb q^m;\pmb\nu)=0\), so that the reduced representation is consistent with the admissible conservation-law manifold. Recent work \cite{Prakash2026} proposed a structure-preserving latent-dynamics framework that embeds the nonlinear conservation law directly into the solution representation, thereby guaranteeing exact conservation. This study also showed enhanced predictive accuracy and model training efficiency when using such conservation-preserving latent representations. Additional theoretical details and comparative analyses are provided in \cite{Prakash2026}.

\section{Probabilistic modeling of latent dynamics solutions}
\label{sec:Section3}

While latent dynamics models provide an alternative to high-fidelity simulations for multi-query approximations, the accuracy of these methods is constrained by the amount of data available and the approximation methods used. Specifically, these models are trained on a finite sample of parameter instances drawn from the full parameter space $\mathcal{V}$, and their performance degrades when evaluated on parameter combinations that differ substantially from those observed during training. In particular, the pointwise prediction error between the high-fidelity solution and the latent dynamics models, given as
\begin{equation}
    \epsilon(\pmb{x}, t, \pmb{\nu}) = \vert \vert \pmb{q} (\pmb{x}, t, \pmb{\nu}) - \pmb{q}^m (\pmb{x}, t, \pmb{\nu}) \vert \vert_2,
\end{equation}
where $\vert \vert \cdot \vert \vert_2$ denotes the Euclidean norm in $\mathbb{R}^m$. Theoretically, the universal approximation theorem for neural networks \cite{Cybenko1989, Hornik1989} establishes that there exists a neural network architecture that would provide highly accurate solutions with $\epsilon \to 0$ if trained on full population data. This theorem is fundamentally an existence result and does not provide guarantees regarding optimization, sample complexity, or generalization performance when training is performed using a finite dataset.
In practice, generating such a large dataset would be impractical since conducting a high-fidelity simulation or experiment for each system parameter instance is computationally expensive. 

Let $\Omega$ represent the space of all possible training datasets obtained by sampling parameter instances from $\mathcal{V}$. We equip $\Omega$ with a $\sigma-$algebra $\mathcal{F}$ and a probability measure $P$ induced by the sampling strategy, for example uniform random sampling over $\mathcal{V}$. Here, $\pmb{\omega} \in \Omega$ denotes a particular realization of the training dataset, comprising specific parameter-solution pairs $\{ (\pmb{\nu}^{(i)}, \pmb{q}^{(i)})\}_{i=1}^{N_{\text{train}}}$, where $N_{\text{train}}$ is the number of training samples drawn from the full population. The probability space $(\Omega, \mathcal{F}, P)$ therefore characterizes uncertainty associated with finite training-data selection. The learned approximate solution depends on this probabilistic training dataset selection and is redefined as $\pmb{q}^m (\pmb{x}, t, \pmb{\nu} ; \pmb{\omega})$. Therefore, $\pmb{q}^m (\pmb{x}, t, \pmb{\nu} ; \cdot) : \Omega \to \mathbb{R}^m$ is $\mathcal{F}-$measurable for each $(\pmb{x}, t, \pmb{\nu})$, and the map $(\pmb{x}, t, \pmb{\nu} ; \pmb{\omega}) \to \pmb{q}^m (\pmb{x}, t, \pmb{\nu} ; \pmb{\omega})$ defines a random field on $\mathcal{X} \times \mathcal{T} \times \mathcal{V} \times \Omega$. This probabilistic definition of modeled solution field enables quantification of uncertainty in model predictions due to limited data, which is introduced through sparse selection of system parameter instances in this article.

\subsection{Modeling random solution fields using variational latent neural fields}

Given this definition of the random solution field, the solution at each spatio-temporal and system parameter instance $\pmb{q}^m(\pmb{x} = \pmb{x}_0, t = t_0, \pmb{\nu} = \pmb{\nu}_0; \pmb{\omega})$ is a random variable. This random variable can have an arbitrary multimodal distribution. Instead of directly representing this multimodal distribution using parametric or nonparametric modeling methods, we employ ideas from modern generative models to draw samples from the approximated solution distribution. While drawing samples from this distribution may not provide us with an explicit probability distribution, it allows us to estimate the population statistical moments through repeated sampling. In this article, we motivate the proposed generative model from theoretical underpinnings of variational autoencoders (VAEs) \cite{Kingma2019} and extend them for modeling solution fields. Unlike existing latent neural field approaches \cite{Du2024, Park2024, Lan2026} that are designed to represent spatio-temporal solutions, the proposed approach is developed specifically for handling parameterized systems.


The random solution field is expressed using an INR-based representation,
\begin{equation}
    \pmb{q}^m(\pmb{x}, t, \pmb{\nu}; \pmb{\omega}) = \pmb{d}_{\theta} (\tilde{\pmb{q}} (t, \pmb{\nu}; \pmb{\omega}), \pmb{x})
    \label{eq:VLNF_majorformula}
\end{equation}
where $\pmb{d}_{\theta} : \mathbb{R}^r \times \mathcal{X} \to \mathbb{R}^m$ is a neural network decoder with parameters $\theta$. The latent vector is a random field represented as
\begin{equation}
    \tilde{\pmb{q}} (t, \pmb{\nu}) \sim \mathcal{N}\Big(\pmb{\mu} (t, \pmb{\nu}), \sigma^2 (\pmb{\nu}) \pmb{I} \Big).
    \label{eq:latentvec_def}
\end{equation}
where $\pmb{\mu}: \mathcal{T} \times \mathcal{V} \to \mathbb{R}^r$ is the mean function, $\sigma^2: \mathcal{V} \to \mathbb{R}_+$ determines the variance, and $\pmb{I} \in \mathbb{R}^{r \times r}$ is the identity matrix. In the definition of the latent vector, $\pmb{\mu} (t, \pmb{\nu})$ and $\sigma^2 (\pmb{\nu})$ can be chosen as an MLP hypernetwork. However, with this selection, there is no mechanism to differentiate whether the system parameter are in-distribution or out-of-distribution. This distinction is important as we would expect in-distribution system parameters to have a low uncertainty, while exhibiting a high uncertainty when evaluated for out-of-distribution scenarios. To account for correlation between neighboring parameter instances, we replace the independent Gaussian latent-state model by a Gaussian-process prior whose covariance structure is determined by distances in parameter space. This Gaussian process prior for $\tilde{\pmb{q}}$ is expressed as
\begin{equation}
    \tilde{\pmb{q}}(t, \pmb{\nu}) \sim \mathcal{GP}\Big(\pmb{\mu}(t, \pmb{\nu}), \mathcal{K}(\pmb{\nu}; \pmb{\nu}')\Big),
\end{equation}
where $\mathcal{K}(\pmb{\nu}; \pmb{\nu}')$ is the covariance kernel. In the present work, correlations are modeled only across parameter space through the covariance kernel, while time enters as a deterministic conditioning variable through the mean function $\pmb{\mu}(t, \pmb{\nu})$. We consider the squared exponential covariance kernel defined as
\begin{equation}
    \mathcal{K}( \pmb{\nu}; \pmb{\nu}') = \sigma_f^2 \exp\Big(-\sum_{i=1}^{d_{\nu}} \frac{\vert \vert \nu_i - \nu_i' \vert \vert^2}{2 l_i^2}  \Big),
\end{equation}
where $l_i$ is the \textit{i}th component of the correlation length vector $\pmb{l} \in \mathbb{R}^{d_{\nu}}$ and $\sigma_f$ is the signal variance. The length-scale parameters $\{l_i\}_{i=1}^{d_{\nu}} \}$ control the rate of correlation decay along each parameter dimension, thereby encoding prior assumptions about smoothness and variability of the latent representation across $\mathcal{V}$. While not studied in this article, other commonly used covariance kernels for Gaussian processes \cite{Rasmussen2005} can also be used in the formulation. The likelihood of the latent vector is 
\begin{equation}
    \pmb{o}(t, \pmb{\nu}) = \tilde{\pmb{q}}(t, \pmb{\nu}) + \pmb{\epsilon}, \quad \pmb{\epsilon} \sim \mathcal{N}(0, \sigma_n^2 \pmb{I}),
\end{equation}
where $\pmb{o}(\cdot, \cdot)$ are the observations of the latent vector assumed to be the true latent vector values with observation noise parameter $\sigma_n$. In this article, we use superscript $(i)$, where $i \in Z^+$ to denote the training data instance. The training system parameters are collected in a vector $\pmb{V} = [\pmb{\nu}^{(1)} \; \pmb{\nu}^{(2)} \; \cdot \cdot \cdot \; \pmb{\nu}^{(N)}]^T \in \mathbb{R}^{N \times d_{\nu}}$, where $N$ is the number of training system parameters. Therefore, the resulting posterior for the latent vector is 
\begin{equation}
    \tilde{\pmb{q}} (\pmb{\nu}) \Big\vert \pmb{o}, \pmb{V} \sim \mathcal{N}(\pmb{\mu}_{GP} (\pmb{\nu}), \pmb{\Sigma}_{GP} (\pmb{\nu})),
\end{equation}
where the posterior distribution is conditioned on the observed latent vectors at training locations $\pmb{V}$. The posterior mean function is given as
\begin{equation}
    \pmb{\mu}_{GP} (\pmb{\nu}) = \pmb{K}(\pmb{\nu}, \pmb{V}) [\pmb{K}(\pmb{V}, \pmb{V}) + \sigma_n^2 \pmb{I} ]^{-1} \pmb{O},
\end{equation}
and the posterior covariance matrix is given as
\begin{equation}
    \pmb{\Sigma}_{GP} (\pmb{\nu}) = \sigma^2_{GP} (\pmb{\nu}) \pmb{I}, \quad  \sigma^2_{GP} (\pmb{\nu}) = \pmb{K}(\pmb{\nu}, \pmb{\nu}) - \pmb{K}(\pmb{\nu}, \pmb{V})[\pmb{K}(\pmb{V}, \pmb{V}) + \sigma_n^2 \pmb{I} ]^{-1} \pmb{K}(\pmb{V}, \pmb{\nu}),
    \label{eq:gp_var}
\end{equation}
where $\pmb{K}(\cdot,\cdot)$ represents the covariance matrix constructed using the covariance kernel $\mathcal{K}$ and $\pmb{O} = [\pmb{o}^{(1)} \; \pmb{o}^{(2)} \; \cdot \cdot \cdot \pmb{o}^{(N)}]^T \in \mathbb{R}^{N \times r}$ is the observation vector. The posterior covariance is approximated as $\pmb{\Sigma}_{GP} \approx \sigma_{GP}^2 \pmb{I}$, which is a diagonal matrix. This diagonal approximation is motivated by the assumption that latent vector components are approximately independent after conditioning on observations, as is often employed in variational inference methods \cite{Kingma2019} for scalable training. The Gram matrix $\pmb{K}(\pmb{V}, \pmb{V}) \in \mathbb{R}^{N \times N}$ encodes pairwise correlations between all training parameter instances, and its regularized inverse conditions the prior on the observed data. The posterior covariance matrix $\pmb{\Sigma}_{GP} (\pmb{\nu})$ decreases near training points and increases in regions poorly covered by training data, thereby providing an intrinsic measure of model confidence across the parameter space. While this Gaussian process representation for the latent vector provides the attractive feature of accounting for uncertainties in out-of-distribution system parameters, this step is associated with a large computational cost. Notably, the cost of computing $[\pmb{K}(\pmb{V},\pmb{V}) + \sigma_n^2 \pmb{I}]^{-1}$ is $\mathcal{O}(N^3)$ and this matrix consumes $\mathcal{O}(N^2)$ memory for storage. Furthermore, even when $[\pmb{K}(\pmb{V},\pmb{V}) + \sigma_n^2 \pmb{I}]^{-1}$ is computed, the cost of evaluating $\pmb{K}(\pmb{\nu}, \pmb{V})[\pmb{K}(\pmb{V}, \pmb{V}) + \sigma_n^2 \pmb{I} ]^{-1} \pmb{K}(\pmb{V}, \pmb{\nu})$ is $\mathcal{O}(N^2)$ which makes Gaussian processes prohibitively expensive during the inference stage when a large number of training samples are used.
In most scenarios, the hyperparameters for these Gaussian processes $(\sigma_f, \pmb{l}, \sigma_n)$ cannot be determined intuitively and therefore, require optimization of maximum likelihood estimates. If such a Gaussian process were to be integrated into this framework, training such a model would be prohibitively expensive as each training epoch would involve at least $\mathcal{O}(N^3)$ operations for estimating the posterior distribution of the latent vector.

We use a separate neural network to approximate $\sigma^2_{GP}$ and $\pmb{\mu}_{GP}$ to amortize the high computational cost of estimating Gaussian process posterior. This posterior covariance is modeled using a Gaussian process variance network (GPVN), $f^{\Sigma-GP}_{\pmb{\eta}}: \mathcal{V} \times \mathbb{R}^{d_{\nu}+2} \to \mathbb{R}$, such that
\begin{equation}
     \sigma^2_{GP} (\pmb{\nu}; \pmb{\tau}) \approx \exp \Big(f^{\Sigma-GP}_{\pmb{\eta}} (\pmb{\nu}; \pmb{\tau}_{GP})\Big)
\end{equation}
where an MLP network uses $\pmb{\tau}_{GP} = [\sigma_n, \pmb{l}, \sigma_f] \in \mathbb{R}^{ d_{\nu} + 2}$ as the hyperparameter vector for the posterior covariance matrix. We learn logarithm of posterior covariance for improved conditioning and preserve positivity of the predicted posterior covariance. GPVN is trained by evaluating \eref{gp_var} on randomly sampled values in the convex hull of $\pmb{\nu}$ and $\pmb{\tau}_{GP}$. Therefore, parameters of GPVN ($\pmb{\eta}$) are obtained by solving a nonlinear regression problem
\begin{equation}
J(\pmb{\eta}) = \underset{i,j}{\mathbb{E}} \Big( \sigma_{GP}^2 (\pmb{\nu}^{(i)}; \pmb{\tau}_{GP}^{(j)}) - \exp \Big(f^{\Sigma-GP}_{\pmb{\eta}} (\pmb{\nu}^{(i)}; \pmb{\tau}_{GP}^{(j)}\Big) \Big)^2,
\label{eq:reg_GPVN}
\end{equation}
where expectations are taken over the training samples generated by random sampling within the expected hyperparameter ranges. Similarly, we approximate the Gaussian process mean function as the Gaussian process mean network (GPMN), defined as
\begin{equation}
\pmb{\mu}_{GP} (t, \pmb{\nu}) \approx   \pmb{f}^{\mu-GP}_{\pmb{\eta}'} (t, \pmb{\nu}),
\label{eq:GPMN_def}
\end{equation}
where $\pmb{f}_{\pmb{\eta}'}^{\mu-GP} : \mathcal{T} \times \mathcal{V} \to \mathbb{R}^r$ is a neural network with parameters $\pmb{\eta}'$. Since the Gaussian process mean function does not provide information about the variance of the latent vector, it is trained in conjunction with the decoder representation in \eref{VLNF_majorformula}. With these neural network approximations of Gaussian process mean and variance functions, the resulting latent vector is written as
\begin{equation}
    \tilde{\pmb{q}}(t, \pmb{\nu}) \sim \mathcal{N}\Big(\pmb{f}^{\mu-GP}_{\pmb{\eta}'} (t, \pmb{\nu}), \exp \Big(f^{\Sigma-GP}_{\pmb{\eta}} (\pmb{\nu}; \sigma_n, \pmb{l}, \sigma_f) \Big) \pmb{I} \Big).
    \label{eq:latentvec_final}
\end{equation}
The resulting latent-state distribution may be interpreted as a variational approximation of the Gaussian process posterior, where both the posterior mean and posterior variance are represented through neural network surrogates.
This latent vector formulation is used in conjunction with the decoder network in \eref{VLNF_majorformula} to approximate the random solution field. This representation of the latent vector combined with the neural field definition in \eref{VLNF_majorformula} is referred to as a variational latent neural field (VLNF). 

\subsection{Training variational latent neural fields}

The latent dynamics modeling framework in \eref{VLNF_majorformula} and \eref{latentvec_final} involves unknown model parameters that must be determined based on available high-fidelity data. We use a training procedure similar to that used for VAEs. This model can be learned by maximizing the log-likelihood of the observed solution distribution. However, log-likelihood is expensive to compute directly as marginal log likelihood estimation is intractable due to the integration over the latent space. Evidence lower-bound (ELBO) \cite{Kingma2019}, derived using Jensen's inequality and an approximate posterior distribution for $\tilde{\pmb{q}}$, provides a tractable lower bound that can be optimized efficiently using stochastic gradient methods. The loss function based on ELBO is given as
\begin{equation}
    \mathcal{L}_{\text{ELBO}} (\pmb{\theta}, \pmb{\eta}' \vert \pmb{q}, \pmb{x}, t, \pmb{\nu}) = - \mathbb{E}_{\tilde{\pmb{q}} \sim q_{\pmb{\eta'}} (\tilde{\pmb{q}} \vert t, \pmb{\nu})} \Big[ \mathbb{E}_{\pmb{x} \sim \mathcal{X}} \Big[\text{log} \; p_{\pmb{\theta}} (\pmb{q} \vert \tilde{\pmb{q}}, \pmb{x})\Big] + \mathcal{D}_{KL} \Big( q_{\pmb{\eta}'}(\tilde{\pmb{q}} \vert t, \pmb{\nu}) \vert \vert p(\tilde{\pmb{q}})\Big),
\end{equation}
where $p_{\pmb{\theta}} (\pmb{q} \vert \tilde{\pmb{q}}, \pmb{x}) $ is the likelihood of the solution given the latent vector and spatial location (modeled by the decoder $\pmb{d}_{\pmb{\theta}}$), $q_{\pmb{\eta}'} (\tilde{\pmb{q}} \vert t, \pmb{
\nu})$ is the approximate posterior distribution given by \eref{latentvec_final}, and $p(\tilde{\pmb{q}}) = \mathcal{N}(0, 1)$ is the standard normal prior. $\mathcal{D}_{KL}$ is the KL-divergence between the approximate posterior latent vector distribution and the prior distribution. The first term encourages the model to reconstruct observed solutions accurately, while the second term regularizes the approximate posterior to remain close to the prior, preventing overfitting and ensuring that the latent representation does not collapse to a deterministic encoder. 

For computational tractability, we assume a Gaussian likelihood $p_{\pmb{\theta}} (\pmb{q} \vert \tilde{\pmb{q}}, \pmb{x}) = \mathcal{N}(\pmb{q} \vert \pmb{d}_{\pmb{\theta}} (\tilde{\pmb{q}}, \pmb{x}), \sigma^2_{obs} \pmb{I})$, where $\sigma^2_{\text{obs}}$ is the observation noise that is absorbed into the loss scaling. Under this assumption, maximizing the expected log-likelihood is equivalent, up to an additive constant and scaling factor, to minimizing a mean-squared reconstruction error. The KL divergence term is derived by noting that for Gaussian distributions $q_{\pmb{\eta}'} (\tilde{\pmb{q}} \vert t, \pmb{\nu}) = \mathcal{N}(\pmb{\mu}, \sigma^2 \pmb{I})$ and $p(\tilde{\pmb{q}}) = \mathcal{N}(\pmb{0}, \pmb{I})$, the KL divergence factorizes as
\begin{equation}
    \mathcal{D}_{KL} (q_{\eta'} \vert \vert p) = \frac{1}{2} \sum_{l=1}^r \Big( \sigma^2 + \mu_l^2 - 1 - \text{log} \sigma^2 \Big).
\end{equation}
For training VLNFs, we have $\mu_l = f^{\mu-GP}_{\pmb{\eta}', l} (t^{(i)}, \pmb{\nu}^{(j)})$ and $\sigma^2 = \exp \Big(f^{\Sigma-GP}_{\pmb{\eta}} (\pmb{\nu}^{(j)}; \pmb{\tau}_{GP}) \Big)$. With these substitutions, the resulting training objective is
\begin{multline}
    J(\pmb{\theta}, \pmb{\eta}', \pmb{\tau}_{GP}) = \underbrace{\underset{i,j,k}{\mathbb{E}} \Big\vert \Big\vert \pmb{q} (\pmb{x}^{(i)}, t^{(j)}, \pmb{\nu}^{(k)}) - \pmb{d}_{\pmb{\theta}} (\tilde{\pmb{q}} (t^{(j)}, \pmb{\nu}^{(k)}; \pmb{\omega}), \pmb{x}^{(i)}) \Big\vert \Big\vert^2}_{\text{Reconstruction-loss}} + \\ \underbrace{\frac{\beta}{2} \; \sum_{k=1}^r \underset{i,j}{\mathbb{E}}  \Big(\exp \Big(f^{\Sigma-GP}_{\pmb{\eta}} (\pmb{\nu}^{(j)}; \pmb{\tau}_{GP}) \Big)+ [f^{\mu-GP}_{\pmb{\eta}', k} (t^{(i)}, \pmb{\nu}^{(j)})]^2 - 1 - f^{\Sigma-GP}_{\pmb{\eta}} (\pmb{\nu}^{(j)}; \pmb{\tau}_{GP}) \Big)}_{\text{Latent vector KL divergence}},
    \label{eq:IRS-UQ_optimization}
\end{multline}
where $\beta$ is the regularization parameter for penalizing the KL divergence term, similar to those used in $\beta-$VAEs \cite{Higgins2017}. During the training, the reparameterization trick \cite{Kingma2019} is employed to separate stochastic sampling from the trainable network parameters, thereby enabling use of standard gradient-based optimization and backpropagation. 

The parameters of the GPVN ($\pmb{\eta}$) are trained at an earlier stage by solving the regression problem in \eref{reg_GPVN}. This pre-training phase is performed offline before the main VLNF training in \eref{IRS-UQ_optimization} and requires only evaluation of the analytical Gaussian-process posterior variance formula on a representative sample of parameter and hyperparameter combinations. On the other hand, the Gaussian process hyperparameters used in GPVN, namely $\sigma_n$, $\pmb{l}$ and $\sigma_f$, are initially unknown and are jointly optimized with the decoder parameters $\pmb{\theta}$ and GPMN parameters $\pmb{\eta}'$ when minimizing the objective function in \eref{IRS-UQ_optimization}. This joint training procedure ensures that the learned mean function $\pmb{f}^{\mu-GP}_{\pmb{\eta}'}$ is optimized for producing solution fields that match the training data when decoded through $\pmb{d}_{\theta}$.

\subsection{Reduced state dynamics modeling for random solution fields}
\label{sec:CommonMethod}

In the previous section, we defined the framework for probabilistic modeling of solution fields, which provides a representation of the latent vector that is a function of temporal coordinates. However, as mentioned in \cite{Prakash2026}, such a functional representation of the latent vector is undesirable for applications involving forecasting problems, as the evaluation of mean function $\pmb{f}_{\pmb{\eta}'}^{\mu-GP} (t, \pmb{\nu})$ for extrapolation to $t > t_{\text{max}}$ (where $t_{\text{max}}$ is the maximum training time) can lead to unreliable predictions due to the lack of an underlying dynamical model. In order to overcome this issue, we instead learn an ODE to govern the dynamics of the GPMN. This approach enables us to provide an estimate of the mean function for scenarios involving time-forecasting, offering more robustness compared to modeling the solution fields directly, as the learned ODE encodes the temporal evolution structure rather than relying on pure interpolation. Therefore, we learn a neural ODE to represent the evolution of the GPMN as 
\begin{equation}
    \frac{d \pmb{f}^{\mu-GP}_{\pmb{\eta}'}}{d t} (t, \pmb{\nu}) = \pmb{g}_{\pmb{\psi}} (\pmb{f}^{\mu-GP}_{\pmb{\eta}'} (t, \pmb{\nu}) , \pmb{\nu}),
\end{equation}
where $\pmb{g}_{\pmb{\psi}} : \mathbb{R}^r \times \mathcal{V} \to \mathbb{R}^r$ is a neural network with parameters $\pmb{\psi}$ that approximates the right-hand side of the ODE governing the latent dynamics. This neural ODE is learned by solving a nonlinear regression problem with the cost function
\begin{equation}
    J_{\text{ODE}} (\pmb{\psi}) = \underset{i,j}{\mathbb{E}} \Big\vert \Big\vert \frac{d \pmb{f}^{\mu-GP}_{\pmb{\eta}'}}{d t} (t^{(i)}, \pmb{\nu}^{(j)}) - \pmb{g}_{\pmb{\psi}} (\pmb{f}^{\mu-GP}_{\pmb{\eta}'} (t^{(i)}, \pmb{\nu}^{(j)}), \pmb{\nu}) \Big\vert \Big\vert^2_2,
    \label{eq:latent_node}
\end{equation}
where the training values for $d \pmb{f}^{\mu-GP}_{\pmb{\eta}'} (t, \pmb{\nu})/d t$ are obtained using automatic differentiation with respect to $t$ applied to this continuous function. Once this neural ODE is trained, the Gaussian process mean for forecasting or inference at arbitrary time $t$ is obtained by integrating the ODE as
\begin{equation}
    \bar{\pmb{f}}^{\mu-GP}_{\pmb{\eta}'}(t, \pmb{\nu}) = \pmb{f}^{\mu-GP}_{\pmb{\eta}'}(t_0, \pmb{\nu}) + \int_{t_0}^t \pmb{g}_{\pmb{\psi}} (\pmb{f}^{\mu-GP}_{\pmb{\eta}'} (t', \pmb{\nu}), \pmb{\nu}) dt',
    \label{eq:NODE_GPMN_Fbar}
\end{equation}
where $t_0$ is the initial time (typically $t_0 = 0$) and $\pmb{f}^{\mu-GP}_{\pmb{\eta}'}(t_0, \pmb{\nu})$ provides the initial condition for the ODE, which can be obtained by evaluating the trained GPMN at the initial time. Note that $\bar{\pmb{f}}^{\mu-GP}_{\pmb{\eta}'}(t, \pmb{\nu})$ denotes the latent trajectory generated by the learned vector field $\pmb{g}_{\pmb{\psi}}$, which could differ slightly from the original trajectory due to time integration errors introduced by the numerical schemes. Using this Gaussian process mean, the final resulting approximation of the latent vector is obtained by sampling from the latent distribution,
\begin{equation}
    \tilde{\pmb{q}}(t, \pmb{\nu}) \sim \mathcal{N}(\bar{\pmb{f}}_{\pmb{\eta'}}^{\mu-GP}, \exp\Big( f_{\pmb{\eta}}^{\Sigma-GP}(t, \pmb{\nu}; \pmb{\tau}_{GP}) \Big)\pmb{I}\Big),
\end{equation}
to get a realization of $\tilde{\pmb{q}}$, denoted as $\tilde{\pmb{q}}_{\omega}$. This realization is then used with the decoder $\pmb{d}_{\pmb{\theta}}$ to provide a realization of the resulting solution field
\begin{equation}
    \pmb{q}^m(\pmb{x}, t, \pmb{\nu}; \pmb{\omega}) = \pmb{d}_{\pmb{\theta}} (\tilde{\pmb{q}}_{\omega} (t, \pmb{\nu}), \pmb{x}).
\end{equation}
By generating multiple realizations through repeated sampling, we can estimate the mean $\mathbb{E}_{\pmb{\omega}} [\pmb{q}^m(\pmb{x}, t, \pmb{\nu}; \pmb{\omega})]$ and variance $\text{Var}_{\pmb{\omega}} [\pmb{q}^m(\pmb{x}, t, \pmb{\nu}; \pmb{\omega})]$ of the predicted solution field. This approach does not incorporate any physical constraints in the model construction, and the resulting random field is obtained solely based on data. In this article, we refer to this method as IRS-UQ, standing for Identified Reduced System with Uncertainty Quantification, as it can be considered as a probabilistic extension of the IRS method proposed in \cite{Prakash2026}. 

Note that if we exclude this low-dimensional dynamics identification step, this proposed variational latent neural field framework extends latent neural operators, such as latent DeepONets \cite{Kontolati2023}, to provide uncertainty estimates in the predicted solution. This exclusion of the ODE-based temporal propagation model, implies that the GPMN is evaluated directly at each query time and the framework reduces to a static latent neural operator that does not enforce a dynamical evolution model. This static variant retains the uncertainty-quantification capability but sacrifices the temporal extrapolation ability provided by the learned ODE vector field.

\section{Physics-informed probabilistic modeling of latent dynamics solutions}
\label{sec:Section4}

IRS-UQ does not incorporate any physical insights into the formulation and may struggle in scenarios where the available measurements are not of high quality. In particular, the results in \cite{Prakash2026} showed that if IRS is trained using sparse and noisy data, then the quality of predictions decreases substantially. We expect similar observations to hold for the IRS-UQ approach when learning solution fields from low-quality measurements, as purely data-driven models may admit multiple latent representations that explain the observed data equally well. Incorporating physical constraints reduces this admissible solution space by introducing additional information derived from the governing equations. 

A physics-informed version of IRS-UQ, denoted as PI-IRS-UQ, can be formulated by adding the nonlinear conservation law constraint to the optimization problem. The simplest approach involves adding this constraint as a soft-penalty term in the optimization objective function, similar to the strategy followed in the widely used physics-informed neural networks (PINNs) \cite{Raissi2019}. The resulting objective function is given as
\begin{multline}
    J(\pmb{\theta}, \pmb{\eta}', \pmb{\tau}_{GP}) = \underbrace{\underset{i,j,k}{\mathbb{E}} \Big\vert \Big\vert \pmb{q} (\pmb{x}^{(i)}, t^{(j)}, \pmb{\nu}^{(k)}) - \pmb{d}_{\pmb{\theta}} (\tilde{\pmb{q}} (t^{(j)}, \pmb{\nu}^{(k)}; \pmb{\omega}), \pmb{x}^{(i)}) \Big\vert \Big\vert^2}_{\text{Reconstruction-loss}} +  \\ \underbrace{\frac{\beta}{2} \; \sum_{k=1}^r \underset{i,j}{\mathbb{E}}  \Big(\exp \Big(f^{\Sigma-GP}_{\pmb{\eta}} (\pmb{\nu}^{(j)}; \pmb{\tau}_{GP}) \Big)+ [f^{\mu-GP}_{\pmb{\eta}', k} (t^{(i)}, \pmb{\nu}^{(j)})]^2 - 1 - f^{\Sigma-GP}_{\pmb{\eta}} (\pmb{\nu}^{(j)}; \pmb{\tau}_{GP}) \Big)}_{\text{Latent vector KL divergence}} +  \\ \underbrace{\lambda \underset{i,j,k}{\mathbb{E}} \Big\vert \Big\vert \frac{\partial \pmb{d}_{\pmb{\theta}} (\tilde{\pmb{q}} (t^{(j)}, \pmb{\nu}^{(k)}; \pmb{\omega}), \pmb{x}^{(i)})}{\partial t} + \nabla \cdot \pmb{f}( \pmb{d}_{\pmb{\theta}} (\tilde{\pmb{q}} (t^{(j)}, \pmb{\nu}^{(k)}; \pmb{\omega}), \pmb{x}^{(i)}); \pmb{\nu}^{(k)}) \Big\vert \Big\vert^2}_{\text{Conservation loss}},
    \label{eq:PI-IRS-UQ_optimization}
\end{multline}
where $\lambda > 0$ is the penalty parameter that governs the enforcement of physical constraint. A larger value of $\lambda$ implies a stricter imposition of the conservation law constraint, while a smaller value results in weaker enforcement. In practice, $\lambda$ often needs to be tuned through cross-validation or heuristic search, and its optimal value can vary significantly across different problems and data quality scenarios \cite{Prakash2026}. This tuning cost
can be prohibitive in large-scale applications.

Once the network architecture is learned, the ODE governing the evolution of the Gaussian process mean is learned similar to IRS-UQ as described in Section 3.3. With this combined formulation, PI-IRS-UQ enables probabilistic modeling of solution fields while utilizing nonlinear conservation laws to improve the robustness for scenarios where measurements are of lower quality. However, as the penalty formulation imposes the nonlinear conservation laws as a soft constraint, there is no guarantee that the conservation is satisfied for spatio-temporal locations and system parameter instances that are not included in the training set. 

\section{Nonlinear conservation law-embedded probabilistic modeling of latent dynamics solutions}
\label{sec:Section5}

The two strategies discussed above do not ensure nonlinear conservation laws are exactly satisfied, especially for spatio-temporal locations and system parameters that are unseen during the model learning process. The results in \cite{Prakash2026} showed that latent dynamics models that exactly satisfy nonlinear conservation laws provide superior accuracy for scenarios with sparse and noisy measurements. Furthermore, these methods are more robust than physics-informed approaches as the latter require careful tuning of the penalty parameters. 

While ECLEIRS provides accurate solution representations even in the presence of sparse and noisy measurements, the method is limited to deterministic solutions and does not provide uncertainties in solution predictions. In this work, we extend ECLEIRS to provide such uncertainty estimates. Similar to IRS-UQ and PI-IRS-UQ, instead of modeling solutions as deterministic fields, we treat the modeled solution as random fields. In particular, these random fields are constructed such that each realization of the solution lies on a lower-dimensional manifold that exactly satisfies nonlinear conservation laws. 

In Section \ref{sec:ECLEIRS}, we discuss a recently developed method in \cite{Prakash2026} for representing deterministic solution fields that satisfy the nonlinear conservation constraints. In Section \ref{sec:ProbRep}, we define random fields that satisfy the nonlinear conservation constraint at each realization. In Section \ref{sec:ECLEIRSUQ}, we formulate a method for modeling random fields that are constrained by nonlinear conservation laws.

\subsection{Deterministic representation of solution constrained to a manifold}
\label{sec:ECLEIRS}

We first discuss a deterministic representation of solutions that exactly satisfy the nonlinear conservation law constraint. We simplify the notation by deriving the method for scalar fields $q \in \mathbb{R}$. The extension to vector-valued solutions $\pmb{q} \in \mathbb{R}^m$ follows analogously by applying the construction component-wise. Nonlinear conservation laws defined in \eref{consv_law} can be rewritten as
\begin{equation}
    \mathcal{R}(\pmb{z}; \pmb{\nu}) = \nabla_{w} \cdot \pmb{z} (\pmb{w}; \pmb{\nu}) = 0
    \label{eq:constraint_consv}
\end{equation}
where $\nabla_w = \frac{\p}{\p t} \hat{t} + \sum_{i=1}^3 \frac{\p}{\p x_i} \hat{x}_i$ is the divergence with respect to the space-time vector $\pmb{w} = [t, \;\pmb{x}]^T \in \mathcal{XT} \subset \mathbb{R}^{d+1}$ and $\pmb{z}(\pmb{w}; \pmb{\nu}) = [q(\pmb{w}, \pmb{\nu}), \; \pmb{f}(q(\pmb{w}, \pmb{\nu}); \pmb{\nu})] : \mathcal{XT} \times \mathcal{V} \to \mathbb{R}^{d+1}$ is the lifted solution-flux vector. This reformulation of the nonlinear conservation laws implies that the solution-flux vector satisfies a space-time divergence-free condition. Based on this definition, the goal is to identify a functional form of deterministic neural fields that lies within the nonlinear conservation law constrained solution space defined in \eref{sol_manifold_def}. For low-dimensional space-time systems ($d + 1 = \{2, 3\}$), classical vector calculus identities enable representation of deterministic divergence-free vector fields $\pmb{z}$ as
\begin{equation}
    \pmb{z}(\pmb{x}, t, \pmb{\nu}) = \nabla_w \times \pmb{a} (\pmb{w}, \pmb{\nu}), \quad \text{and satisfy} \quad \nabla_w \cdot \nabla_w \times \pmb{a} (\pmb{w}, \pmb{\nu}) = 0,
    \label{eq:vec_iden_lowdim}
\end{equation}
where $\pmb{a}: \mathcal{XT} \times \mathcal{V} \to \mathbb{R}^{d+1}$ is the vector potential. This representation guarantees conservation identically because the divergence-free property follows from the differential operator itself. For a higher-dimensional vector field $d + 1 > 3$, which is the case for three-dimensional spatial problems (that is 3D space plus time yields a 4D space-time), the above vector calculus identity does not directly apply. Instead, for such high-dimensional deterministic vector fields, similar identities were identified in \cite{Barbarosie2011, Kelliher2021} using ideas from exterior calculus \cite{Spivak1965, Lee2003}, and were subsequently used in the context of PINNs to impose a mass conservation structure in \cite{Richter2022}. In \cite{Prakash2026}, this identity was used to impose nonlinear conservation law structure when modeling deterministic solution fields using reduced state dynamics approaches. The identity in \eref{vec_iden_lowdim} corresponds to the vanishing of the square of the exterior derivative in the De Rham chain complex. This property reflects the exactness of the differential construction and provides the mathematical foundation for conservation-by-construction representations. In particular, it was shown in \cite{Kelliher2021} that there exists a skew-symmetric matrix field $\pmb{A}(\pmb{w}, \pmb{\nu}) \in \mathbb{R}^{d+1 \times d+1}$, such that $\pmb{z}$ is defined as the row-wise divergence of $\pmb{A}(\pmb{w}, \pmb{\nu})$, given by
\begin{equation}
    \pmb{z}^m (\pmb{w}; \pmb{\nu}) = \begin{bmatrix}
    q^m (\pmb{w}; \pmb{\nu})\\
    \pmb{f}^m (\pmb{w}; \pmb{\nu})\\
    \end{bmatrix} = \begin{bmatrix}
    \nabla_w \cdot \bar{\pmb{A}}_1 (\pmb{w}; \pmb{\nu})\\
     \cdot \\
     \cdot \\
    \nabla_w \cdot \bar{\pmb{A}}_{d+1} (\pmb{w}; \pmb{\nu})\\
    \end{bmatrix},
    \label{eq:Amat_CINR}
\end{equation}
where $\bar{\pmb{A}}_i$ denotes the i\textit{th} row of $\pmb{A}$. This representation exactly satisfies the conservation constraint in \eref{constraint_consv} by virtue of the antisymmetry of $\pmb{A}$ and the commutativity of partial derivatives. 

\subsection{Probabilistic representation of solution constrained to a manifold}
\label{sec:ProbRep}

Until now, we have assumed that the parameterized random field $q(\pmb{x}, t, \pmb{\nu}; \pmb{\omega}): \mathcal{X} \times \mathcal{T} \times \mathcal{V} \times \Omega \to \mathbb{R}$ lies in a Euclidean domain. To formally define random fields constrained to satisfy conservation laws, we first introduce the appropriate function spaces. Let $\mathcal{Z} = L^2 (\mathcal{X T}; \mathbb{R}^{d+1})$ be the infinite-dimensional Hilbert space of solution-flux vector fields defined on a space-time domain $\mathcal{X}\mathcal{T}$, equipped with the $L^2$ inner product. The exact solution-flux manifold $\tilde{\mathcal{M}}_{\nu} \subset \mathcal{Z}$ is the set of all possible solution-flux vector fields $\pmb{z}(\pmb{w})$ that satisfy the conservation law $\mathcal{R}$ such that
    \begin{equation}
    \tilde{\mathcal{M}}_{\nu} = \{\pmb{z} \in \mathcal{Z} \vert \mathcal{R}(\pmb{z}; \pmb{\nu}) = 0\}.
    \label{eq:sol_manifold_prob}
\end{equation}
Thus, $\tilde{\mathcal{M}}_{\nu}$ represents the admissible set of conservation-law-consistent solution-flux vector fields for a fixed parameter instance $\pmb{\nu}$. A random field that exactly satisfies conservation law is defined as a collection of random variables $\{\pmb{z}(\cdot, \pmb{\nu}; \pmb{\omega}) : \Omega \to \mathcal{Z} \}_{\pmb{w} \in \mathcal{X T}}$ on a probability space $(\Omega, \mathcal{F}, P)$ such that for every realization $\pmb{\omega} \in \Omega$, the sample path belongs to the manifold $\tilde{\mathcal{M}}_{\nu}$ almost surely as 
\begin{equation}
    P(\{ \pmb{\omega} \in \Omega: \pmb{z} (\cdot, \pmb{\nu}; \pmb{\omega}) \in \tilde{\mathcal{M}}_{\nu} \}) = 1.
    \label{eq:manifold_def}
\end{equation}
Consequently, uncertainty is restricted to physically admissible directions within the conservation-law manifold.

This random field is constructed by defining a deterministic embedding $\pmb{d}_{\theta}: \tilde{\mathcal{Z}} \times \mathcal{X} \to \tilde{\mathcal{M}}_{\nu}$, where $\tilde{\mathcal{Z}} \subset \mathbb{R}^r$ is the latent space and the latent vector $\tilde{\pmb{q}}: \mathcal{T} \times \mathcal{V} \times \Omega \to \tilde{\mathcal{Z}}$ is defined on the underlying probability space. Therefore, a method is designed to ensure that the range of the neural field decoder is restricted to $\tilde{\mathcal{M}}_{\nu}$, that is $\text{Ran}(\pmb{d}_{\theta}(\tilde{\pmb{z}}, \cdot ) \subset \tilde{\mathcal{M}}_{\nu}$ for all $\tilde{\pmb{z}} \in \mathcal{Z}$. 

\subsection{Latent dynamics modeling for probabilistic solution constrained to a manifold}
\label{sec:ECLEIRSUQ}

In the previous subsection, we defined the solution space that comprises the random solution-flux field, such that each instance of the random field satisfies the nonlinear conservation law constraint. The objective of the present subsection is to construct a probabilistic latent dynamics model whose realizations are restricted to the admissible manifold $\tilde{\mathcal{M}}_{\nu}$ defined in \eref{manifold_def}. Extending the vector calculus identity for $d + 1 = \{2, 3\}$ to random fields, implies that a random solution-flux vector field when represented as
\begin{equation}
    \pmb{z}(\pmb{w}, \pmb{\nu}; \pmb{\omega}) = \nabla_w \times \pmb{a} (\pmb{w}, \pmb{\nu}; \pmb{\omega}), \quad \text{satisfies} \quad \nabla_w \cdot \nabla_w \times \pmb{a} (\pmb{w}, \pmb{\nu}; \pmb{\omega}) = 0,
    \label{eq:vec_iden_lowdim}
\end{equation}
where $\pmb{a} (\pmb{w}, \pmb{\nu}; \pmb{\omega})$ is a random vector potential field, and thus satisfies the conservation structure. Similarly, for a higher-dimensional vector field $d + 1 > 3$, we can extend the result of \cite{Barbarosie2011, Kelliher2021} to random fields such that each realization satisfies this constraint. We enable this extension by defining a skew-symmetric random matrix field $\pmb{A}(\pmb{w}, \pmb{\nu}; \pmb{\omega}): \mathcal{X T} \times \mathcal{V} \times \Omega \to \mathbb{R}^{d+1 \times d+1}$, where each realization $\pmb{\omega}$ gives a matrix field. With this construction, random solution-flux vector field $\pmb{z}^m (\pmb{w}; \pmb{\nu})$, satisfying the conservation law constraint, is defined as
\begin{equation}
    \pmb{z}^m (\pmb{w}, \pmb{\nu}; \pmb{\omega}) = \begin{bmatrix}
    q^m (\pmb{w}, \pmb{\nu}; \pmb{\omega})\\
    \pmb{f}^m (\pmb{w}, \pmb{\nu}; \pmb{\omega})\\
    \end{bmatrix} = \begin{bmatrix}
    \nabla_w \cdot \bar{\pmb{A}}_1 (\pmb{w}, \pmb{\nu}; \pmb{\omega})\\
     \cdot \\
     \cdot \\
    \nabla_w \cdot \bar{\pmb{A}}_{d+1} (\pmb{w}, \pmb{\nu}; \pmb{\omega})\\
    \end{bmatrix},
    \label{eq:Amat_CINR_random}
\end{equation}
where $\bar{\pmb{A}}_i (\pmb{w}, \pmb{\nu}; \pmb{\omega})$ is the $i$th row of the skew-symmetric random matrix field $\pmb{A} (\pmb{w}, \pmb{\nu}; \pmb{\omega}) \in \mathbb{R}^{(d+1) \times (d+1)}$. The $\frac{d (d+1)}{2}$ independent components of $\pmb{A} (\pmb{w}, \pmb{\nu}; \pmb{\omega})$, represented as a random vector field $\pmb{a} (\pmb{w}, \pmb{\nu}; \pmb{\omega}) \in \mathbb{R}^{\frac{d (d+1)}{2}}$, are modeled as
\begin{equation}
    \pmb{a} (\pmb{w}, \pmb{\nu}; \pmb{\omega})= \pmb{d}_{\pmb{\theta}} (\tilde{\pmb{q}} (t, \pmb{\nu}; \pmb{\omega}), \pmb{x}),
    \label{eq:ECLEIRS_cons}
\end{equation}
where $\pmb{d}_{\pmb{\theta}}$ is the MLP-based decoder network and $\tilde{q}$ is given in \eref{latentvec_final} as also used for IRS-UQ and PI-IRS-UQ. Unlike the decoders used in IRS-UQ and PI-IRS-UQ, the decoder in \eref{ECLEIRS_cons} parametrizes the high-dimensional vector potential of space-time divergence-free solution-flux vector rather than the solution itself. For a $1$-D spatial-domain system,
\begin{equation}
    \pmb{A} (\pmb{w}, \pmb{\nu}; \pmb{\omega})= \begin{bmatrix}
    0 & a (\pmb{w}, \pmb{\nu}; \pmb{\omega})\\
    -a (\pmb{w}, \pmb{\nu}; \pmb{\omega}) & 0 \\
    \end{bmatrix},
\end{equation}
implying that $\pmb{a} (\pmb{w}, \pmb{\nu}; \pmb{\omega})$ is a scalar represented as $a (\pmb{w}, \pmb{\nu}; \pmb{\omega}) \in \mathbb{R}$ and 
\begin{equation}
    \pmb{z}^m (\pmb{w}, \pmb{\nu}; \pmb{\omega})= \begin{bmatrix}
    q^m \\
    f^m \\
    \end{bmatrix} = \begin{bmatrix}
    \frac{\p a}{\p x_1} \\
    - \frac{\p a}{\p t} \\
    \end{bmatrix}.
    \label{eq:z_m_1D}
\end{equation}
Similarly, for a $2$-D spatial domain system, 
\begin{equation}
    \pmb{A} (\pmb{w}, \pmb{\nu}; \pmb{\omega}) = \begin{bmatrix}
    0 & a_1 & a_2 \\
    -a_1 & 0 & a_3 \\
    -a_2 & -a_3 & 0 
    \end{bmatrix},
\end{equation}
such that $\pmb{a} (\pmb{w}, \pmb{\nu}; \pmb{\omega}) = [a_1 \; a_2 \; a_3]^T \in \mathbb{R}^3$ and the dependence on $(\pmb{w}, \pmb{\nu}; \pmb{\omega})$ has been suppressed in each component. Therefore, the solution-flux random vector is obtained as
\begin{equation}
    \pmb{z}^m (\pmb{w}, \pmb{\nu}; \pmb{\omega})= \begin{bmatrix}
    q^m \\
    f^m_1 \\
    f^m_2 \\
    \end{bmatrix} = \begin{bmatrix}
    \frac{\p a_1}{\p x_1} + \frac{\p a_2}{\p x_2} \\
    -\frac{\p a_1}{\p t} + \frac{\p a_3}{\p x_2} \\
    -\frac{\p a_2}{\p t} - \frac{\p a_3}{\p x_1} \\
    \end{bmatrix}.
    \label{eq:z_m_2D}
\end{equation}
Lastly, for a system with a $3$-D spatial domain,
\begin{equation}
    \pmb{A} (\pmb{w}, \pmb{\nu}; \pmb{\omega}) = \begin{bmatrix}
    0 & a_1 & a_2  & a_3 \\
    -a_1 & 0 & a_4 & a_5 \\
    -a_2 & -a_4 & 0 & a_6 \\
    -a_3 & -a_5 & -a_6 & 0
    \end{bmatrix},
\end{equation}
such that $\pmb{a} (\pmb{w}, \pmb{\nu}; \pmb{\omega}) = [a_1 \; a_2 \; a_3 \; a_4 \; a_5 \; a_6 ]^T \in \mathbb{R}^6$ and
\begin{equation}
    \pmb{z}^m (\pmb{w}, \pmb{\nu}; \pmb{\omega})= \begin{bmatrix}
    q^m \\
    f^m_1 \\
    f^m_2 \\
    f^m_3 \\
    \end{bmatrix} = \begin{bmatrix}
    \frac{\p a_1}{\p x_1} + \frac{\p a_2}{\p x_2} + \frac{\p a_3}{\p x_3}  \\
    -\frac{\p a_1}{\p t} + \frac{\p a_4}{\p x_2} + \frac{\p a_5}{\p x_3} \\
    -\frac{\p a_2}{\p t} - \frac{\p a_3}{\p x_1} + \frac{\p a_6}{\p x_3}\\
    -\frac{\p a_3}{\p t} - \frac{\p a_5}{\p x_1} - \frac{\p a_6}{\p x_2}\\
    \end{bmatrix}.
    \label{eq:z_m_3D}
\end{equation}
Through algebraic manipulation, it can be verified that these satisfy $\nabla_w \cdot \pmb{z} (\pmb{w}, \pmb{\nu}; \pmb{\omega}) = 0$ by the skew-symmetry of $\pmb{A}$. Therefore, every realization generated using this representation automatically belongs to $\tilde{\mathcal{M}}_{\nu}$. As ECLEIRS-UQ representation enforces a space-time divergence-free solution-flux field by construction, exact conservation is guaranteed for the modeled solution-flux pair, while consistency between the modeled flux and the constitutive physical flux depends on the available flux data and the training objective.

Following the definition of $\tilde{\pmb{q}} (t, \pmb{\nu}; \pmb{\omega})$ in \eref{latentvec_final}, the problem reduces to determining the unknown parameters by solving a nonconvex optimization problem
\begin{multline}
    J(\pmb{\theta}, \pmb{\eta}', \pmb{\tau}_{GP}) = \underbrace{\underset{i,j,k}{\mathbb{E}} \vert \vert \pmb{z} (\pmb{x}_k, t_i, \pmb{\nu}_j) - \pmb{z}^m (\tilde{\pmb{q}} (t_j, \pmb{\nu}_k; \pmb{\omega}), \pmb{x}_i) \vert \vert^2}_{\text{Solution-flux vector reconstruction-loss}} +  \\ \underbrace{\frac{\beta}{2} \; \sum_{k=1}^r \underset{i,j}{\mathbb{E}}  \Big(\exp \Big(f^{\Sigma-GP}_{\pmb{\eta}} (\pmb{\nu}^{(j)}; \pmb{\tau}_{GP}) \Big)+ [f^{\mu-GP}_{\pmb{\eta}', k} (t^{(i)}, \pmb{\nu}^{(j)})]^2 - 1 - f^{\Sigma-GP}_{\pmb{\eta}} (\pmb{\nu}^{(j)}; \pmb{\tau}_{GP}) \Big)}_{\text{Latent vector KL divergence}},
    \label{eq:ECLEIRS-UQ_optimization}
\end{multline}
where $\pmb{z}^m (\pmb{w}, \pmb{\nu}; \pmb{\omega})$ is chosen as \eref{z_m_1D}, \eref{z_m_2D} and \eref{z_m_3D} for systems that are spatially $1$-D, $2$-D and $3$-D respectively. Unlike IRS-UQ and PI-IRS-UQ, this optimization process searches only over physically admissible solution representations, reducing the effective hypothesis space relative to unconstrained formulations. Once this network architecture is learned, the ODE governing the dynamics of the GPMN is learned similar to IRS-UQ as shown in Section \ref{sec:CommonMethod}. As this approach can be considered as a probabilistic extension of the ECLEIRS framework proposed in \cite{Prakash2026}, we refer to it as ECLEIRS-UQ. Similar to ECLEIRS and PI-IRS-UQ, this method also requires data for both solutions and flux evaluations during training. While ECLEIRS-UQ allows sampling from solution-flux distribution such that each realization satisfies the nonlinear conservation law exactly, we show in \textit{Theorem 1} that the resulting sample expectation of predictions also satisfy the nonlinear conservation law. 

\bigskip
\noindent \textbf{Theorem 1:} \textit{Assume that $\pmb{z}(\pmb{w}, \pmb{\nu}; \pmb{\omega})$ is square integrable and sufficiently regular so that the expectation operator and the space-time divergence operator commute. Consider the random solution-flux vector $\pmb{z}: \mathcal{X} \mathcal{T} \times \mathcal{V} \times \Omega \to \tilde{\mathcal{M}}_{\nu}$, where $\tilde{\mathcal{M}}_{\nu}$ is the solution manifold, defined in \eref{sol_manifold_prob}, that satisfies the exact conservation law given in \eref{constraint_consv}. The resulting mean of the solution-flux vector also satisfies the exact conservation constraint.
}

\noindent \textbf{Proof:} The random solution-flux vector $\pmb{z}: \mathcal{X} \mathcal{T} \times \mathcal{V} \times \Omega \to \tilde{\mathcal{M}}_{\nu}$ satisfies the conservation law, implying 
\begin{equation}
    \nabla_w \cdot \pmb{z} (\pmb{w}, \pmb{\nu}; \pmb{\omega}) = 0.
\end{equation}
Let $\pmb{\mu}_z: \mathcal{X} \mathcal{T} \times \mathcal{V} \to \mathcal{R}^{d+1}$ be the mean of predicted solution-flux vector field $\pmb{z}$ defined as $\pmb{\mu}_z = \mathbb{E}[\pmb{z}] = \int_{\Omega} \pmb{z} (\cdot ; \pmb{\omega}) p (\pmb{z}(\cdot, \pmb{\omega})) d \pmb{\omega}$, where $p(\cdot)$ is the probability distribution of $\pmb{z}$. Taking the space-time divergence of this mean, and using the property that expectation and space-time divergence operators commute, we get
\begin{equation}
    \nabla_w \cdot \pmb{\mu}_z = \mathbb{E}[\nabla_w \cdot \pmb{z}] = 0.
\end{equation}
The equality $\mathbb{E}[\nabla_w \cdot \pmb{z}] = 0$ follows from $\nabla_w \cdot \pmb{z} = 0$ and the integrability assumption.
Therefore, the expected solution-flux field also satisfies the nonlinear conservation law exactly. 

\section{Computational implementation}
\label{sec:Section6}

\begin{figure}[t]
    \centering
    \includegraphics[width=\textwidth,trim={0.2cm 1.3cm 0cm 0.5cm},clip]{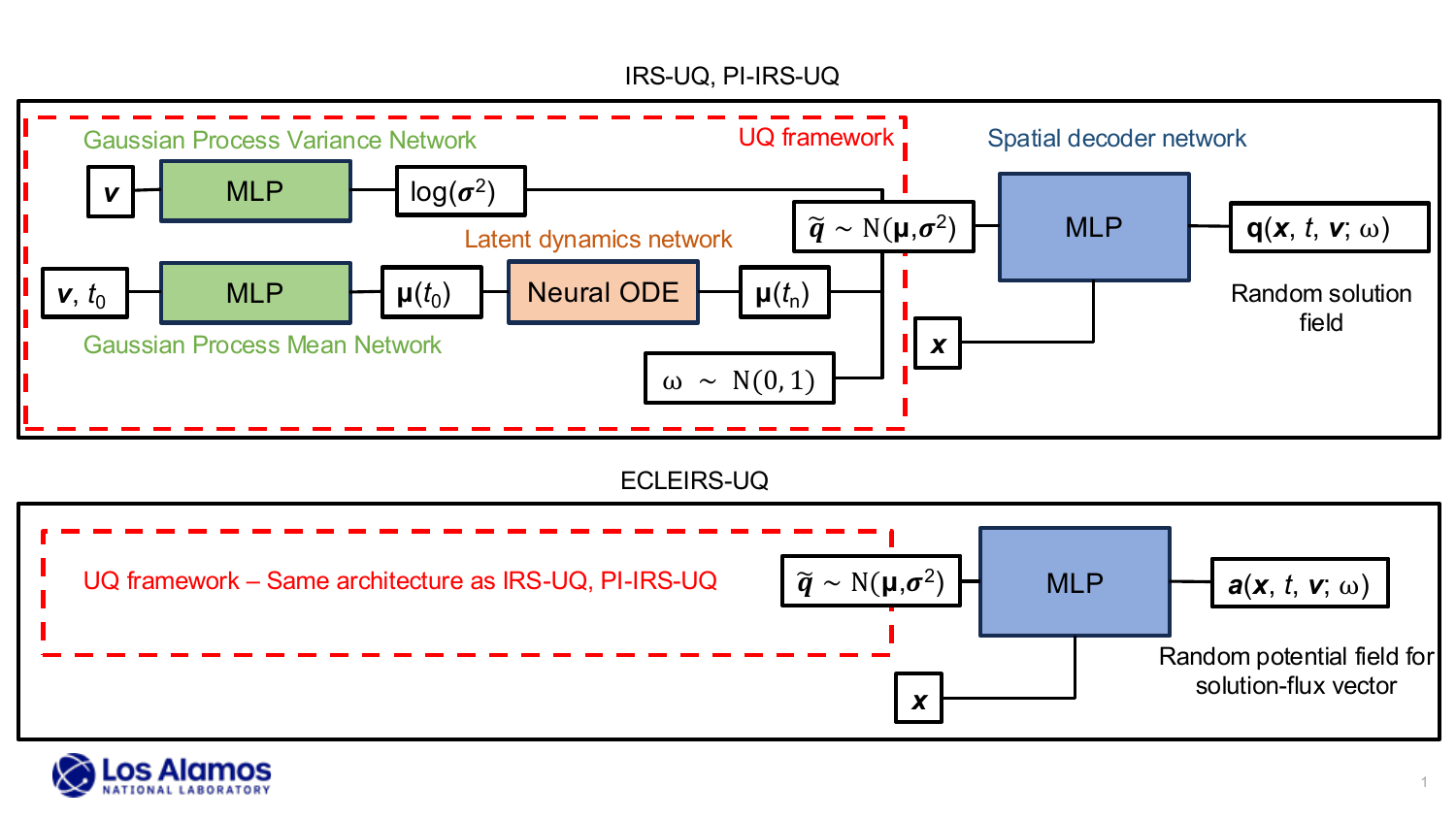}
    \caption{Neural network architectures for IRS-UQ, PI-IRS-UQ and ECLEIRS-UQ. The diagram illustrates the three main components: (1) Gaussian Process Variance Network (GPVN) $f_{\pmb{\eta}}^{\Sigma-GP}$ that predicts latent variance, (2) Gaussian Process Mean Network (GPMN) $\pmb{f}_{\pmb{\eta}'}^{\mu-GP}$ coupled with a neural ODE $\pmb{g}_{\pmb{\psi}}$ for latent dynamics, and (3) decoder network $\pmb{d}_{\pmb{\theta}}$ that maps latent representations to solution (IRS-UQ, PI-IRS-UQ) or vector potential fields (ECLEIRS-UQ). For ECLEIRS-UQ, an additional transformation via space-time divergence converts the vector potential to the space-time divergence-free solution-flux vector. }
    \label{fig:NNArchitect}
\end{figure}

The neural network architectures for IRS-UQ, PI-IRS-UQ and ECLEIRS-UQ are compared in \figref{NNArchitect}. For all three modeling approaches, model development and testing involve two stages: offline and online. The offline stages encompasses all computationally expensive operations, such as data generation by running several high-fidelity simulations or experiments, data pre-processing involving data cleanup and assembling into the relevant dataset format, and model learning involving solving the nonlinear optimization problem to identify model weights while also performing hyperparameter tuning for accurate results on the training set. While this stage is very expensive, it is typically performed only once during the modeling process. The online stage is a computationally inexpensive step that involves deploying the trained model. This step can be performed multiple times at low computational cost, making these latent dynamics models attractive for multi-query applications.

\subsection{Offline stage - Model learning}

The offline stages for learning the different latent dynamics approaches are very similar. In this article, we follow the segregated approach for model training, similar to the one in \cite{Prakash2026} with some important changes to the implementation. The primary motivation for this segregated strategy is to decompose a highly coupled optimization problem into a sequence of smaller subproblems, thereby improving training stability and reducing the computational complexity. In this segregated approach, we treat the training of GPVN, the neural field decoder, and the equations for GPMN evolution as three distinct model training steps, which are performed sequentially. More details of each of these steps are listed below:
\begin{enumerate}
    \item \textit{Training the GPVN:} The first step involves learning GPVN by solving the optimization problem in \eref{reg_GPVN}. Given the data for system parameters $\pmb{\nu}$ assembled as $\pmb{V}$, the ground truth variance is computed using \eref{gp_var} for randomly sampled hyperparameters ($\sigma_f$, $\sigma_n$ and $\pmb{l}$) from a bounding box of the respective values. 
    \item \textit{Training GPMN and decoder:} The GPVN trained in the previous step is frozen and integrated into the latent dynamics model as shown in \eref{latentvec_final}. With this integrated model and given the data for $\pmb{q}$ (and $\pmb{f}(\cdot, \pmb{\nu})$ for PI-IRS-UQ and ECLEIRS-UQ), we solve the optimization problems: \eref{IRS-UQ_optimization} for IRS-UQ, \eref{PI-IRS-UQ_optimization} for PI-IRS-UQ and \eref{ECLEIRS-UQ_optimization} for ECLEIRS-UQ to obtain the parameters for GPMN and decoder.  
    \item \textit{Training neural ODE for GPMN evolution:} The learned GPMN from the previous step is used to compute $\partial f_{\pmb{\eta}'}^{\mu-GP}/ \partial t$ using automatic differentiation, which is then used to solve the nonlinear regression problem in \eref{latent_node} for obtaining the equations governing the evolution of the latent vector mean. This process is the same for all the three modeling approaches.
\end{enumerate}

\subsection{Online stage - Model testing}

Once the latent dynamics model is learned in the offline stage, the model can be deployed in a multi-query application. The key steps in this stage are listed below:
\begin{enumerate}
    \item In this article, we assume that the initial condition is parameterized by $\pmb{\nu}$. Therefore, the initial condition of the mean of the latent vector is obtained by evaluating $\pmb{f}_{\pmb{\eta}'}^{\mu-GP} (t = t_0, \pmb{\nu})$. 
    \item With this initial condition for the mean of the latent vector, the dynamics of the latent vector are propagated in time by numerically integrating \eref{NODE_GPMN_Fbar} using a suitable time integration scheme to provide $\bar{\pmb{f}}_{\pmb{\eta}'}^{\mu-GP} (t, \pmb{\nu})$ for desired time instances.
    \item The time-marched mean of the latent vector is used along with the GPVN to generate samples of the latent vector as in \eref{latentvec_final}. These samples of the latent vector are used alongside the decoder to provide the solution at different spatial query locations. Note that as the latent vector is a random variable, each sample will provide a realization of the solution. We obtain the estimate of the mean and standard deviation of the solution, by computing several realizations and taking the mean and standard deviations of them.
\end{enumerate}

If the goal of the model is to explore the parameter space without evaluating the dynamics forecasting capability, then the GPMN can be directly evaluated using \eref{GPMN_def} without any time-integration in Step 2 of the online stage. Irrespective of the selection of time-integration scheme or direct evaluation of the GPMN, ECLEIRS-UQ is designed such that each realization of the solution-flux vector satisfies the nonlinear conservation law exactly. All the models discussed above can also be evaluated in specific spatial regions providing targeted inexpensive inference when modeling large scale dynamics. For example, when evaluating boundary drag force, the model needs to be only evaluated at the boundaries and not over the full spatial domain. This query-based evaluation is a key advantage of the implicit neural representation approach, as it allows flexible spatial resolution without retraining.

\section{Results}
\label{sec:Section7}

In this section, we demonstrate the applicability of the proposed methods for three numerical experiments: 1) 1-D advection problem, 2) 2-D Euler problem and 3) 2-D shallow water problem. These benchmark problems evaluate the proposed methods across increasing levels of complexity, ranging from a linear scalar conservation law to nonlinear systems of conservation laws with multiple interacting wave structures. For each of these three problems, we evaluate the performance of the three modeling approaches: IRS-UQ, PI-IRS-UQ and ECLEIRS-UQ. We assess these models for accurate prediction of the solution, capturing the uncertainty for in-distribution and out-of-distribution testing points and satisfaction of underlying conservation laws.  

\begin{table}[t]
    \centering
    \caption{Hyperparameters in the neural network architectures for different problems, where $n_h$ is the number of hidden layers and $n_n$ is the number of neurons in each hidden layer.}    
    \begin{tabular}{|c|c|c|c|c|}
        \hline
         \textbf{Test case} & \textbf{GPMN} ($f_{\pmb{\eta}'}^{\mu-GP}$) & \textbf{GPVN} ($f_{\pmb{\eta}'}^{\Sigma-GP}$) &\textbf{Decoder} ($\pmb{d}_{\pmb{\theta}}$) & \textbf{Dynamics} ($\pmb{g}_{\pmb{\psi}}$)  \\
        \hline
         1-D advection & $n_h = 3, n_n = 100$ & $n_h = 3, n_n = 200$ & $n_h = 4, n_n = 100$ & $n_h = 4, n_n = 200$ \\
         \hline
         2-D Euler & $n_h = 3, n_n = 100$ & $n_h = 3, n_n = 200$ & $n_h = 4, n_n = 100$ & $n_h = 4, n_n = 100$ \\
         \hline         
         2-D shallow water & $n_h = 3, n_n = 100$ & $n_h = 3, n_n = 200$ & $n_h = 4, n_n = 100$ & $n_h = 4, n_n = 100$ \\
         \hline
    \end{tabular}
    \label{tab:hyper_tab}
\end{table}

The hyper-parameters for the neural networks used for representing GPMN, GPVN, decoder and dynamics for each numerical example is shown in \tabref{hyper_tab}. All these sub-networks use SIREN activation function \cite{Sitzmann2020}. The resulting model training optimization problems are solved using ADAM-W optimizer \cite{Kingma2017, AdamW2019} to identify the unknown weight and biases of the network. For all the test cases, these optimizers have a starting learning rate of 0.0005 and is decreased by a factor of $0.3$ if convergence loss asymptotes. Lastly, the dimensionality of the latent space is set to $d_{\nu}+2$ for all the numerical experiments in this study. Using comparable network architectures across all three modeling approaches reduces architectural bias and allows performance differences to be attributed primarily to the underlying modeling framework rather than differences in network capacity.

We assess the performance of the models by comparing the reference results to the expectation of prediction
\begin{equation}
    \pmb{q}^m_{\text{mean}} (\pmb{x}, t, \pmb{\nu}) =  \mathbb{E}_{\pmb{\omega}} [ \pmb{q}^m(\pmb{x}, t, \pmb{\nu}; \pmb{\omega}) ],
\end{equation}
where the expectation is with respect to each realization ($\pmb{\omega}$) of the solution. This mean solution along with the predicted mean flux satisfies the conservation structure as proved in Theorem 1.
The resulting error defined as
\begin{equation}
    \pmb{\epsilon} (\pmb{x}, t, \pmb{\nu}) = \vert \pmb{q}(\pmb{x}, t, \pmb{\nu}) - \pmb{q}^m_{\text{mean}}  (\pmb{x}, t, \pmb{\nu}) \vert,
\end{equation}
where $\vert \cdot \vert$ implies the absolute value. We also compare relative error
\begin{equation}
    \pmb{\epsilon}_r (\pmb{\nu}) = \frac{\vert\vert \pmb{q}(\pmb{x}, t, \pmb{\nu}) - \pmb{q}^m_{\text{mean}}  (\pmb{x}, t, \pmb{\nu}) \vert \vert^2_2}{\vert\vert \pmb{q}(\pmb{x}, t, \pmb{\nu}) \vert \vert^2_2},
\end{equation}
where $\vert\vert\cdot\vert\vert_2$ is taken with respect to all spatial and temporal coordinates. This metric evaluates the quality of the reconstructed parameterized solution manifold, and therefore provides a direct measure of predictive accuracy across the parameter space. We also calculate the conservation error 
\begin{equation}
    \epsilon_{\text{consv}} (\pmb{\nu}) = \mathbb{E}_{\pmb{x}, t} \Big[ \Big\vert \Big\vert \frac{\partial \pmb{q}}{\partial t} + \nabla \cdot \pmb{f}(\pmb{q}, \pmb{\nu}) \Big\vert \Big\vert_2 \Big]
\end{equation}
and evaluate the performance of different methods in satisfying the underlying nonlinear conservation law. To assess the uncertainty estimation capability of the models, we compute the standard deviation of the prediction
\begin{equation}
    \pmb{q}^m_{\text{SD}} (\pmb{x}, t, \pmb{\nu}) = \sqrt{\mathbb{E}_{\pmb{\omega}} [ (\pmb{q}^m(\pmb{x}, t, \pmb{\nu}; \pmb{\omega}))^2 ] - (\pmb{q}^m_{\text{mean}} (\pmb{x}, t, \pmb{\nu}))^2}.
\end{equation}
The standard deviation provides a local measure of predictive uncertainty and is used to assess whether the model can correctly identify regions in space-time and parameter space  where predictions are less reliable. In particular, we compare the correlation between standard deviation and errors, to evaluate the UQ capability of the model. The correlation coefficient for each parameter $\pmb{\nu}$ is computed as
\begin{equation}
    r(\pmb{\nu}) = \frac{\sum_{\pmb{x}, t}\Big(\pmb{\epsilon}(\pmb{x}, t, \pmb{\nu}) - \mathbb{E}_{\pmb{x}, t} [\pmb{\epsilon}(\pmb{x}, t, \pmb{\nu})]\Big)\Big(\pmb{q}^m_{\text{SD}}(\pmb{x}, t, \pmb{\nu}) - \mathbb{E}_{\pmb{x}, t} [\pmb{q}^m_{\text{SD}}(\pmb{x}, t, \pmb{\nu})]\Big) }{\sqrt{\sum_{\pmb{x}, t}\Big(\pmb{\epsilon}(\pmb{x}, t, \pmb{\nu}) - \mathbb{E}_{\pmb{x}, t} [\pmb{\epsilon}(\pmb{x}, t, \pmb{\nu})]\Big)^2 \sum_{\pmb{x}, t}\Big(\pmb{q}^m_{\text{SD}}(\pmb{x}, t, \pmb{\nu}) - \mathbb{E}_{\pmb{x}, t} [\pmb{q}^m_{\text{SD}}(\pmb{x}, t, \pmb{\nu})]\Big)^2 }}.
\end{equation}
The correlation coefficient evaluates whether regions associated with larger prediction errors are also assigned larger uncertainty estimates. Therefore, a higher correlation coefficient closer to one implies a better uncertainty estimation capability of the model, whereas a value of zero implies worse capability. We also compute the mean coverage offered by the two times standard deviation, given as 
\begin{equation}
    \text{cov}_{2\sigma} (\pmb{\nu}) = \mathbb{E}_{x,t} \Big[ \begin{cases}
    1, &  \pmb{q}^m_{\text{mean}} - 2 \pmb{q}^m_{\text{SD}} \leq \pmb{q} \leq \pmb{q}^m_{\text{mean}} + 2 \pmb{q}^m_{\text{SD}}  \\
    0, & \text{otherwise}
\end{cases} \Big].
\end{equation}
This coverage metric allows assessing if predicted uncertainty bounds covers the ground truth solution. Under the assumption of approximately Gaussian predictive uncertainty, an ideally calibrated model should achieve empirical coverage close to the nominal confidence level associated with a two-standard-deviation prediction interval. Deviations from this value indicate either under-confident or over-conservative uncertainty estimates. The correlation coefficient and coverage metric therefore provide complementary information. The former assesses whether uncertainty is assigned to the correct regions, while the latter evaluates whether the magnitude of the uncertainty estimates is appropriately calibrated. Therefore, it is important to look at both the correlation and coverage parameters together for assessing the uncertainty estimation capability of different modeling approaches. Throughout the numerical experiments, uncertainty estimates are generated using Monte Carlo sampling of the latent distribution, where the latent states are sampled ten times. The resulting computational overhead is modest relative to the offline training cost and remains substantially smaller than the cost of repeated high-fidelity simulations for data generation.

\subsection{1-D scalar advection problem} 

\begin{figure}[t]
    \centering
    \subfigure[\label{fig:STnRatio_0p2}]{\includegraphics[width=0.32\textwidth, trim={0.2cm 0cm 1.0cm 0.5cm},clip]{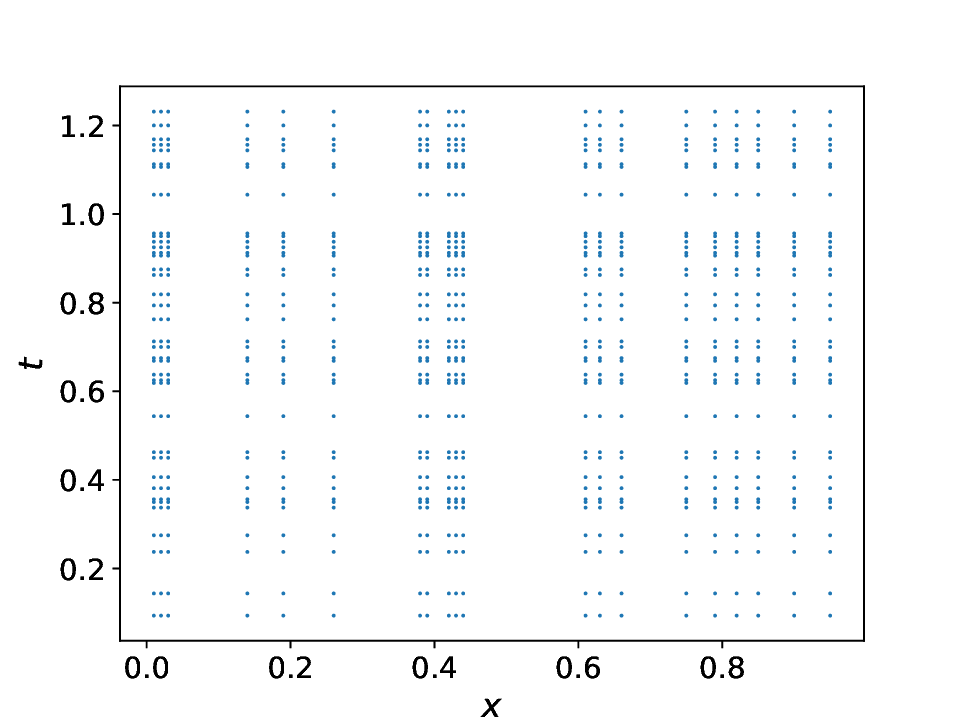}}
    \subfigure[\label{fig:STnRatio_0p6}]{\includegraphics[width=0.32\textwidth, trim={0.2cm 0cm 1.0cm 0.5cm},clip]{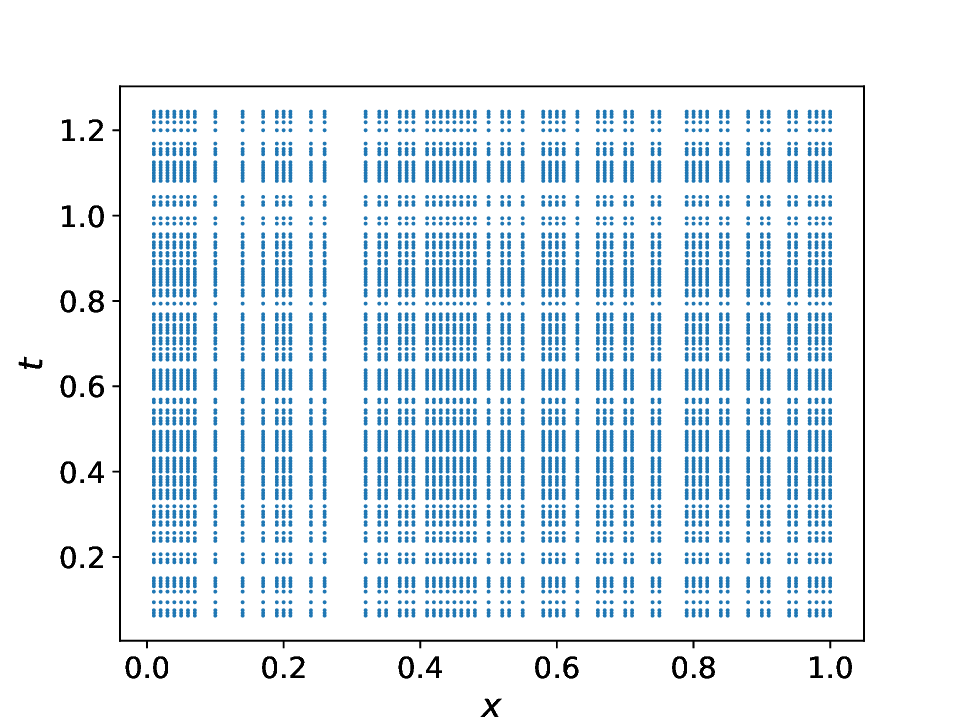}}
    \subfigure[\label{fig:STnRatio_1p0}]{\includegraphics[width=0.32\textwidth, trim={0.2cm 0cm 1.0cm 0.5cm},clip]{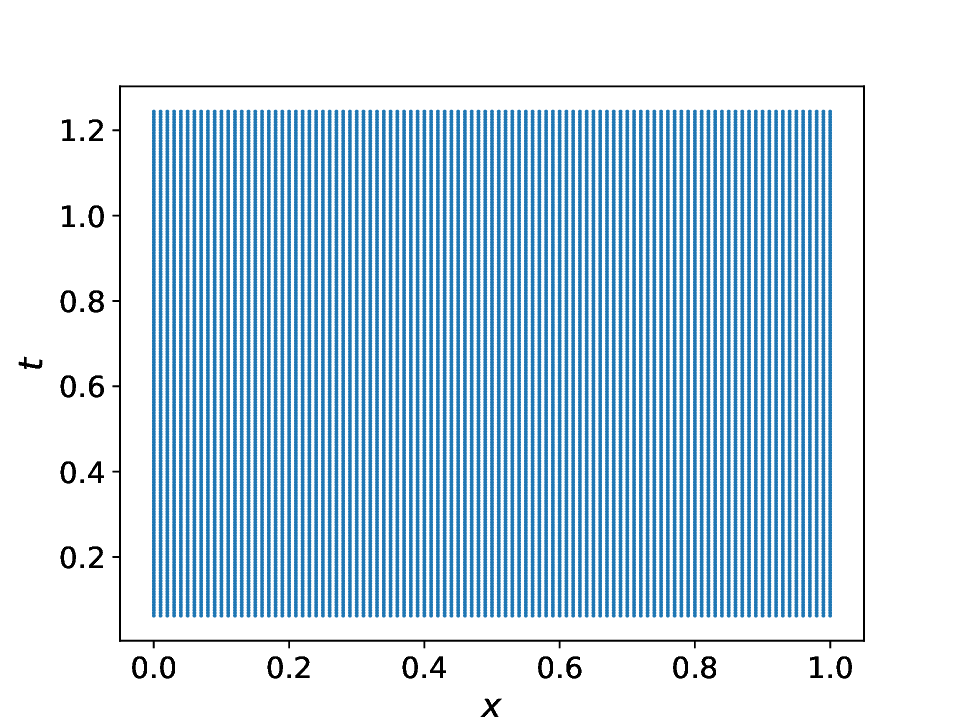}}
    \vspace{-3mm}
    \caption{1-D advection problem: Space-time sensor locations for training different
reduced-state dynamics models. Retaining  (a) $20\%$, (b) $60\%$ and (e) $100\%$ sparsity in  spatial
locations and time instances for training. Note that $x\%$ sparsity in both space and time implies $x^2/100 \%$ sparsity in the entire dataset. The figure is replicated from the author's prior work \cite{Prakash2026}.}
    \label{fig:STnRatio}
\end{figure}

The 1-D scalar advection problem models the transport of scalar $q$ using the conservation law of the form
\begin{equation}
    \frac{\p q}{\p t} + \frac{\p (c q)}{\p x} = 0,
\end{equation}
where $c$ is the advection velocity. This PDE is subject to initial scalar field of the form
\begin{equation}
    q(t = 0) = \begin{Bmatrix}
        q_0, & x \leq x_0\\
        0, & x > x_0
    \end{Bmatrix}.
\end{equation}
With these initial conditions, the system is parameterized as $\pmb{\nu} = [c, q_0, x_0] \in \mathbb{R}^3$. Furthermore, periodic boundary conditions are used for this problem. The simulations used for data generation were conducted by using a  WENO-JS \cite{Jiang1996} spatial discretization and third-order total variation diminishing (TVD) Runge-Kutta for temporal discretization \cite{Shu1988}. A Courant-Fredrichs-Lewy (CFL) number of 0.5 is used to conduct these simulations. These simulations were conducted for the same number of time steps for all parameters, thereby resulting in different total simulation time. The details about the parameter set used for training and validating the model is shown in \tabref{Advec_dataset}. Note the validation parameter set contains parameter combinations that are significantly outside the convex hull of training parameters. The selection of such an expanded validation set allows for a comprehensive analysis of the proposed framework for scenarios with significant extrapolation. In this extrapolation regime, the predictions are not expected to match the reference results, but the framework should yield a high standard deviation in the results indicating lower confidence in the prediction. We also assess the performance of different modeling approaches for scenarios where the data quality is degraded, which is simulated by sparsely sampling the domain in space and time (as shown in \figref{STnRatio}) and adding Gaussian noise with zero mean and $\sigma_N$ standard deviation. 

\begin{table}[t]
    \centering
        \caption{Details of the learning and validation datasets for 1-D advection problem. The learning dataset is further randomly divided between training ($75 \%$) and testing set ($25 \%$) to ensure that the model is not overfitted.}
    \begin{tabular}{|c|c|c|}
    \hline
         \textbf{Dataset Name} & \textbf{Number of data points} & \textbf{Parameter values} \\
         & ($n_x \times n_t$) & ($c \times q_0 \times x_{0}$) \\
    \hline
         Training & $101 \times 190$ & $c \in \{0.8, 0.9, 1.0, 1.1, 1.2 \}$ \\
        & & $q_0 \in \{0.9, 0.95, 1.0, 1.05, 1.1\}$ \\
        & & $x_{0} \in \{0.1, 0.15, 0.2, 0.25, 0.3\}$ \\
         \hline
         Validation 
         & $101 \times 190$ & Halton sampling, $c \in [0.4, 1.6]$ \\
        & & Halton sampling, $q_0 \in [0.65, 1.35]$ \\
        & & Halton sampling, $x_0 \in [0.05, 0.4]$ \\
        \hline
    \end{tabular}
    \label{tab:Advec_dataset}
\end{table}

\begin{figure}[t]
    \centering
    \subfigure[\label{fig:STnRatio_0p6} Relative error ($\epsilon_{r}$)]{\includegraphics[width=0.32\textwidth, trim={3cm 1.0cm 5cm 2cm},clip]{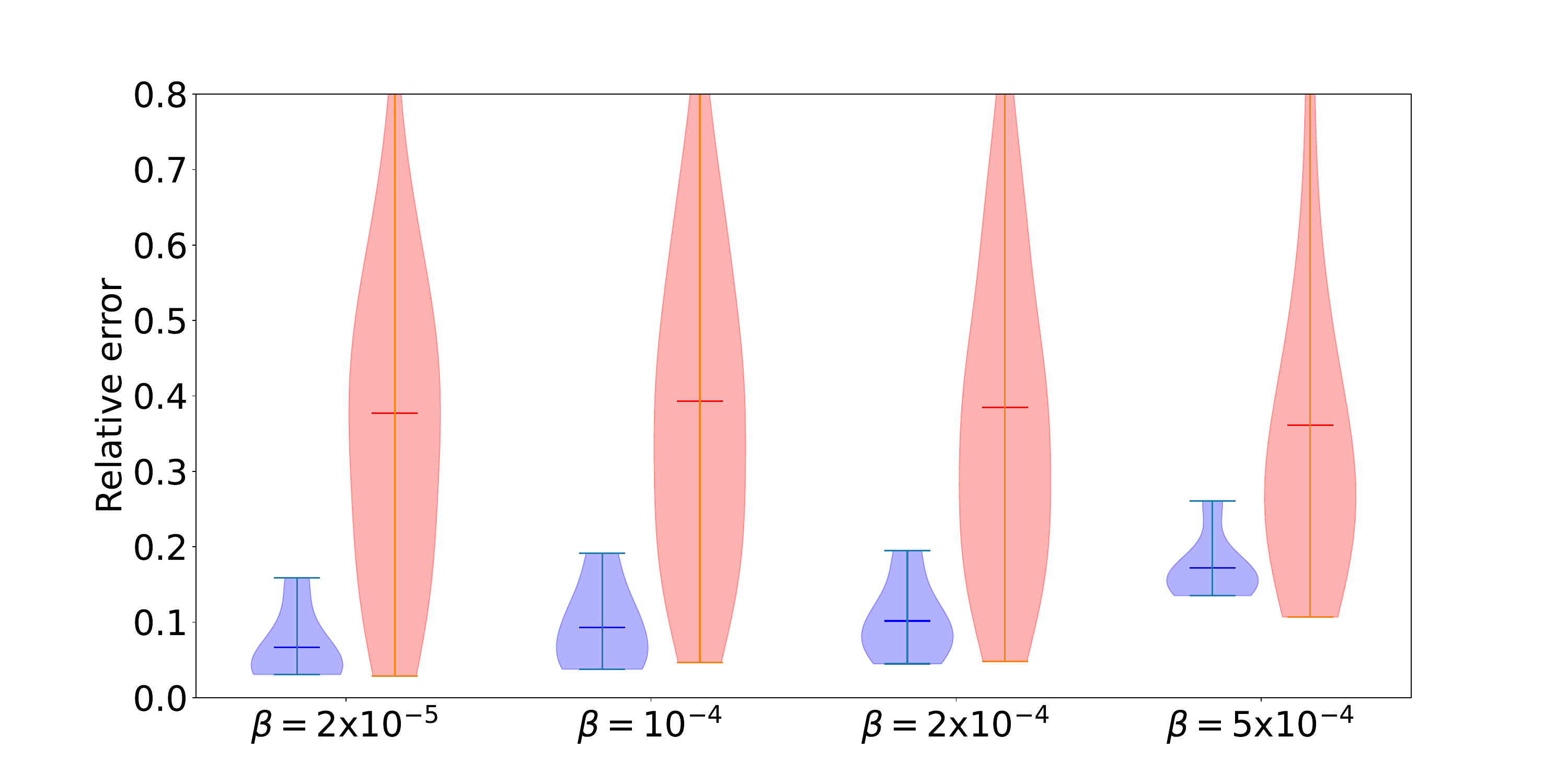}}
    \subfigure[\label{fig:STnRatio_1p0} Correlation Coefficient ($r$)]{\includegraphics[width=0.32\textwidth, trim={3cm 1.0cm 5cm 2cm},clip]{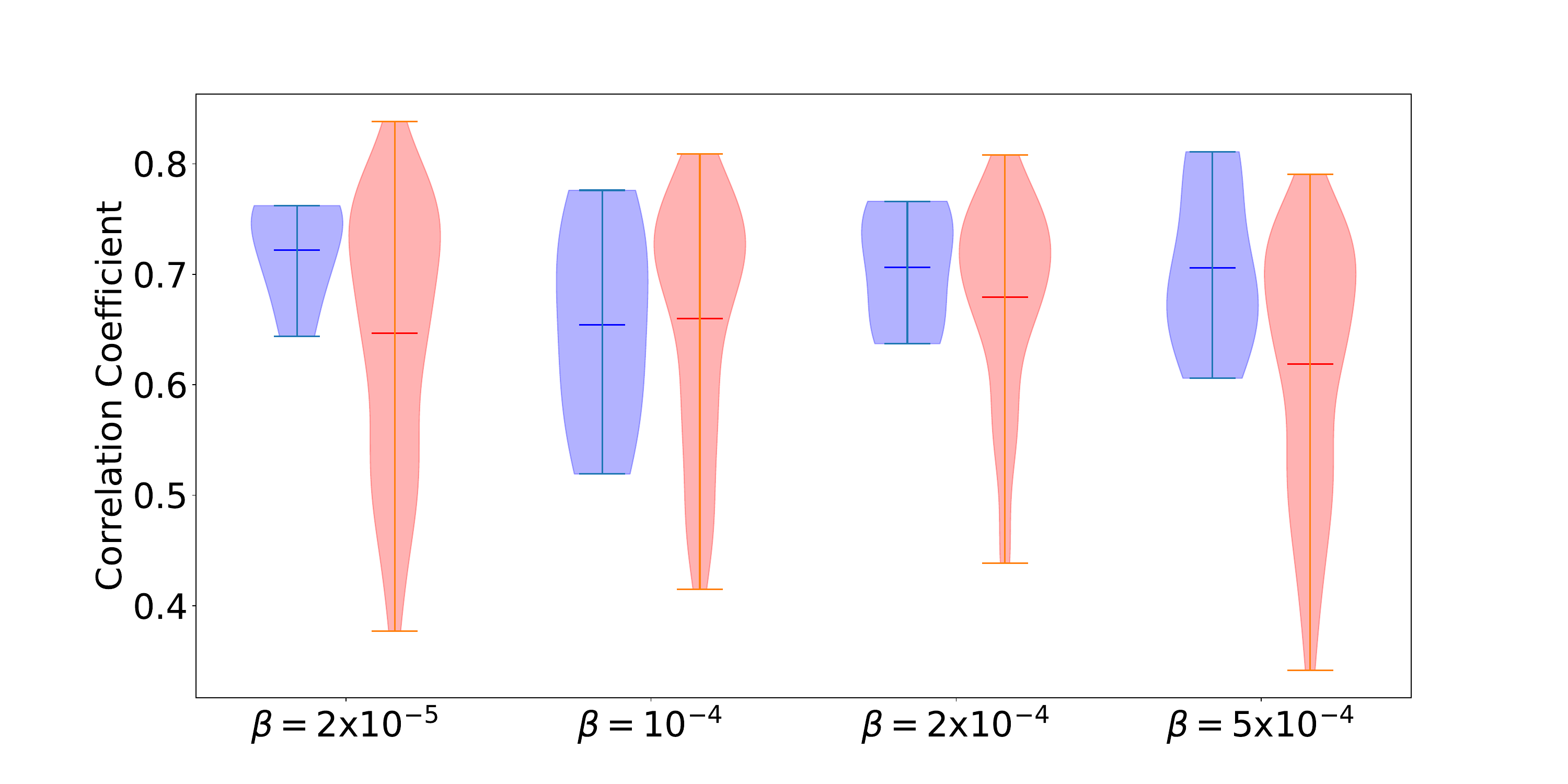}}
    \subfigure[\label{fig:STnRatio_0p6} Coverage ($\text{cov}_{2\sigma}$)]{\includegraphics[width=0.32\textwidth, trim={3cm 1.0cm 5cm 2cm},clip]{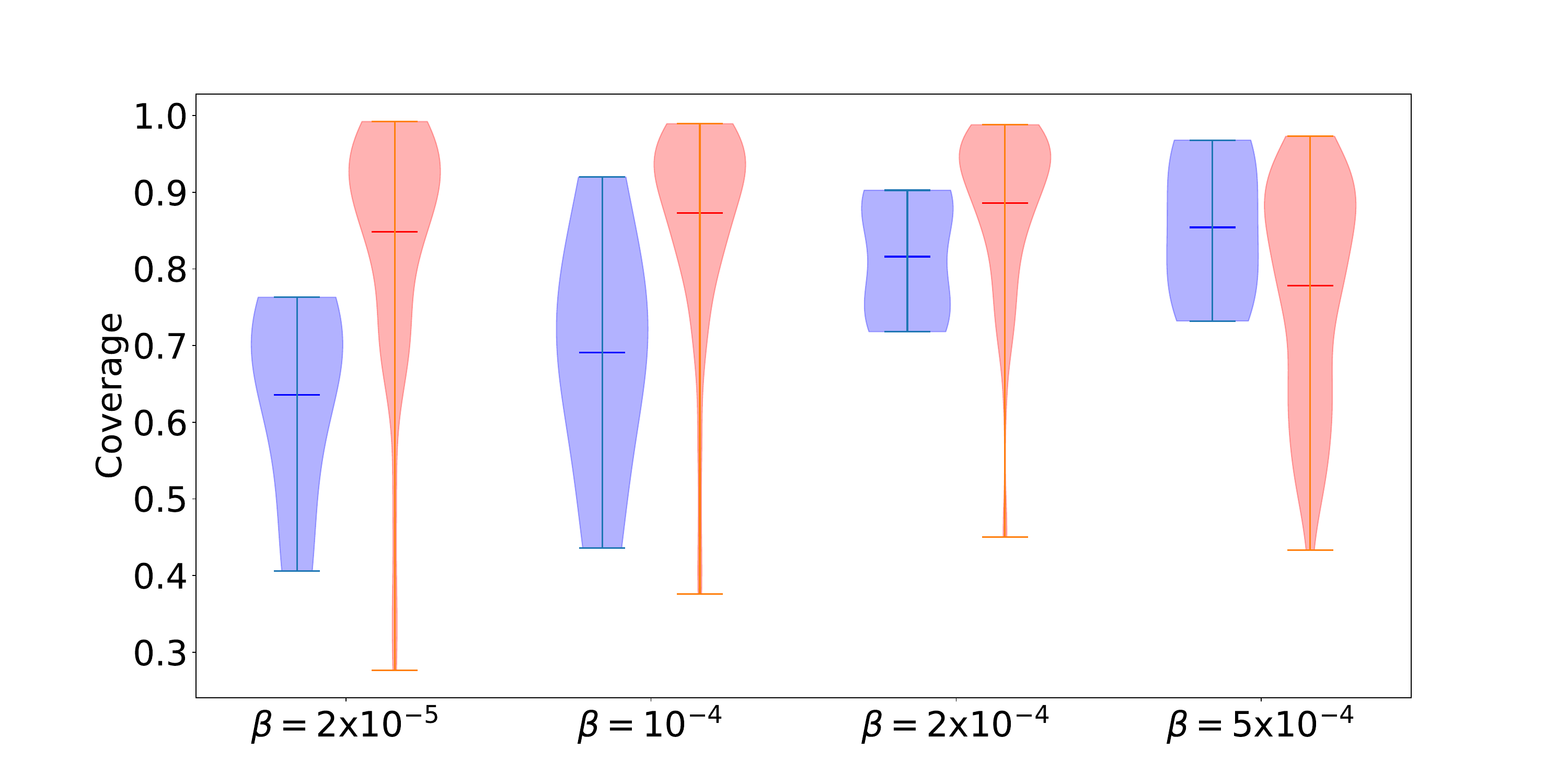}}
    \vspace{-3mm}
    \caption{1-D advection problem: Violin plot showing the distribution with respect to the system parameters for ECLEIRS-UQ trained on clean full resolution data with different values of KL-divergence parameter ($\beta$). The results are shown in blue for the interpolation parameters, while they are in red for extrapolation parameters.}
    \label{fig:Violin_advec_ecleirs_compbeta}
\end{figure}

The first step involves identifying the optimal KL-divergence factor $\beta$ that provides robust uncertainty estimates while not compromising the accuracy of the predictions. The variation of error and uncertainty metrics with $\beta$ for ECLEIRS-UQ is given in \figref{Violin_advec_ecleirs_compbeta}. We observe that the mean errors for the interpolation parameters increase with the increase in $\beta$, especially for $\beta = 5 \times 10^{-4}$ which has visibly high errors. Interestingly, the larger errors for the extrapolation parameters decrease slightly, especially at the highest value of $\beta$, although this comes at the cost of increase in the lower extrapolation errors. We do not observe a clear trend in the variation of correlation coefficient and mean values remain similar for all $\beta$ values for both interpolation and extrapolation parameters. Conversely, we observe a clear trend of increase in mean coverage for interpolation parameters with an increase in $\beta$. Mean coverage for the extrapolation parameter appears to lower for the highest $\beta$ value. These results indicate there is tradeoff between accuracy and coverage estimation for interpolation parameters with an increase in $\beta$. This tradeoff is consistent with the role of the KL-divergence regularization term. Increasing $\beta$ places greater emphasis on latent-space regularization, encouraging the posterior distribution to remain closer to the prior. As a consequence, uncertainty estimates become more conservative and coverage generally improves, while the reconstruction accuracy may deteriorate because the latent representation becomes less flexible. Through this analysis, we observe that $\beta = 10^{-4}$ appears to provide a well-balanced result with lower errors while also providing good uncertainty metrics. Therefore, we use this $\beta$ value for rest of the tests. Note that the same trend was also observed for IRS-UQ and PI-IRS-UQ and therefore, we use the same value of $\beta$ for consistency between different modeling approaches. 

\begin{figure}[t]
    \centering
    \subfigure[\label{fig:} {$\nu = [0.89, 1.0, 0.25]$}]{\includegraphics[width=0.32\textwidth, trim={0.2cm 0cm 1cm 0.5cm},clip]{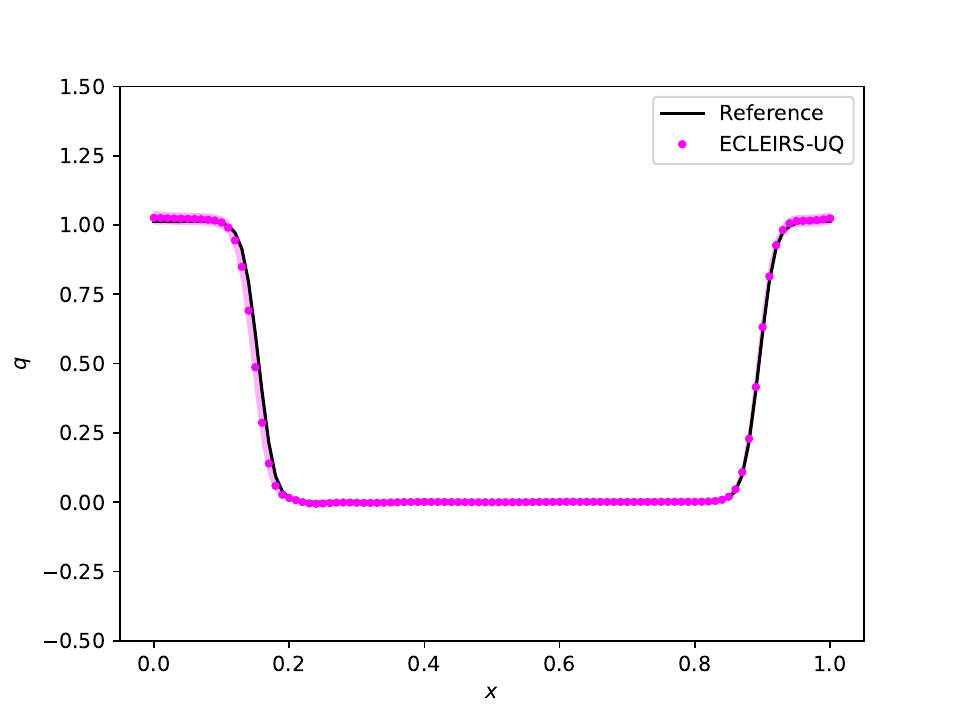}}
    \subfigure[\label{fig:} {$\nu = [0.7, 1.12, 0.19]$}]{\includegraphics[width=0.32\textwidth, trim={0.2cm 0cm 1cm 0.5cm},clip]{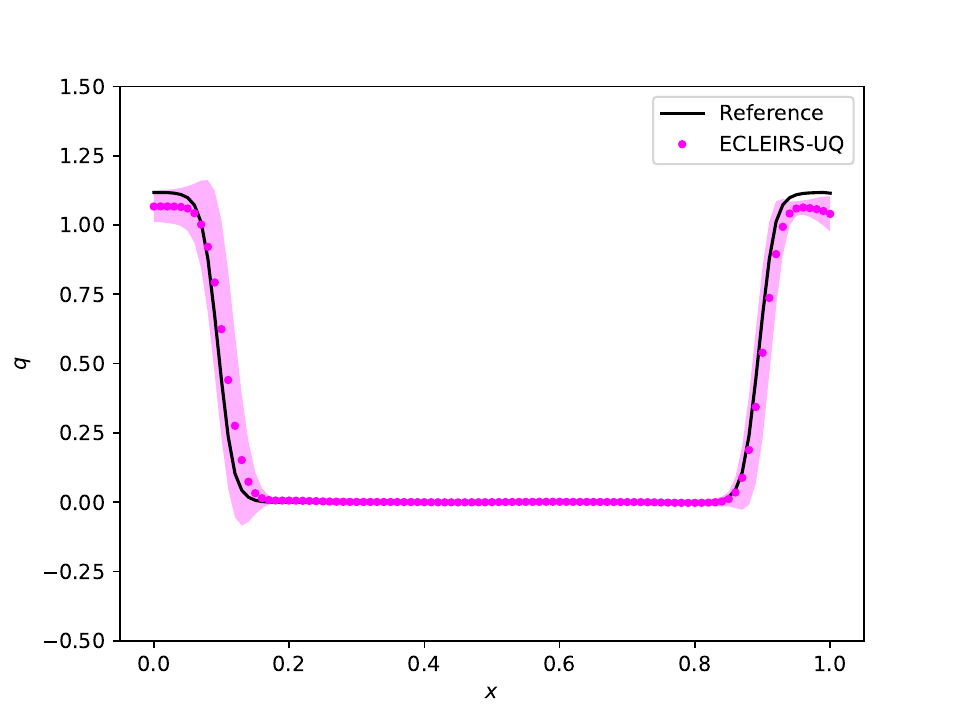}}
    \subfigure[\label{fig:} {$\nu = [0.4, 0.65, 0.05]$}]{\includegraphics[width=0.32\textwidth, trim={0.2cm 0cm 1cm 0.5cm},clip]{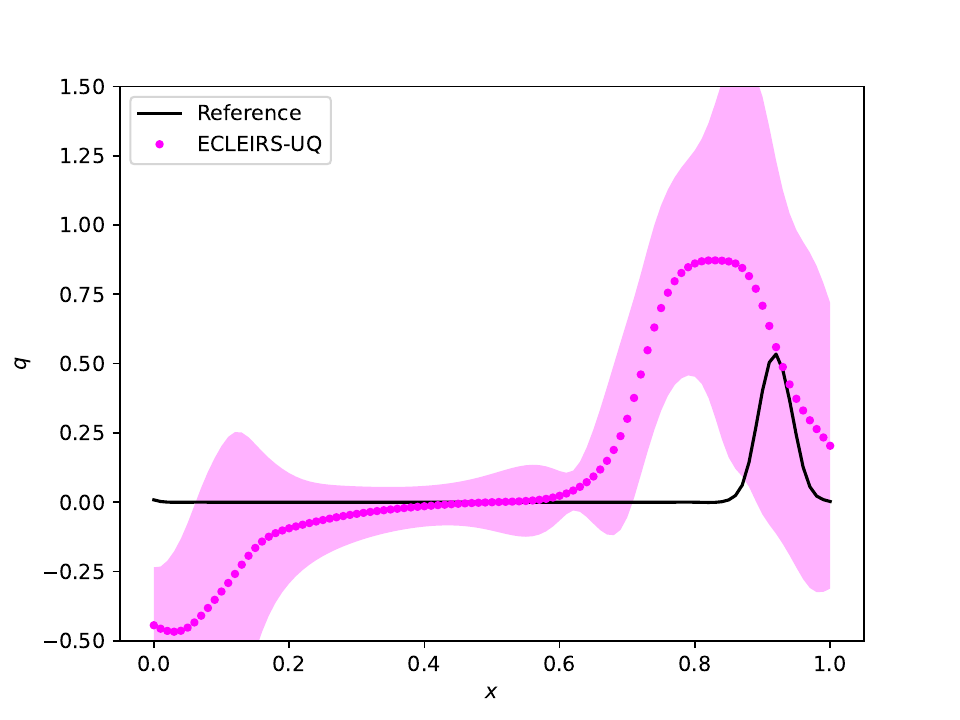}}  
    \vspace{-3mm}
    \caption{1-D advection problem: Solution prediction by ECLEIRS-UQ trained on full resolution data without noise for (a) interpolation parameter (at $t = 1$s), (b) slight extrapolation parameter (at $t = 1.29$s) and (c) significant extrapolation parameter (at $t = 2.25$s). The pink circles depict the mean prediction $q^m_{\text{mean}}$, while the shaded region is $2 q^m_{\text{SD}}$.}
    \label{fig:Solution_advec}
\end{figure}

\begin{figure}[t]
    \centering
    \begin{picture}(0,20)    
    \put(-215,0){Reference}
    \put(-150,0){Mean ($q^m_{\text{mean}}$)}    
    \put(-50,0){SD ($q^m_{\text{SD}}$)}    
    \put(50,0){Error ($\epsilon$)}    
    \put(140,0){Coverage ($\text{cov}_{2\sigma}$)}    
    \end{picture}    
    \subfigure[\label{fig:STnRatio_0p6} {$\nu = [0.89, 1.0, 0.25]$} ]{\includegraphics[width=\textwidth,trim={0cm 62cm 0cm 62cm},clip]{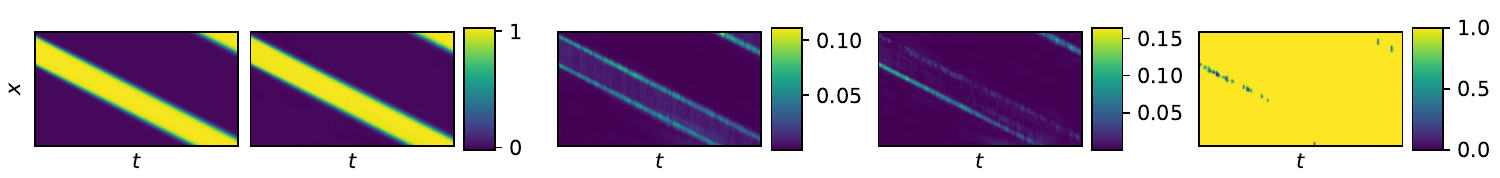}}
    \subfigure[\label{fig:STnRatio_0p6} {$\nu = [0.7, 1.12, 0.19]$} ]{\includegraphics[width=\textwidth,trim={0cm 62cm 0cm 62cm},clip]{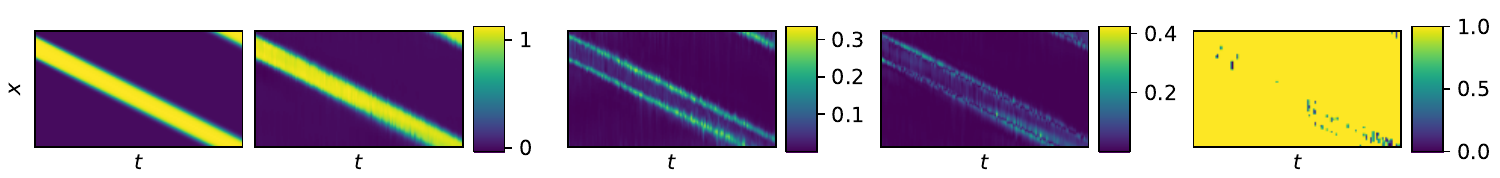}}
    \subfigure[\label{fig:STnRatio_0p6} {$\nu = [0.4, 0.65, 0.05]$} ]{\includegraphics[width=\textwidth,trim={0cm 62cm 0cm 62cm},clip]{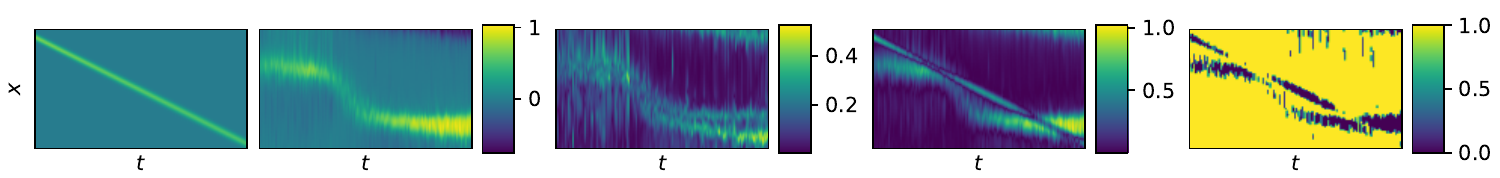}}
    \caption{1-D advection problem: Space-time contour of solution prediction by ECLEIRS-UQ trained on full resolution data without noise for (a) an interpolation parameter, (b) a slight extrapolation parameter and (c) a significant extrapolation parameter.}
    \label{fig:STContour_1Dadvection}
\end{figure}
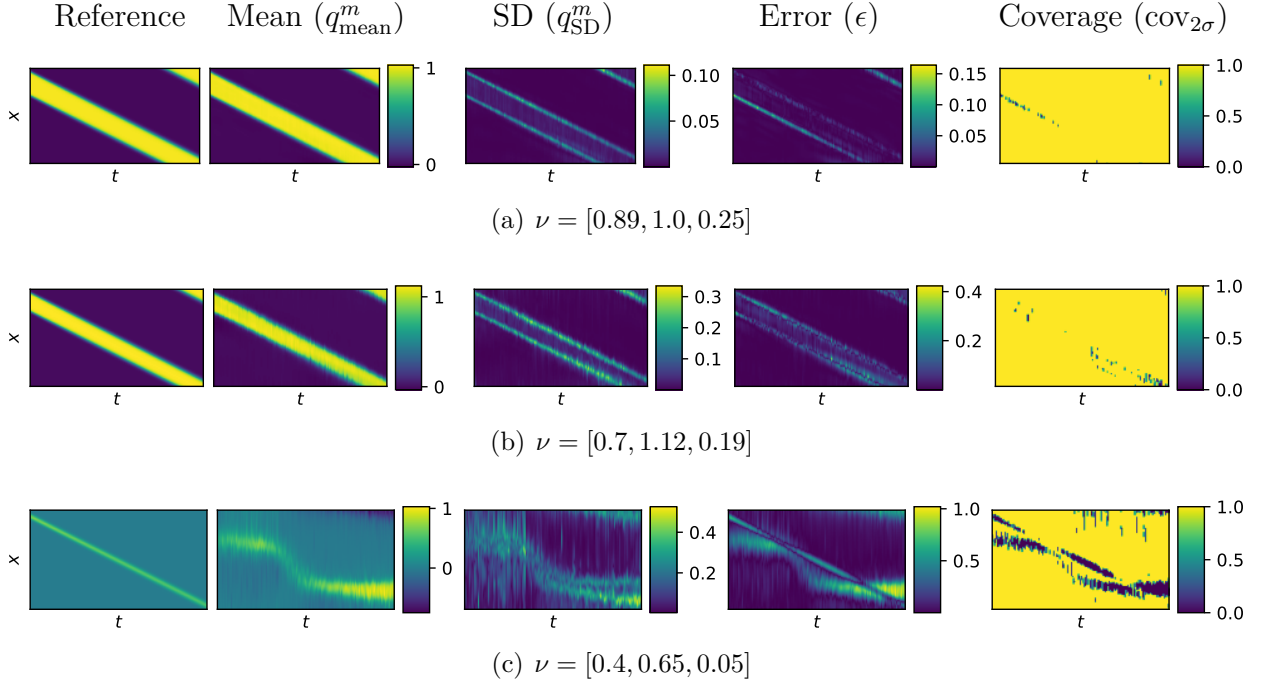

The predicted solution  using ECLEIRS-UQ trained on full-resolution data without noise is shown in \figref{Solution_advec}. For the interpolation parameter, the predicted solution is close to the reference data, while exhibiting low uncertainties. Similarly, predicted solution for the slight extrapolation parameter is similar to the reference, while some errors exist. In this case, the predicted uncertainty in solution prediction covers the ground truth very well. For the significant extrapolation parameter, we observe that the predicted solution deviates significantly from the reference results, while also providing large uncertainty in prediction. The results demonstrate one of the primary advantages of the proposed probabilistic framework. Although the mean prediction becomes increasingly inaccurate as the parameter moves farther from the training manifold, the associated uncertainty estimates increase accordingly, providing a useful indicator of reduced model confidence.

The space-time contours of the solution prediction by ECLEIRS-UQ for both interpolation and extrapolation parameters is shown in \figref{STContour_1Dadvection}. For the interpolation parameter, we observe that the prediction is close to the ground truth results, which is expected as the validation parameter is within the convex hull of testing parameter. Furthermore, we observe that the standard deviation prediction is low and well correlated with the error. We also observe that the coverage is equal to one throughout the domain, indicating that it is highly likely that the reference solution lies within the uncertainty bounds. The low standard deviation observed for the interpolation parameter is consistent with the availability of nearby training examples in parameter space, resulting in a highly localized posterior distribution in the latent space.
For the slight extrapolation parameter, we observe the mean prediction is close to the reference solution but deviates more than the interpolation parameters. In this case, the standard deviation is higher than the interpolation parameters, and correlated well with the errors, which are also higher in this case. Similarly, the coverage is equal to one throughout the space-time domain. The increase in standard deviation relative to the interpolation case indicates that the model correctly recognizes the reduced information available in the extrapolation regime and responds by increasing predictive uncertainty.
 For the significant extrapolation parameter, we observe that the prediction does not agree with the reference results at all and leads to high errors. This behavior is expected as the model is trained on substantially different set of system parameters. Despite the lack of accuracy, we observe that the standard deviation of the result also increase accordingly and correlates moderately well with the errors. While the coverage is not as expansive as other two cases, we do observe adequate coverage for majority of the domain, indicating the uncertainty estimates for highly inaccurate results will likely cover the reference solution. In this scenario, the uncertainty estimates remain informative because they expand substantially in regions where the prediction error is largest.

\begin{figure}[t]
    \centering
    \subfigure[\label{fig:STnRatio_0p6} Relative error ($\epsilon_{r}$)]{\includegraphics[width=0.49\textwidth, trim={2cm 1.5cm 5cm 2cm},clip]{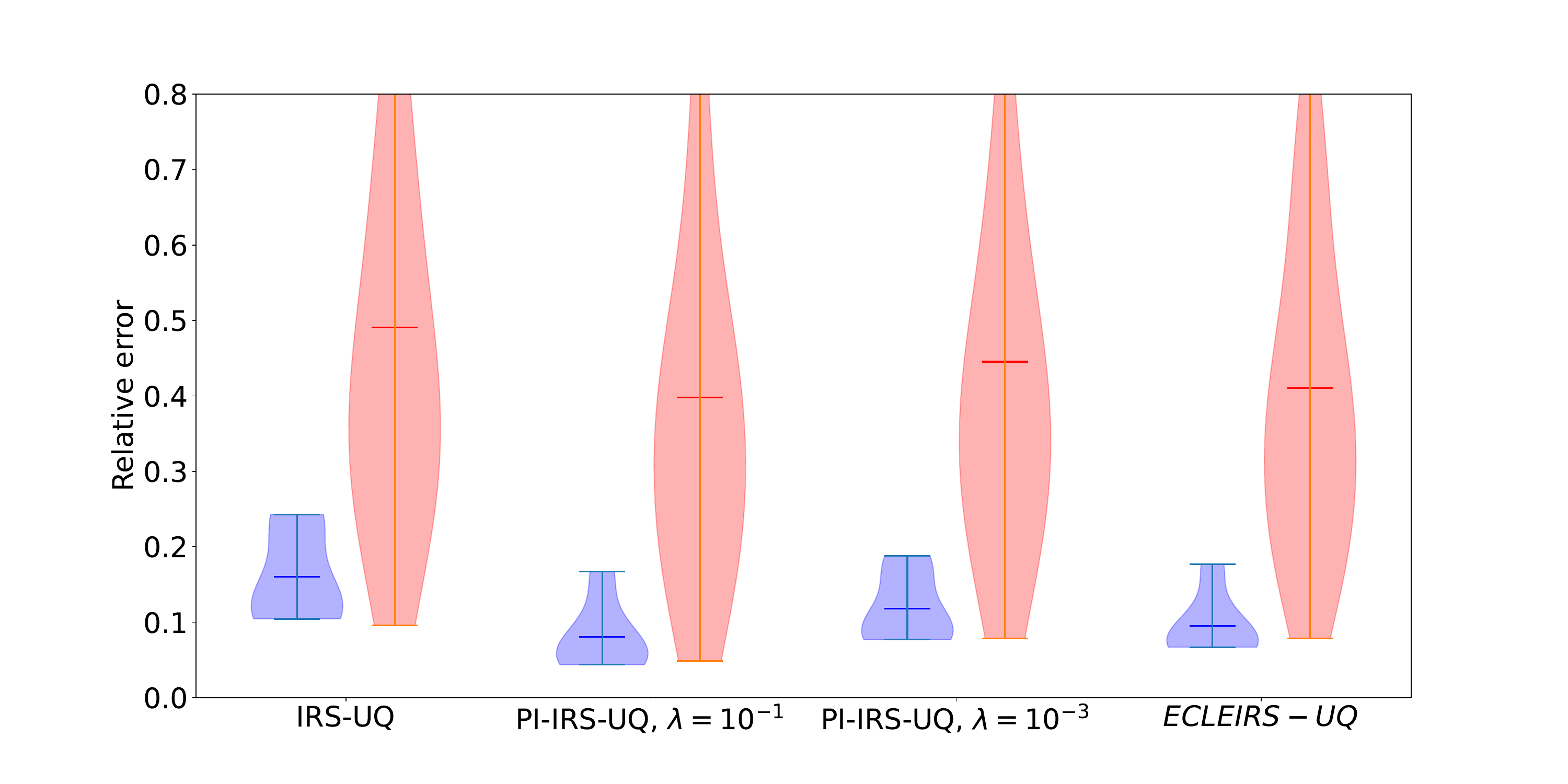}}
    \subfigure[\label{fig:STnRatio_1p0} Mean conservation error ($\epsilon_{\text{consv}}$)]{\includegraphics[width=0.49\textwidth, trim={2cm 1.5cm 5cm 2cm},clip]{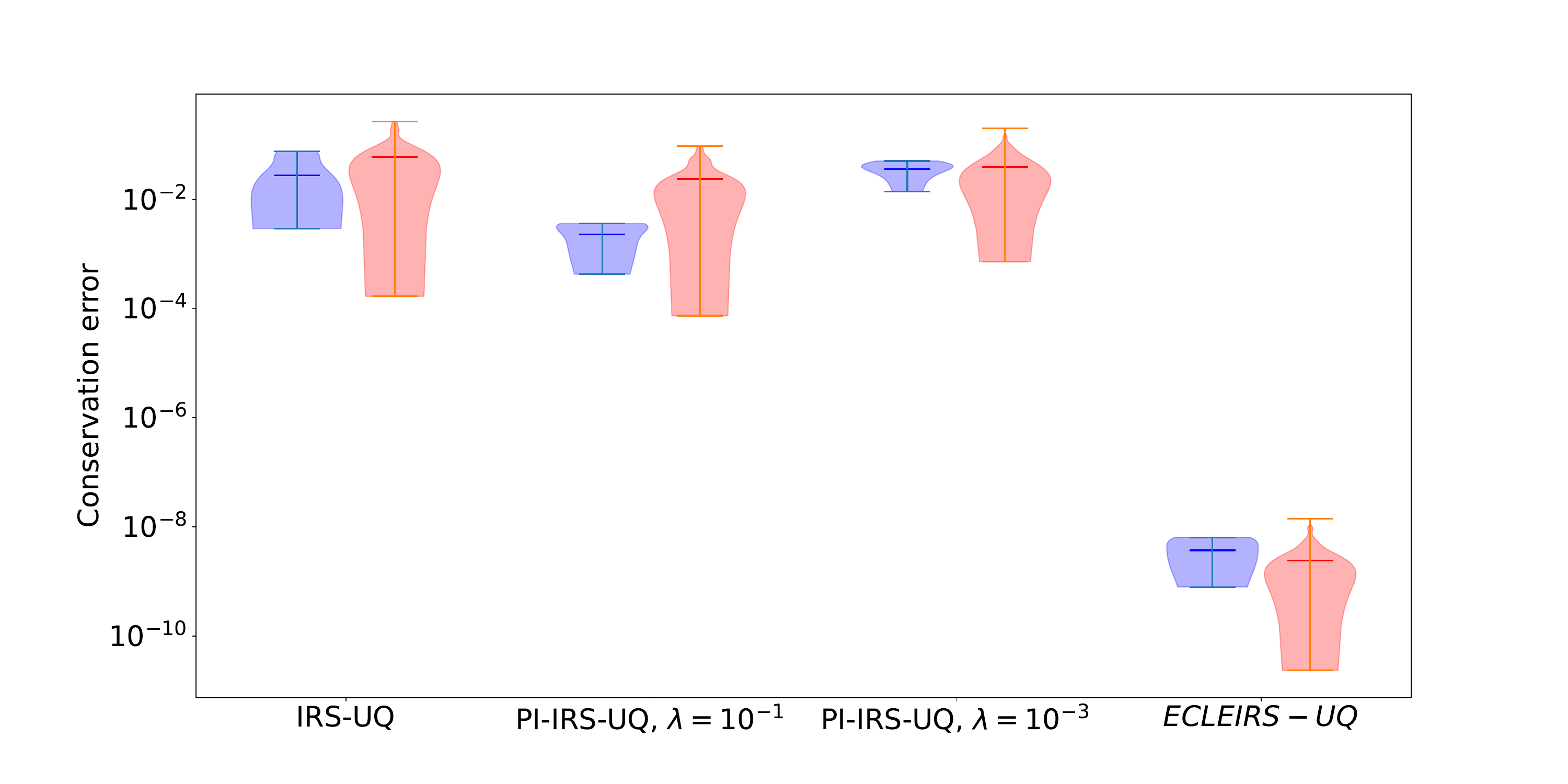}}

    \subfigure[\label{fig:STnRatio_0p6} Correlation Coefficient ($r$)]{\includegraphics[width=0.49\textwidth, trim={2cm 1.5cm 5cm 2cm},clip]{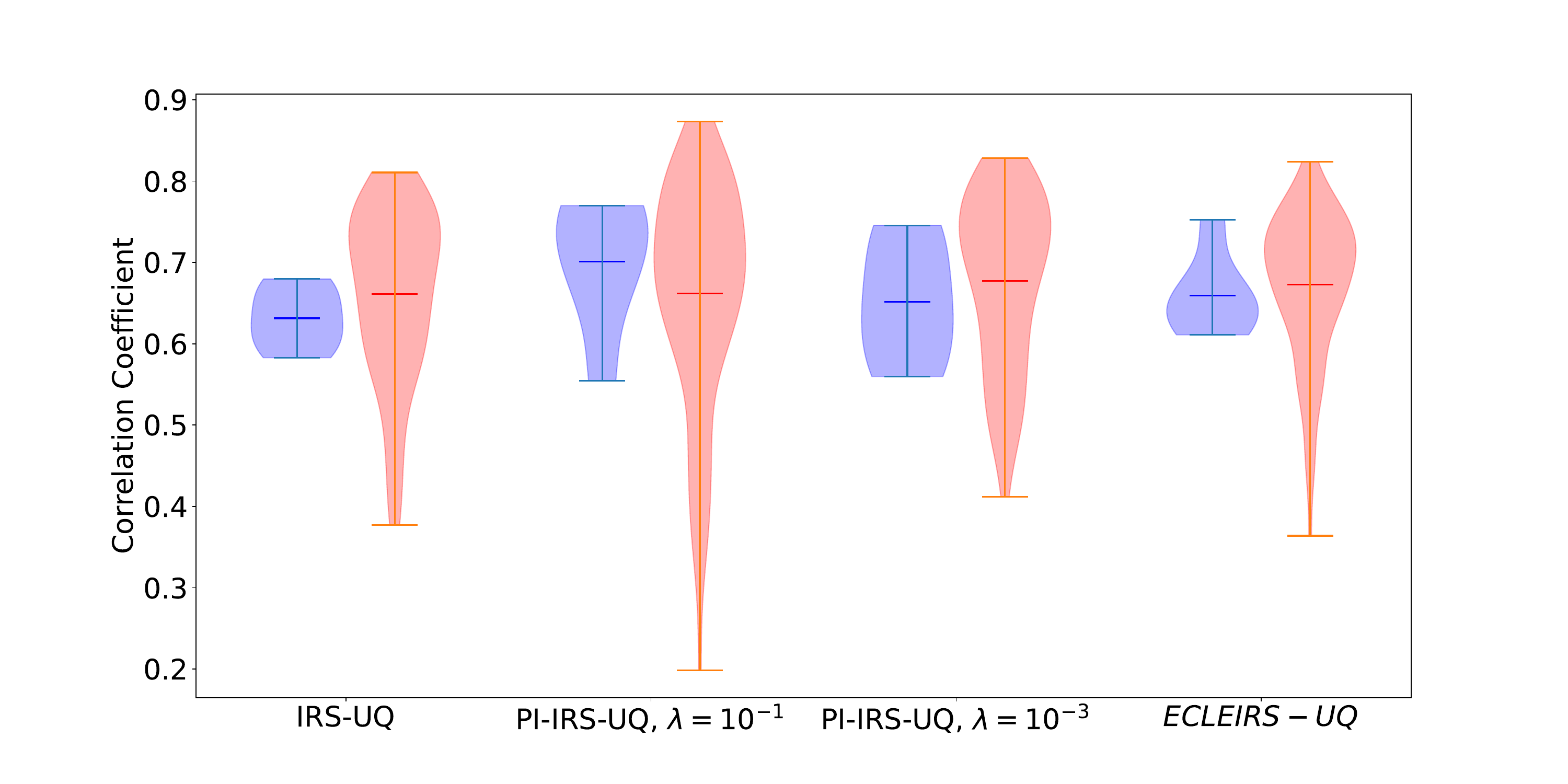}}
    \subfigure[\label{fig:STnRatio_1p0} Coverage ($\text{cov}_{2\sigma}$)]{\includegraphics[width=0.49\textwidth, trim={2cm 1.5cm 5cm 2cm},clip]{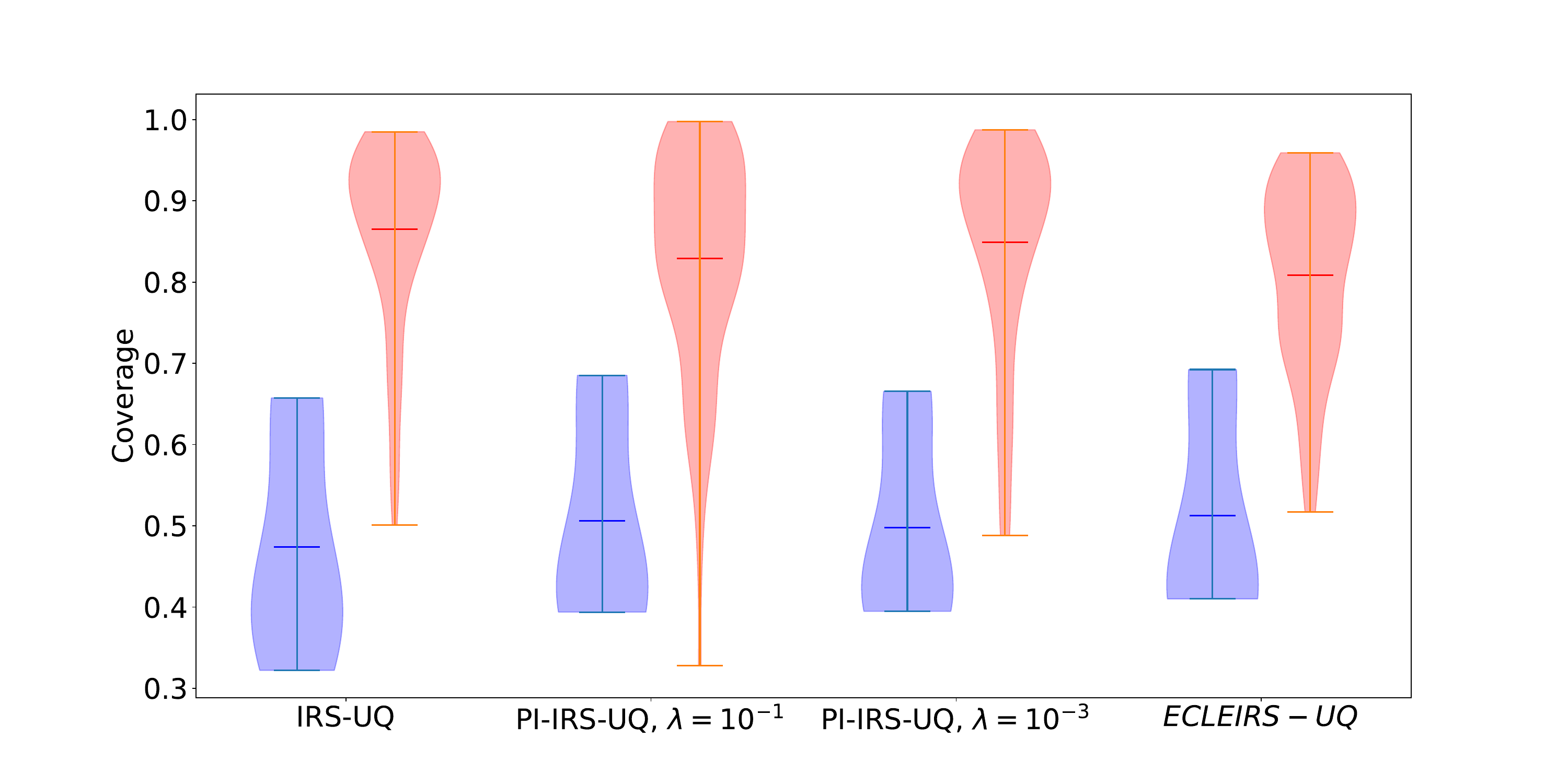}}    
    \vspace{-3mm}
    \caption{1-D advection problem: Violin plot showing the distribution of metrics with respect to the system parameters for different modeling approaches for randomly distributed $4\%$ spatio-temporal sparse training data ($20\%$ sparsity in space and $20\%$ sparsity in time) with added unit normal noise with standard deviation of $0.2$. The results are shown in blue for the interpolation parameters, while they are in red for extrapolation parameters.}
    \label{fig:Violin_advec}
\end{figure}

\begin{figure}[h!]
    \centering
    \subfigure[\label{fig:STnRatio_0p6} Relative error ($\epsilon_{r}$)]{\includegraphics[width=0.32\textwidth, trim={3cm 1.5cm 5cm 2cm},clip]{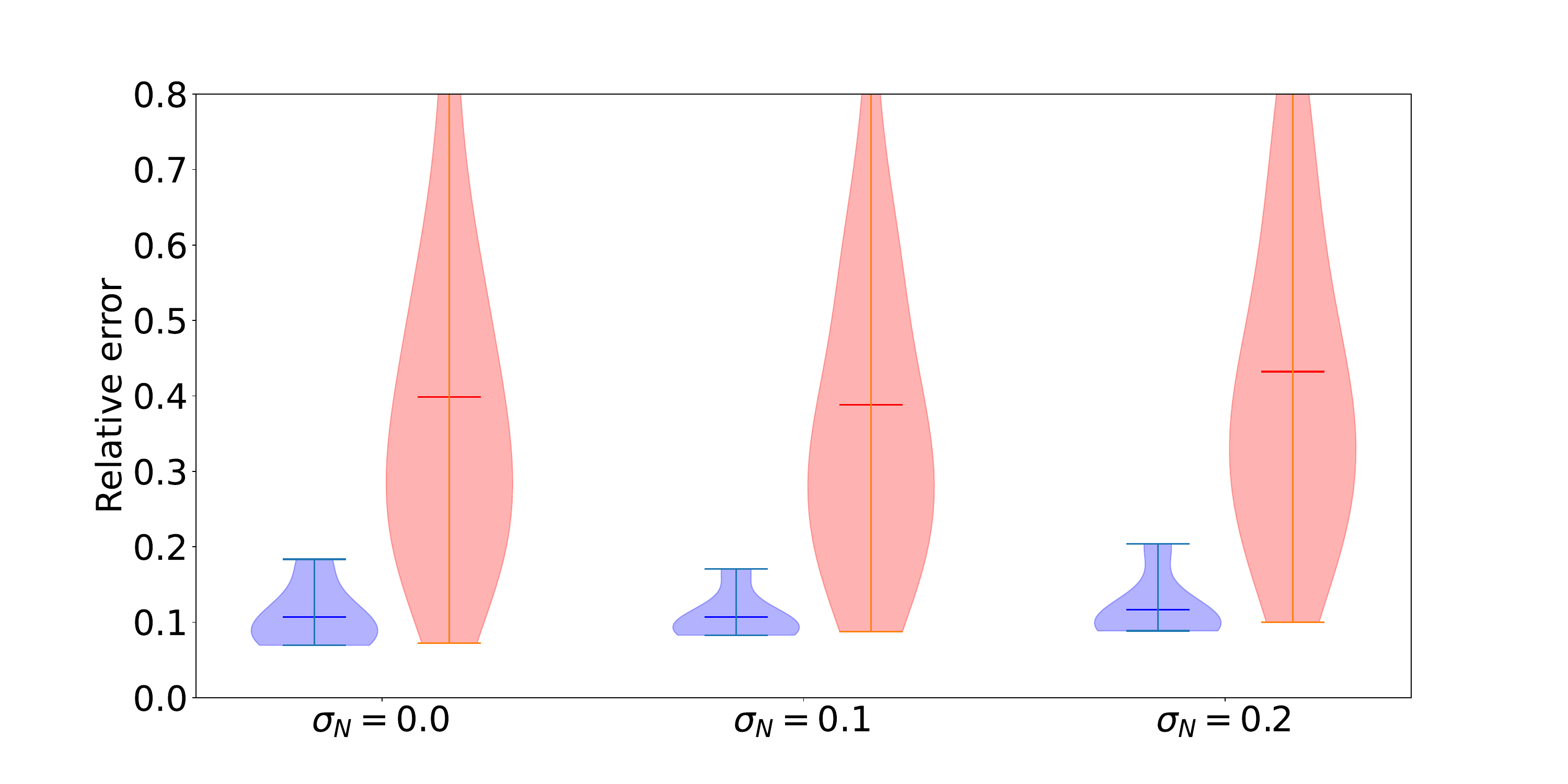}}
    \subfigure[\label{fig:STnRatio_1p0} Correlation Coefficient ($r$)]{\includegraphics[width=0.32\textwidth, trim={3cm 1.5cm 5cm 2cm},clip]{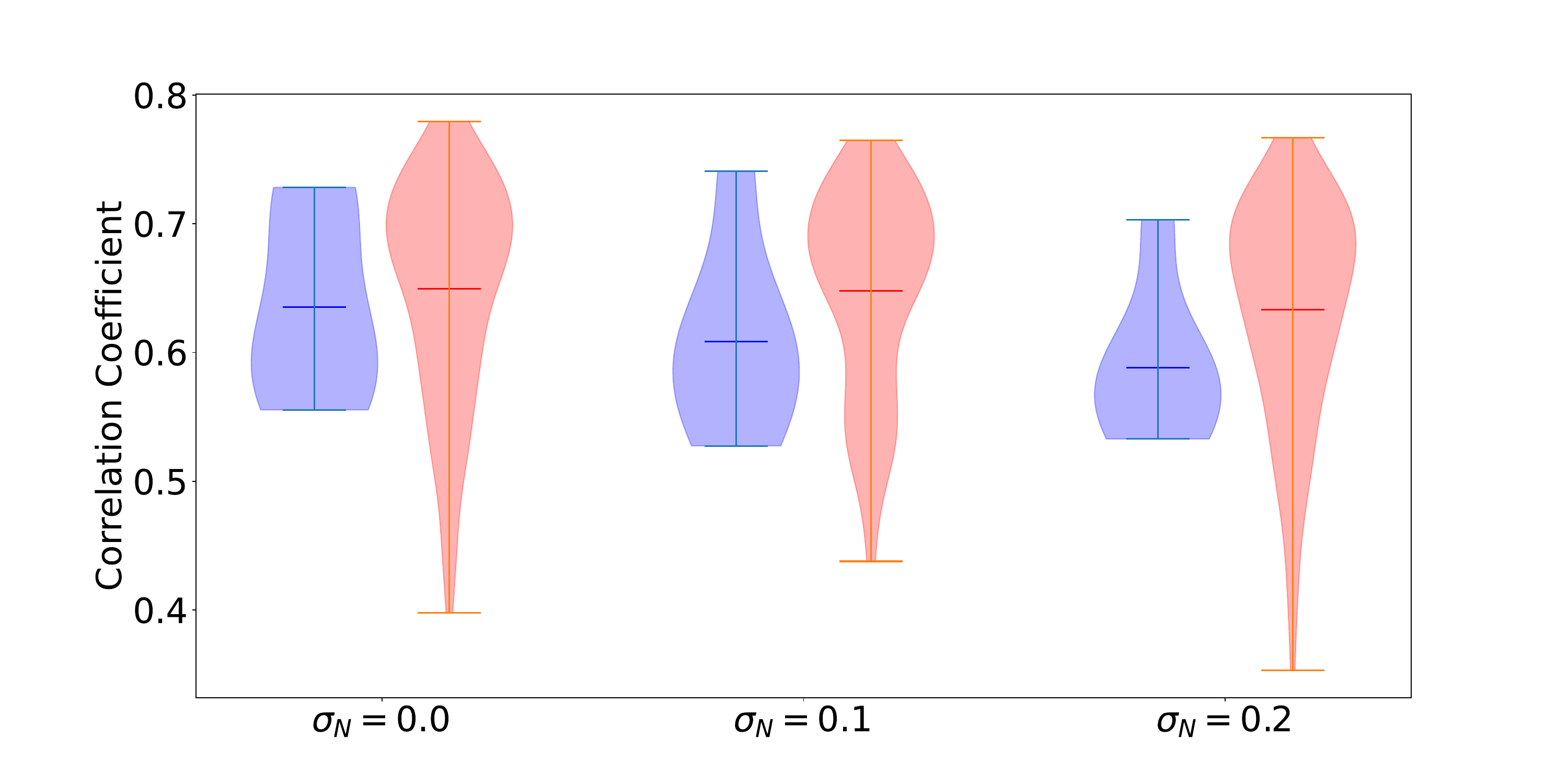}}
    \subfigure[\label{fig:STnRatio_0p6} Coverage ($\text{cov}_{2\sigma}$)]{\includegraphics[width=0.32\textwidth, trim={3cm 1.5cm 5cm 2cm},clip]{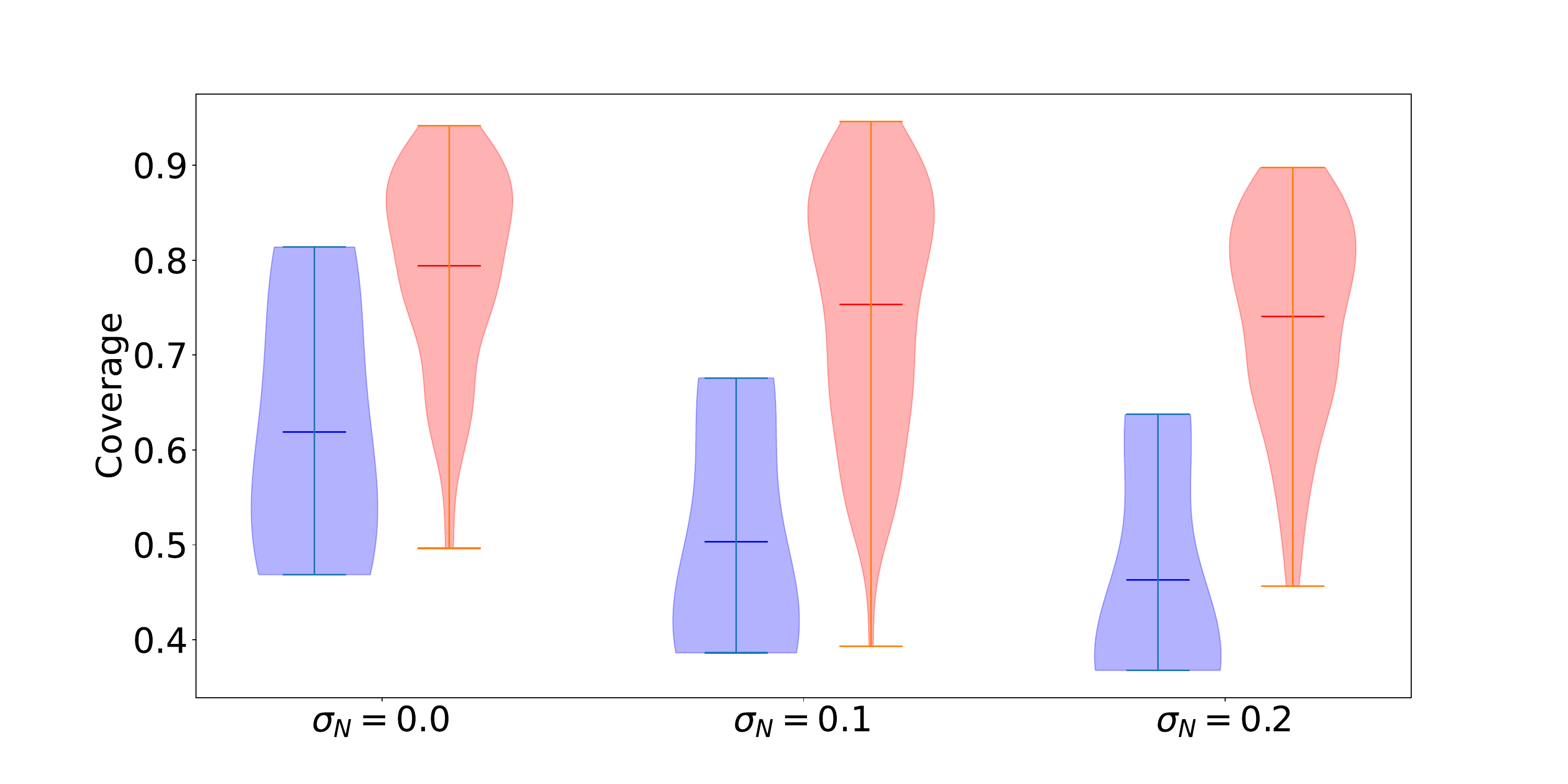}}
    \vspace{-3mm}
    \caption{1-D advection problem: Violin plot showing the distribution with respect to the system parameters for ECLEIRS-UQ trained on randomly distributed $4\%$ spatio-temporal sparse training data ($20\%$ sparsity in space and $20\%$ sparsity in time) with different levels of added noise ($\sigma_n$). The results are shown in blue for the interpolation parameters, while they are in red for extrapolation parameters.}
    \label{fig:Violin_advec_ecleirs_compnoise}
\end{figure}

The violin plots showing the distribution of different result metrics with respect to the system parameters for different modeling approaches trained with sparse and noisy data are shown in \figref{Violin_advec}. The results indicate that PI-IRS-UQ and ECLEIRS-UQ have better accuracy than IRS-UQ in the presence of sparse and noisy training data for both interpolation and extrapolation parameters. While PI-IRS-UQ with $\lambda = 10^{-1}$ has better accuracy than ECLEIRS-UQ, the difference is small and it comes at a high offline-stage expense of tuning these models as also discussed in detail in \cite{Prakash2026}. ECLEIRS-UQ also exhibits the lowest conservation error, at the level
expected for single-precision arithmetic, for both interpolation and extrapolation parameter regimes. The other models exhibit significantly higher conservation errors for both interpolation and extrapolation regimes, indicating that these methods do not guarantee conservation in the testing dataset. All the modeling approaches have similar performance when it comes to both UQ metrics: correlation coefficient and coverage. This result indicates that the introduction of physical constraints does not adversely affect UQ capability. Instead, the primary benefit of the physics-constrained approaches is improved predictive accuracy and conservation-law satisfaction. 
The results show that all models exhibit, on an average, a high correlation coefficient for both interpolation and extrapolation parameters. The high mean correlation of $\approx 0.7$ indicates the model is performing well in identifying regions of both low and high errors in the space-time domain. High average coverage across parameters for extrapolation is also an encouraging sign for the model performance. While the coverage for the interpolation parameter is low, it is not very concerning as the errors are also low for interpolation. Therefore, low errors that are not covered by standard deviation are not detrimental to how the model prediction is analyzed in a deployment scenario. Conversely, having high errors during extrapolation that are well covered are considerably more important as it provides a conservative estimate of how the model fails. 

The distribution of prediction error, correlation coefficient and coverage for ECLEIRS-UQ trained on sparse data with different levels of added noise is shown in \figref{Violin_advec_ecleirs_compnoise}. The results indicate that the performance of ECLEIRS-UQ does not degrade significantly with an increase in added noise. There is a slight increase in the relative errors and a slight decrease in coverage for the interpolation parameters but this increase is not significant. These results are encouraging because they indicate robustness in solution prediction and uncertainty estimation capabilities with reduced quality of data. This robustness is particularly important for practical applications where measurement uncertainty is unavoidable and may exceed the noise levels considered in the present benchmark problem.

\begin{figure}[t]
    \centering
    \includegraphics[width=0.4
    \linewidth]{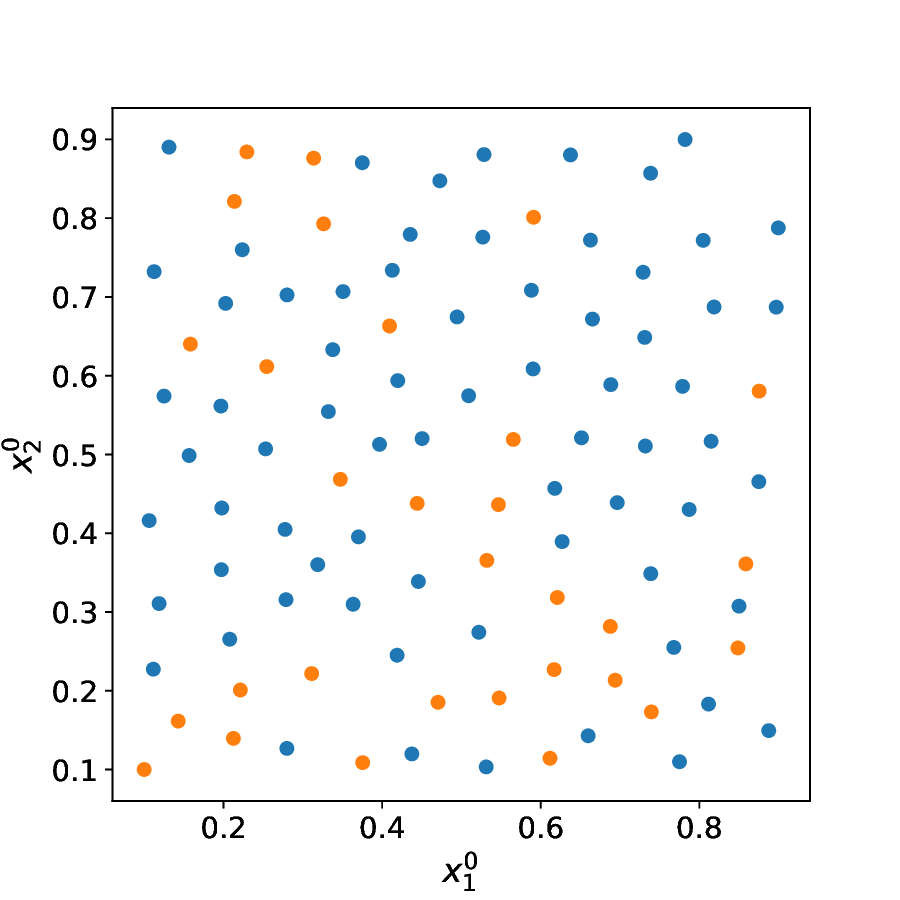}
    \caption{The selection of parameter points used in the learning dataset (blue markers) and validation dataset (orange markers). The figure is replicated from author's prior work \cite{Prakash2026}.}
    \label{fig:Euler2D_paramsel}
\end{figure}

\begin{figure}[t]
    \centering
    \subfigure[1\% sparsity\label{fig:STnRatio_0p01_Euler}]{\includegraphics[width=0.32\textwidth, trim={0.2cm 0cm 1.0cm 0.5cm},clip]{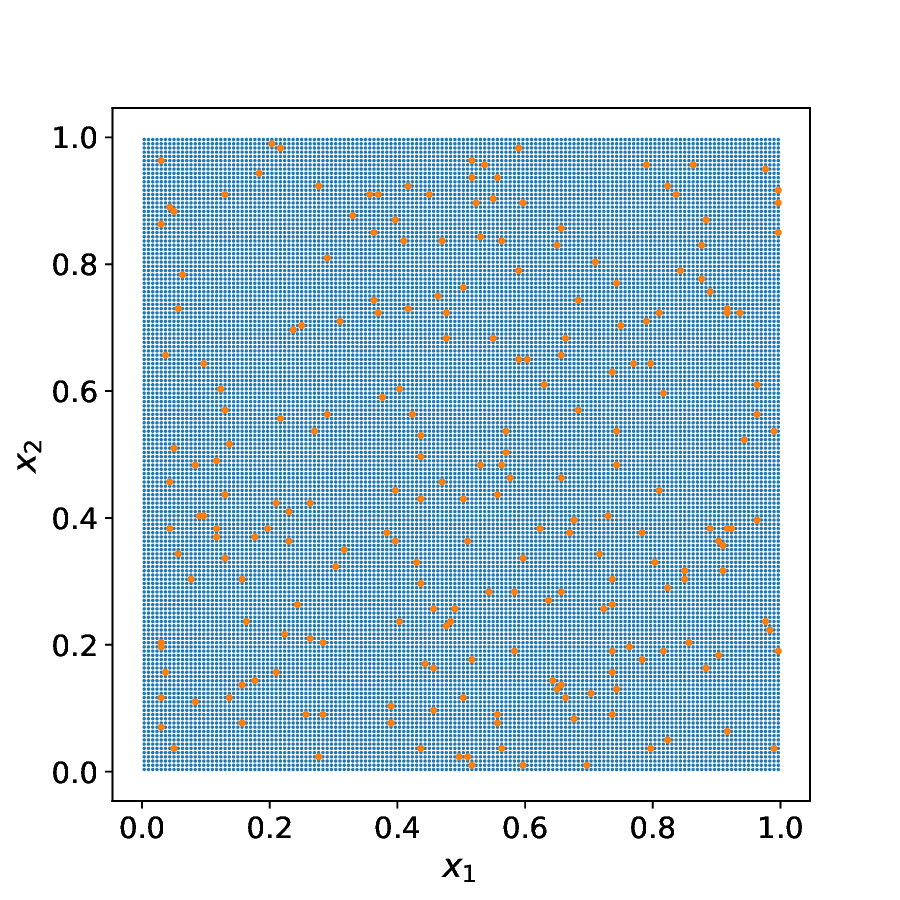}}
    \subfigure[5\% sparsity\label{fig:STnRatio_0p05_Euler}]{\includegraphics[width=0.32\textwidth, trim={0.2cm 0cm 1.0cm 0.5cm},clip]{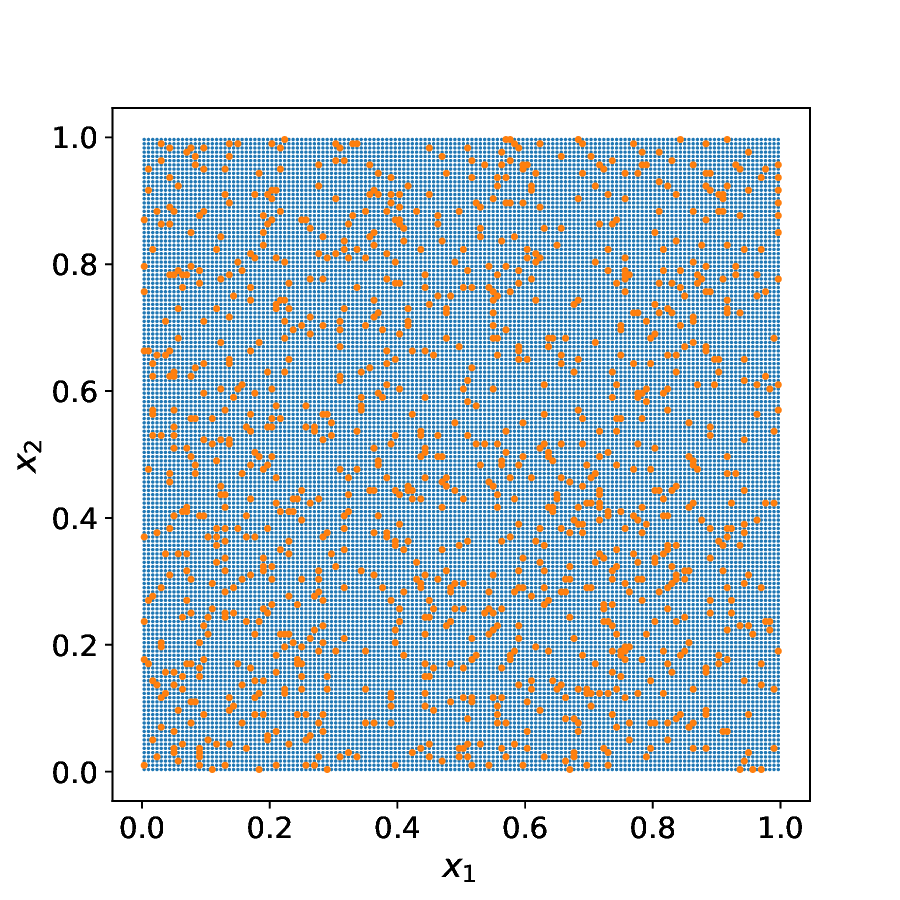}}
    \subfigure[20\% sparsity\label{fig:STnRatio_0p2_Euler}]{\includegraphics[width=0.32\textwidth, trim={0.2cm 0cm 1.0cm 0.5cm},clip]{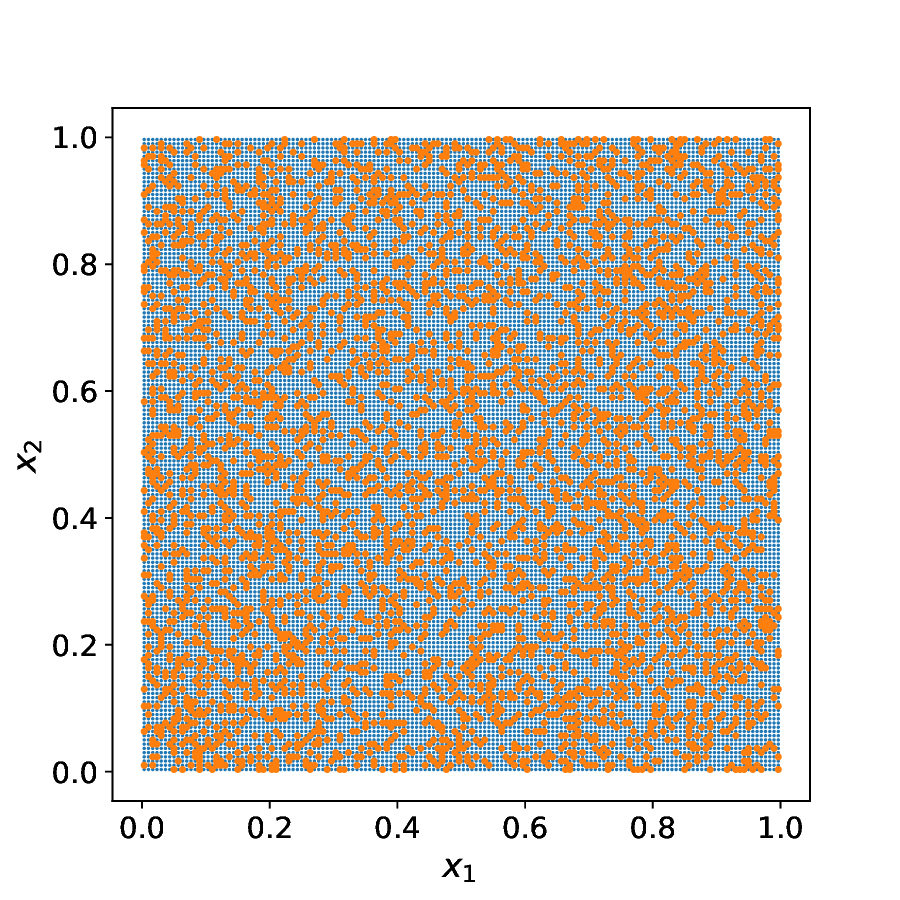}}
    \vspace{-3mm}
    \caption{2-D Euler problem: Spatial locations of data used to learn different reduced state dynamics models at different spatial sparsity. The blue markers denote the spatial resolution of high-fidelity simulation and the orange marker denote the sparse data locations used for learning the model. Note that marker size for sparse points is 5 times greater than the marker size of high-fidelity spatial locations, thereby the selected data is sparser than how it appears here. The figure is replicated from the authors' prior work \cite{Prakash2026}.}
    \label{fig:STnRatio_Euler}
\end{figure}

\begin{figure}
    \centering
    \subfigure[Relative error ($\epsilon_{r}$)]{\includegraphics[width=0.32\textwidth, trim={3cm 1.0cm 5cm 2cm},clip]{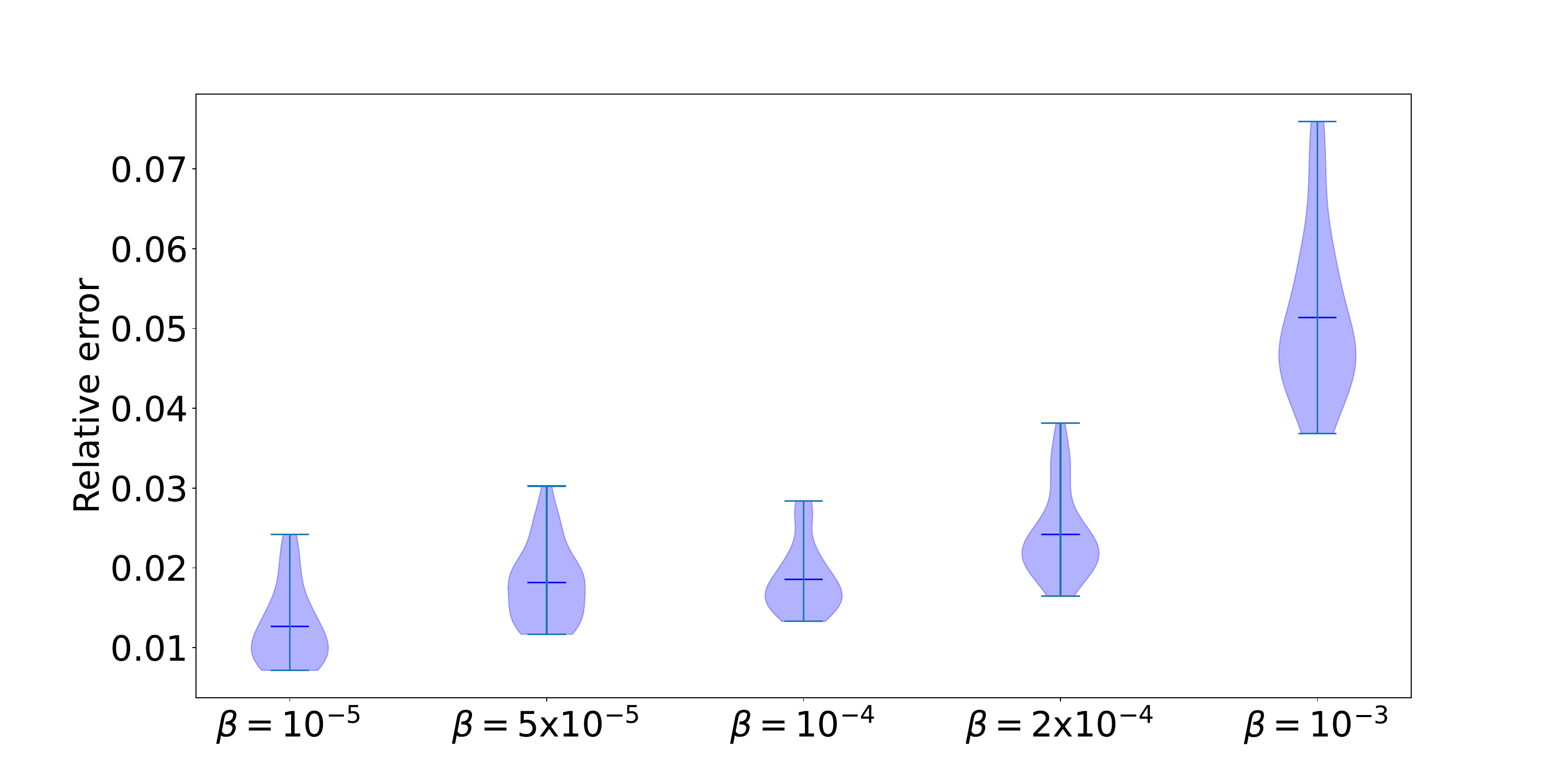}}
    \subfigure[Correlation Coefficient ($r$)]{\includegraphics[width=0.32\textwidth, trim={3cm 1.0cm 5cm 2cm},clip]{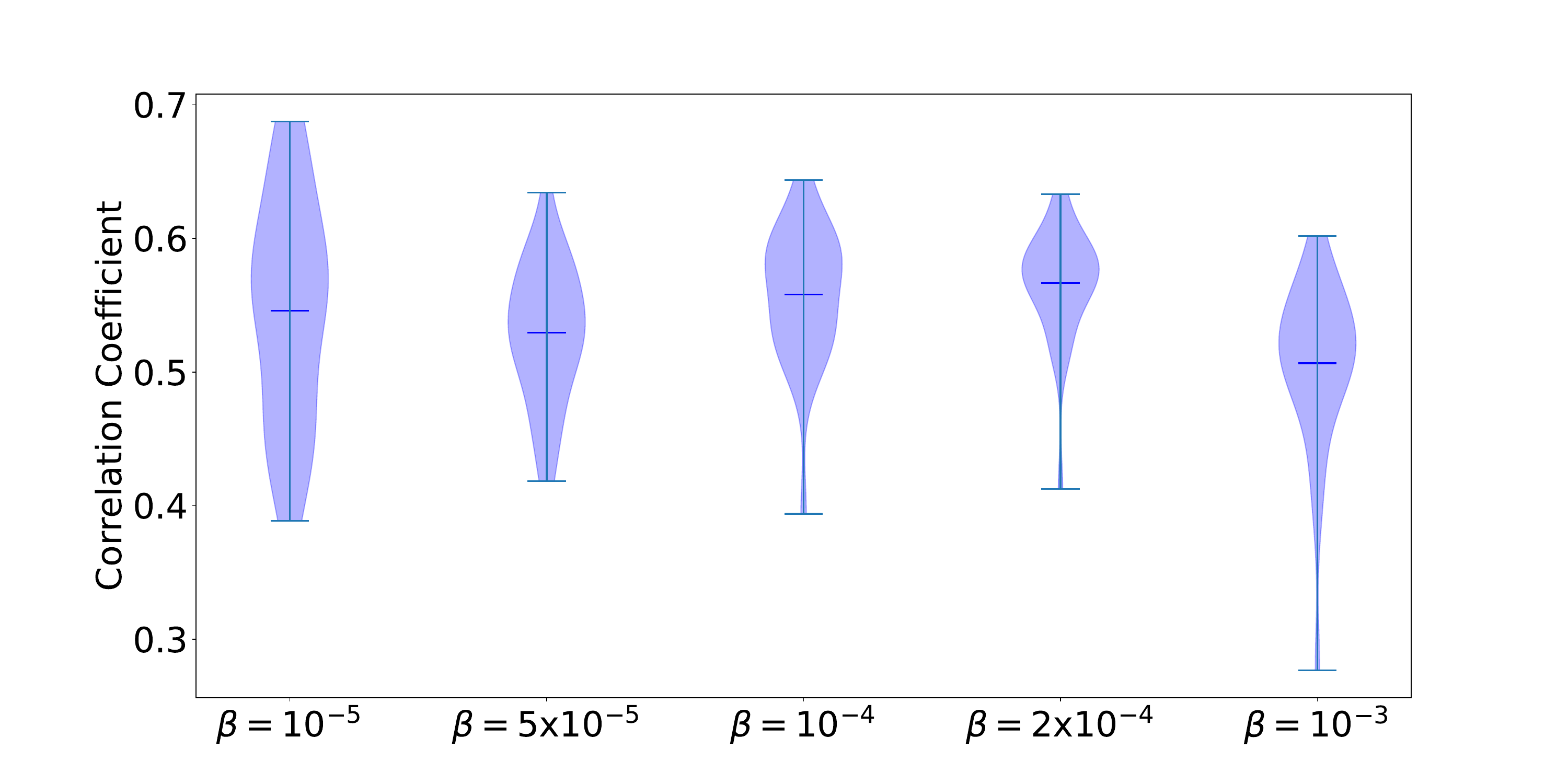}}
    \subfigure[Coverage ($\text{cov}_{2\sigma}$)]{\includegraphics[width=0.32\textwidth, trim={3cm 1.0cm 5cm 2cm},clip]{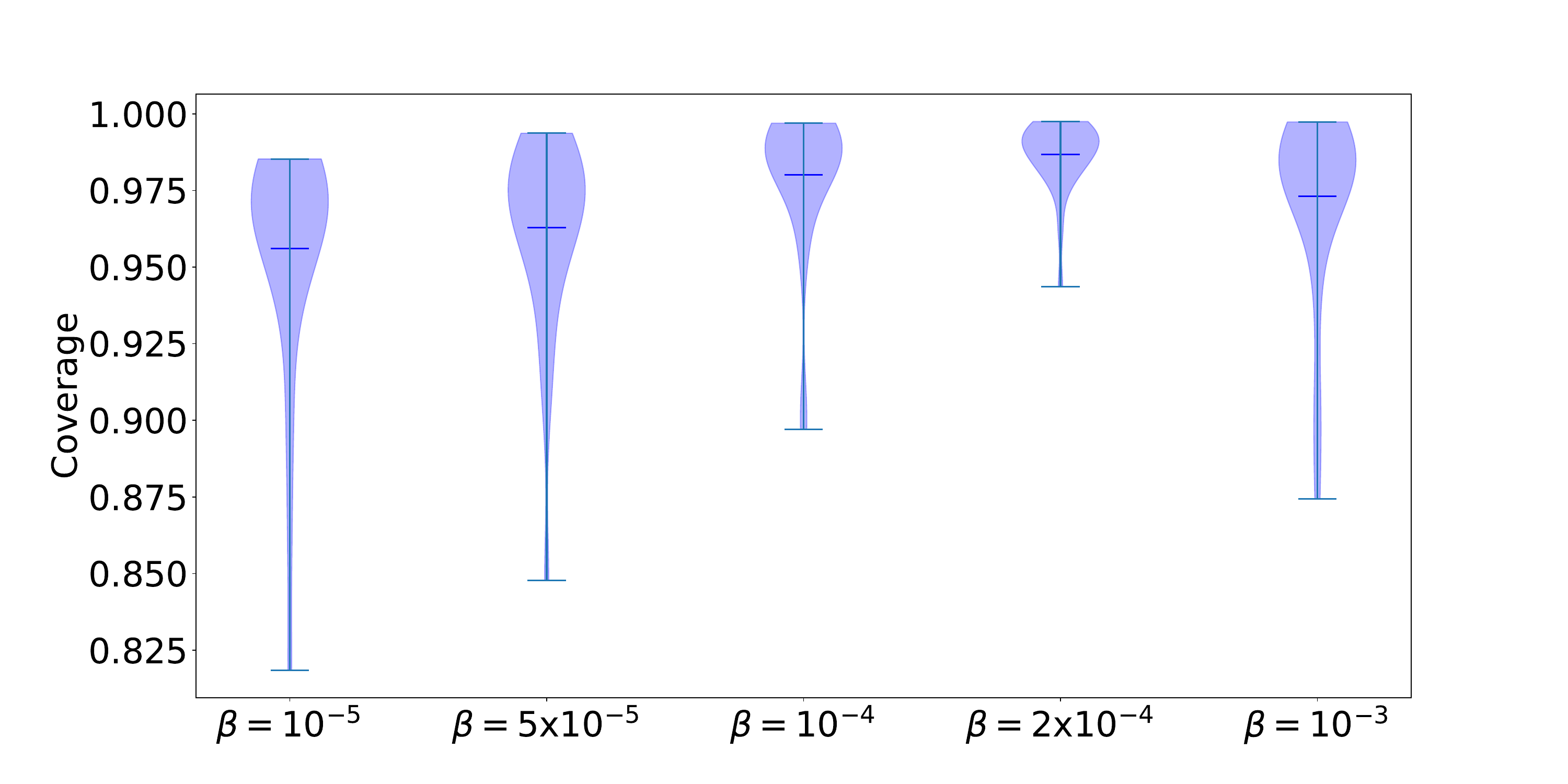}}
    \vspace{-3mm}
    \caption{2-D Euler problem: Violin plot showing the distribution with respect to the system parameters for ECLEIRS-UQ trained on $5\%$ spatial and $20\%$ temporal sparse clean data with different values of KL-divergence parameter ($\beta$).}
    \label{fig:Violin_Euler_ecleirs_compbeta}
\end{figure}

\subsection{2-D Euler equation} 

We consider the evolution of density fields governed by the set of Euler equations with the following conservation form:
\begin{equation}
    \frac{\p \pmb{q}}{\p t} + \nabla \cdot \bm{f} (\bm{q}) = 0,
\end{equation}
where $\pmb{q} = [\rho, \; \rho \bm{u}, \; \rho E]^T$ and $\pmb{f} (\pmb{q}) = [\rho \pmb{u},\;  \rho \bm{u} \otimes \bm{u} + p \bm{I},\; (E + p) \bm{u}]^T$ are the solution and flux vectors respectively. An ideal gas equation of state with $\gamma = 1.4$ is considered to couple the equations. The simulation setup is similar to 2-D version of classical Sod-Tube problem, where the initial density and pressure jumps are propagated over time $t \in [0,2]$. The simulation is parameterized with the initial density field separates the two materials with the material interface corner placed at $\bm{\nu} = [x^0_1, x^0_2]$. The domain is partitioned as $\Omega = \Omega_1\cup\Omega_2$, where $\Omega_1 \coloneqq \{ \pmb{x}\textnormal{ : } x_1\leq x_1^0\text{ and } x_2\leq x_2^0\}$ and $\Omega_2 \coloneqq \{ \pmb{x}\textnormal{ : } x_1> x_1^0\text{ or } x_2> x_2^0\} = \Omega\backslash\Omega_1$. Additional details about the simulation setup can be found in \cite{Prakash2026}. This simulation code is used to generate density fields for 100 parametric combinations shown in \figref{Euler2D_paramsel}. From this generated data, $70\%$ of parameters are used to train proposed models, whereas the rest $30\%$ of parameters are used to validate modeling methods. In this example we model the evolution of density fields, therefore, conservation is only enforced in the continuity equation. Exact conservation in other equations, such as momentum and energy, can also be enforced but this enforcement comes at the cost of loss of consistent prediction of momentum and energy flux as also briefly discussed in \cite{Prakash2026}. Similar to the 1-D advection problem, we assess the performance of different modeling approaches for scenarios where the data quality is degraded. We simulate these scenarios by sparsely sampling the domain in space and time (shown in \figref{STnRatio_Euler}).

\begin{figure}[t!]
    \centering
    \begin{picture}(0,20)    
    \put(-215,0){Reference}
    \put(-140,0){Mean ($q^m_{\text{mean}}$)}    
    \put(-40,0){SD ($q^m_{\text{SD}}$)}    
    \put(60,0){Error ($\epsilon$)}    
    \put(140,0){Coverage ($\text{cov}_{2\sigma}$)}    
    \end{picture}    
    \subfigure[\label{fig:STnRatio_0p6} IRS-UQ]{\includegraphics[width=\textwidth,trim={0cm 37.8cm 0cm 38cm},clip]{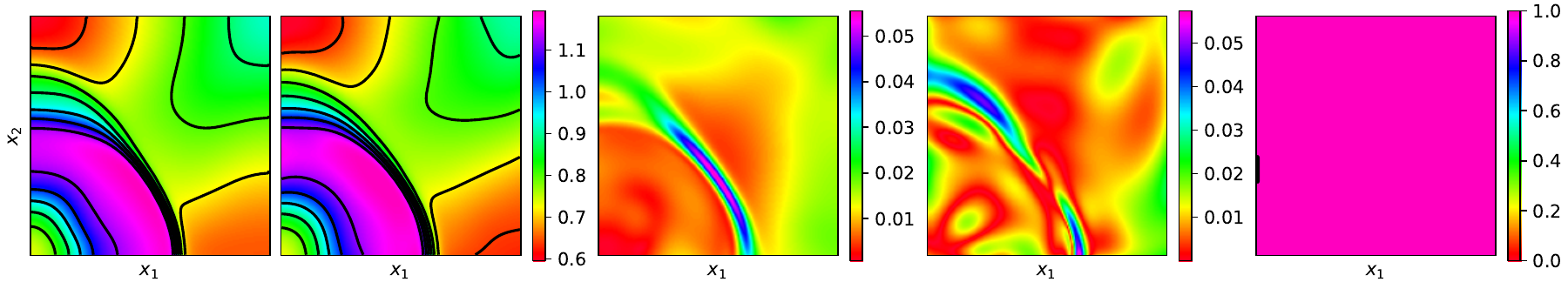}}
    \subfigure[\label{fig:STnRatio_0p6} PI-IRS-UQ, {$\lambda = 10^{-3}$} ]{\includegraphics[width=\textwidth,trim={0cm 37.8cm 0cm 38cm},clip]{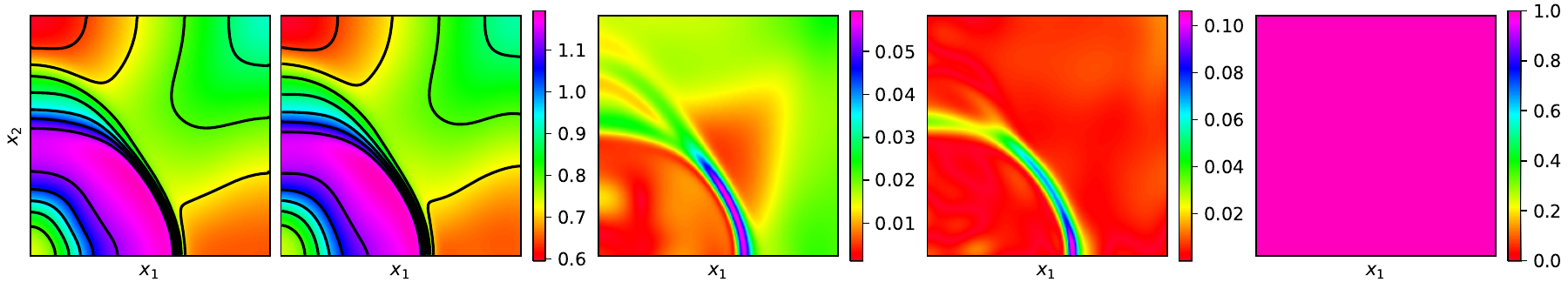}}
    \subfigure[\label{fig:STnRatio_0p6} PI-IRS-UQ, {$\lambda = 10^{-1}$} ]{\includegraphics[width=\textwidth,trim={0cm 37.8cm 0cm 38cm},clip]{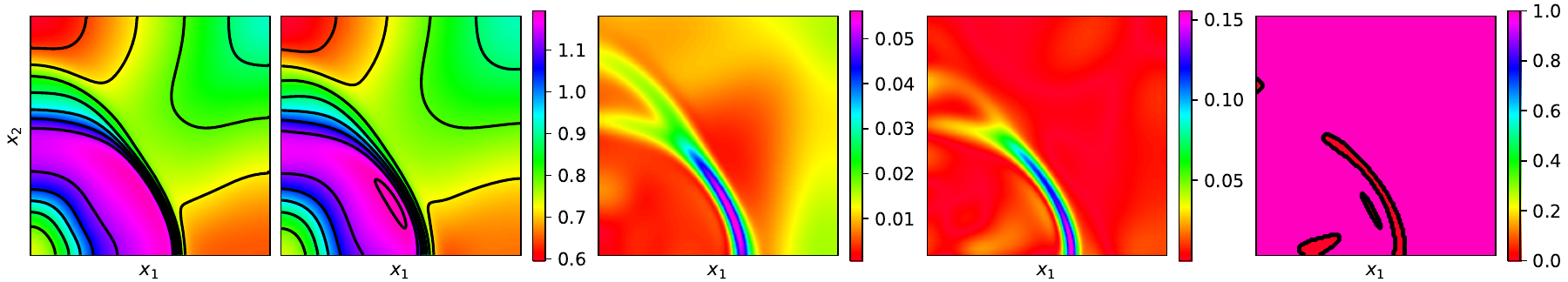}}
    \subfigure[\label{fig:STnRatio_0p6} ECLEIRS-UQ ]{\includegraphics[width=\textwidth,trim={0cm 37.8cm 0cm 38cm},clip]{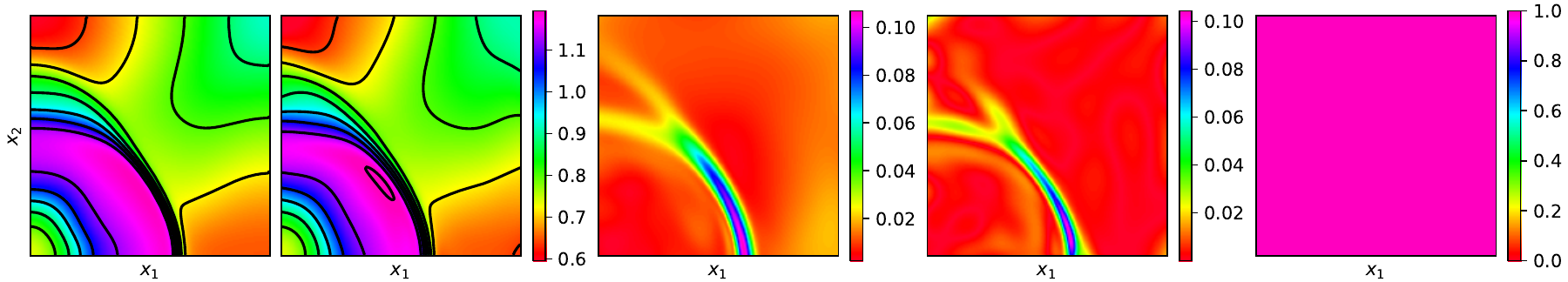}}    
    \caption{2-D Euler problem: Contour of density prediction at $\pmb{\nu}=[0.35, 0.47]$ for different models trained on $20\%$ spatial and $100\%$ temporal sampling of data without added noise.}
    \label{fig:Contour_2DEuler}
\end{figure}

We first evaluate the variation in the results with the KL-divergence term. The results for ECLEIRS-UQ with different $\beta$ are shown in \figref{Violin_Euler_ecleirs_compbeta}. At lower values of $\beta$, we observe that the ECLEIRS-UQ model predicts the solution very accurately. 
While the accuracy is maintained until $\beta = 2 \times 10^{-4}$, the error increases significantly with a further increase in $\beta$. Conversely, we observe an increase in correlation coefficient with an increase in $\beta$ until $\beta = 2\times 10^{-4}$, and a further increase leads to a reduction in the correlation coefficient. Similar trends are observed for coverage with the model trained with $\beta = 1 - 2 \times 10^{-4}$ having the highest coverage. These results indicate that the model trained with $\beta = 10^{-4}$ offers the best compromise between prediction accuracy and good uncertainty estimates. A further decrease in $\beta$ increases accuracy but at the cost of worse performance for uncertainty metrics. Similarly, an increase in $\beta$ from this ideal value provides slightly better uncertainty metrics but also leads to a reduction in prediction accuracy. Similar trends were also observed for IRS-UQ and PI-IRS-UQ. Based on this analysis, we choose $\beta = 10^{-4}$ for the rest of the numerical tests.

Contours of predicted density field for different modeling approaches are compared in \figref{Contour_2DEuler}. As the models compared in this figure are trained on good quality data, that is $20\%$ spatial and $100\%$ temporal sampling of data and no added noise, all models perform very well and provide density predictions that matches closely match the reference data. For all the models, maximum errors are observed close to the shock edge. This behavior is expected because discontinuities and steep gradients are inherently difficult to approximate using smooth neural-network representations. We see that while IRS exhibits a low error, the correlation between uncertainty estimates (given by SD) and errors are observed to be low. However, the uncertainty estimates encapsulate the errors well, thereby leading to nearly perfect coverage. The correlation between uncertainty estimates and errors are better for PI-IRS-UQ and appear to improve with higher $\lambda$. ECLEIRS-UQ appears to yield very high correlation between the errors and uncertainty estimates and also a perfect coverage of the solution. This observation indicates that ECLEIRS-UQ more effectively concentrates uncertainty in regions where the solution is difficult to predict, rather than uniformly inflating uncertainty bounds throughout the domain. While ECLEIRS-UQ give the best uncertainty estimates, other methods also perform well. The main takeaway from this result is that proposed method for UQ in latent dynamics model performance appears to work well, irrespective of whether the physics constraint is enforced weakly or strictly by construction.

\begin{figure}[t]
    \centering
    \subfigure[Relative error ($\epsilon_r$)]{\includegraphics[width=0.32\textwidth, trim={2cm 1.5cm 5cm 2cm},clip]{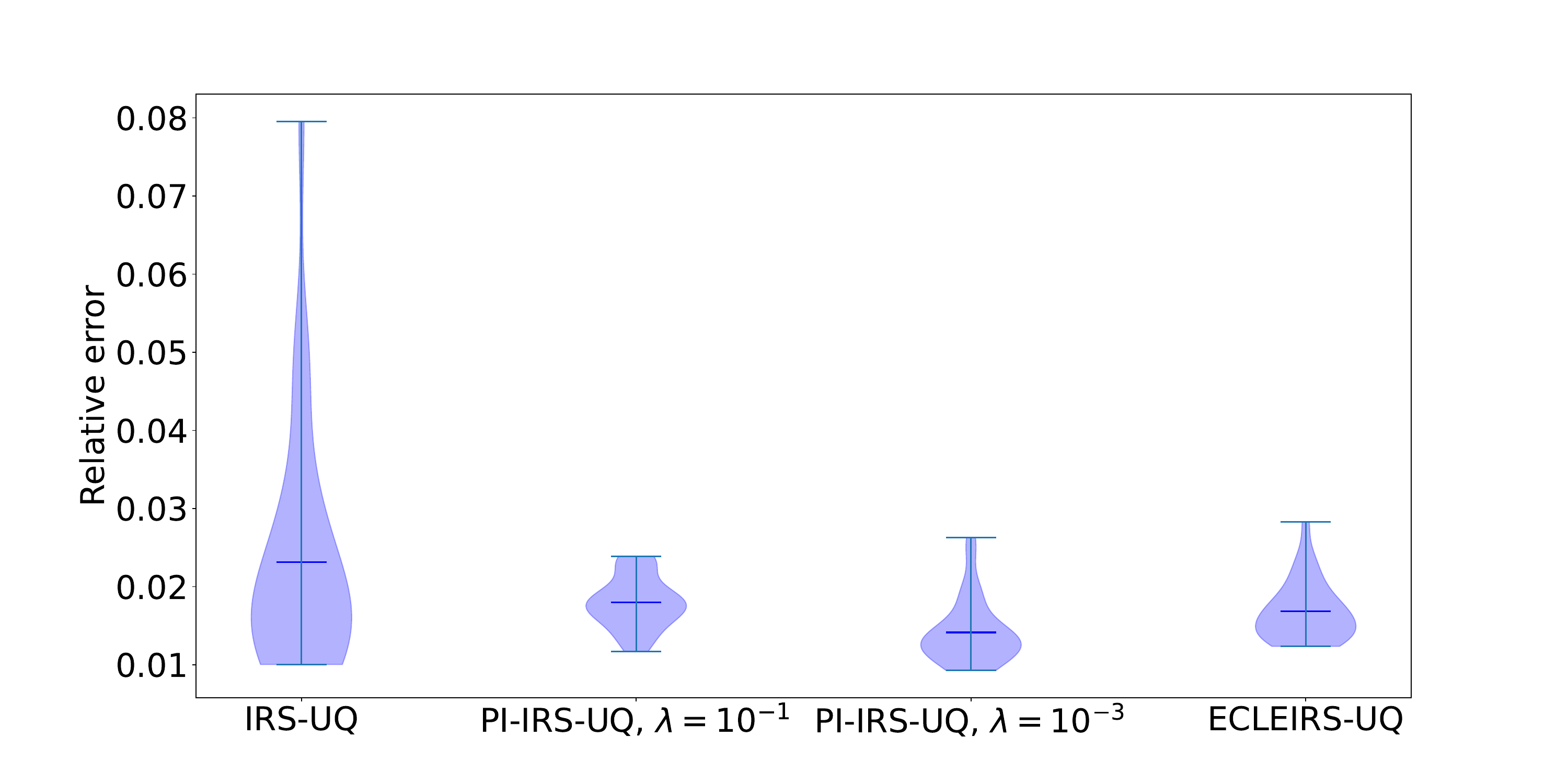}}
    \subfigure[Correlation Coefficient ($r$)]{\includegraphics[width=0.32\textwidth, trim={2cm 1.5cm 5cm 2cm},clip]{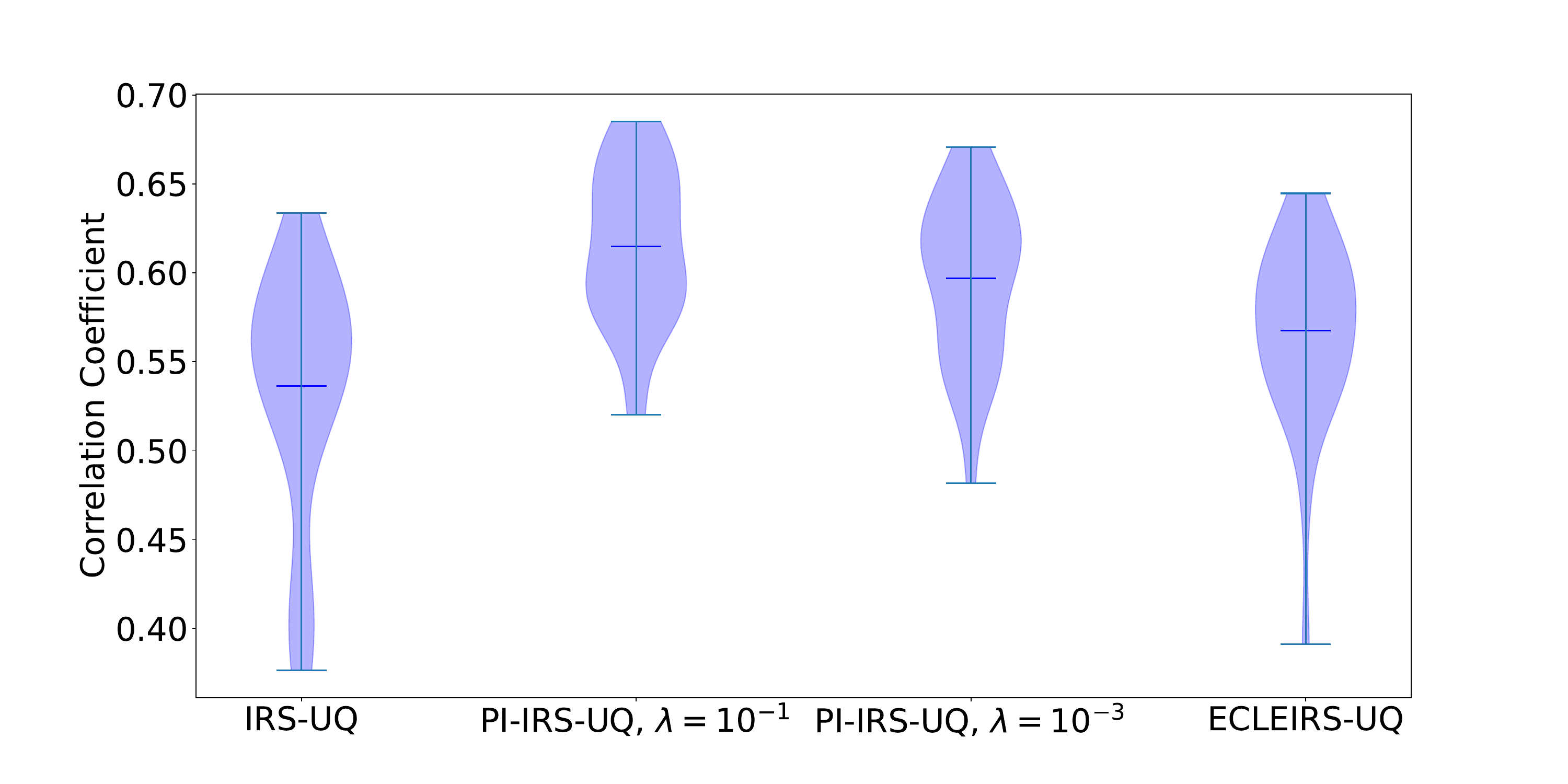}}
    \subfigure[Coverage ($\text{cov}_{2\sigma}$)]{\includegraphics[width=0.32\textwidth, trim={2cm 1.5cm 5cm 2cm},clip]{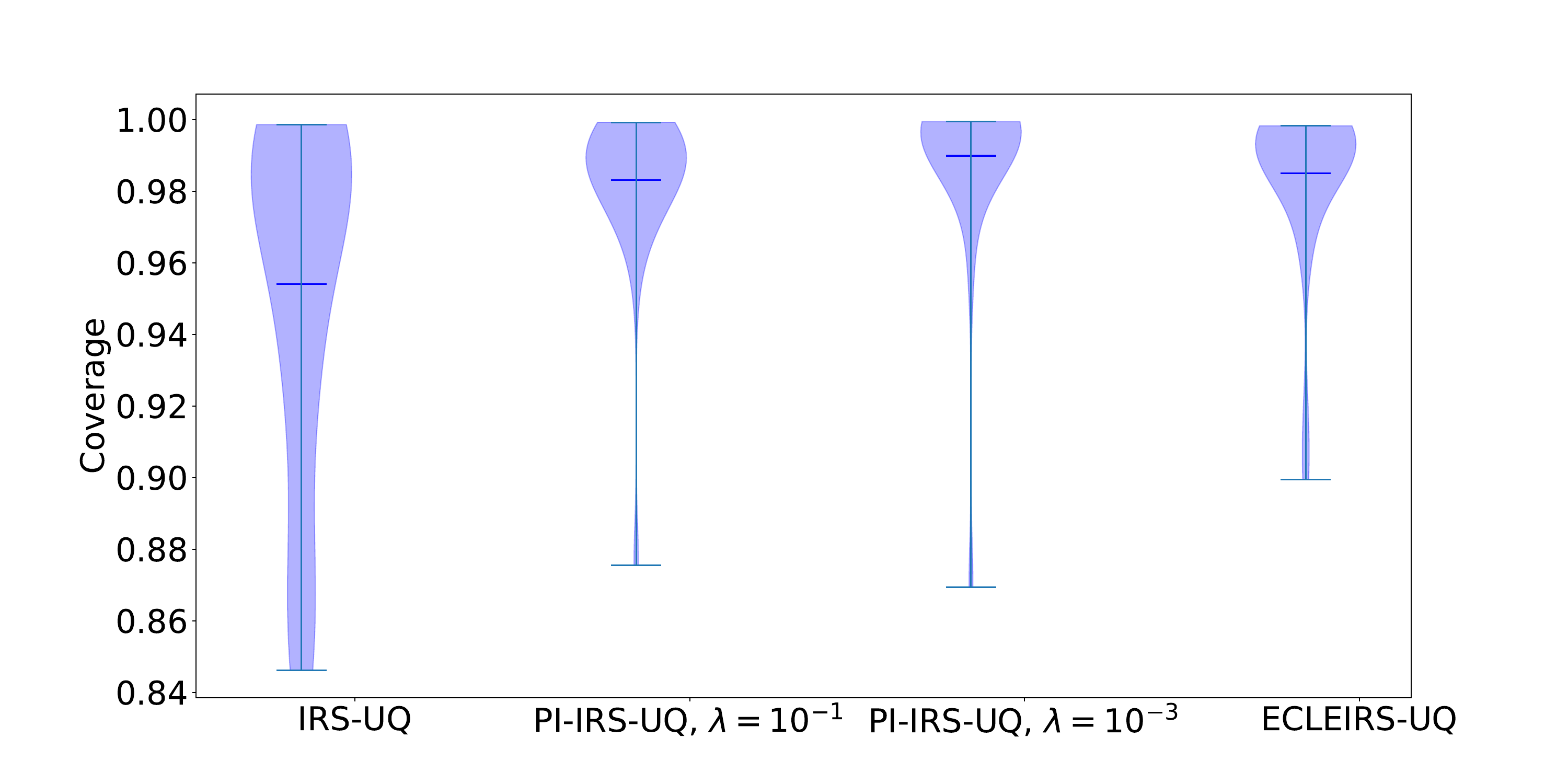}}
    \vspace{-3mm}
    \caption{2-D Euler problem: Violin plot showing the distribution of relative errors and uncertainty estimates with respect to the system parameters for different modeling approaches for randomly distributed $5\%$ spatial and $100\%$ temporal sampling of data with no added noise.}
    \label{fig:Violin_Euler_clean}
\end{figure}

The performance of different modeling approaches is evaluated for the validation dataset and presented as violin plots. The violin plots of relative error and uncertainty estimates for models trained on $5\%$ spatial and $100\%$ temporal sampling of data with no added noise are shown in \figref{Violin_Euler_clean}. The results indicate that IRS-UQ yields the lowest accuracy, while PI-IRS-UQ and ECLEIRS-UQ exhibits a high accuracy. We observe a high coverage, with an average of over $95\%$, for all the modeling approaches indicating that the predicted uncertainty bounds cover the solution uncertainty very well. These results indicate that uncertainty estimates we get are likely conservative in the prediction which is ideal for out-of-distribution scenarios. Furthermore, we observe a medium correlation between standard deviation and errors, which is lower than those observed for the 1-D advection problem.
This lower value is attributed to decrease in correlations away from the shock interface region as also observed from the contour plots. A key reason for this reduction in correlation is the sparsely sampling spatial and temporal locations used to train these models. This behavior is studied in more detail by comparing the model performance for ECLEIRS-UQ for different levels of spatial sparsity while keeping the temporal sparsity fixed.

\begin{figure}[t]
    \centering
    \subfigure[Relative error ($\epsilon_r$)]{\includegraphics[width=0.32\textwidth, trim={2cm 1.5cm 5cm 1.5cm},clip]{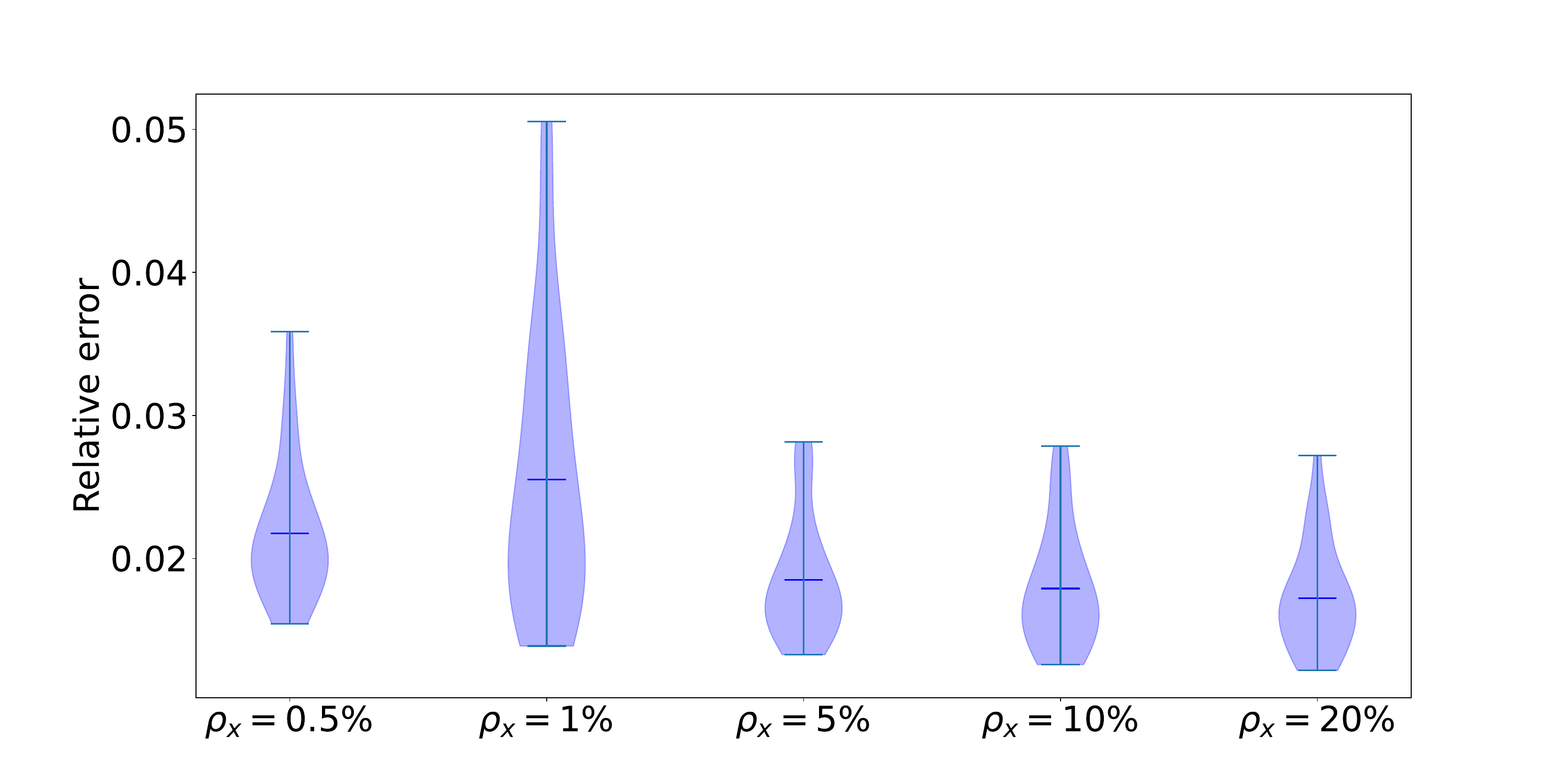}}
    \subfigure[Correlation Coefficient ($r$)]{\includegraphics[width=0.32\textwidth, trim={2cm 1.5cm 5cm 1.5cm},clip]{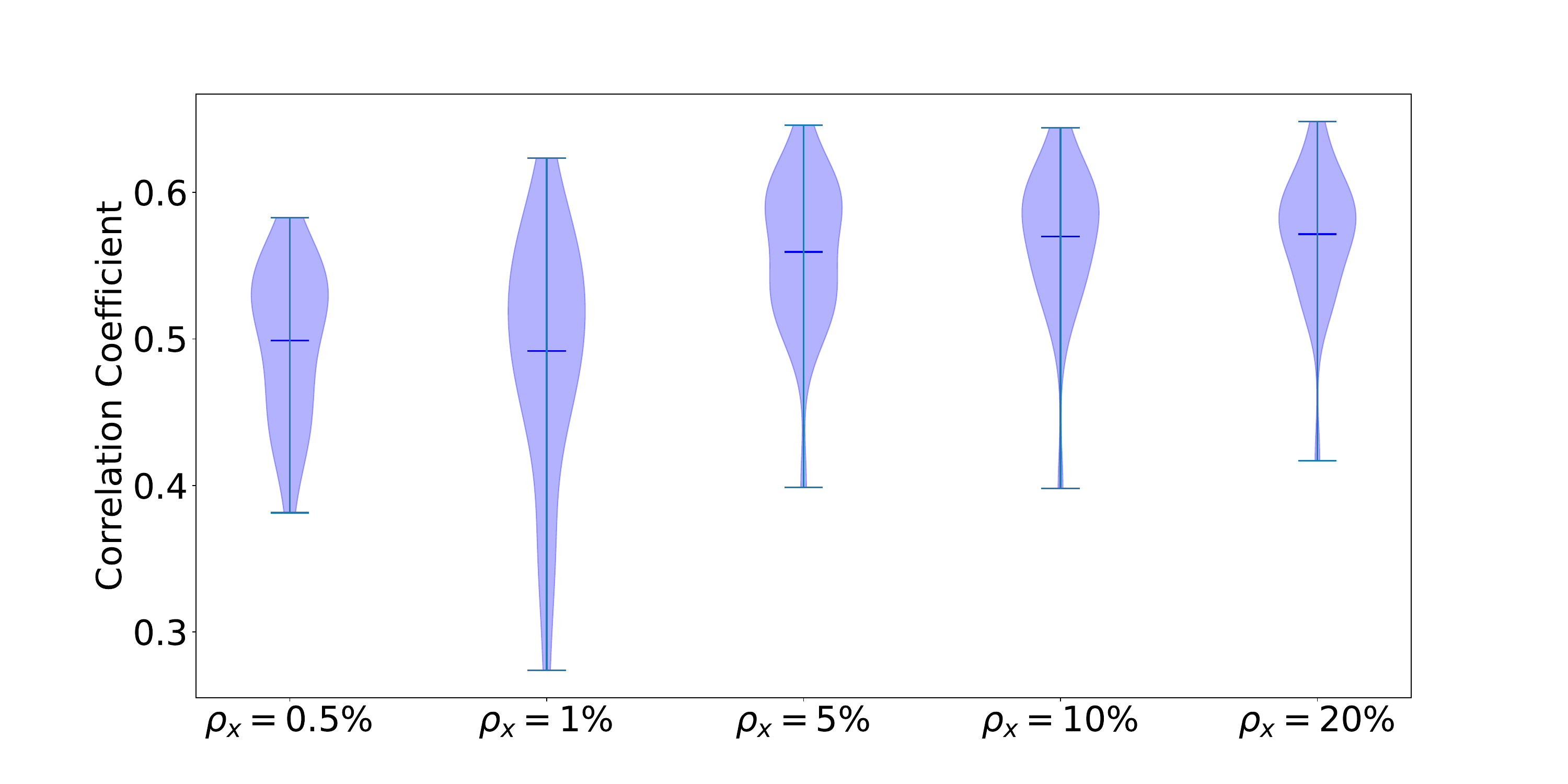}}
    \subfigure[Coverage ($\text{cov}_{2\sigma}$)]{\includegraphics[width=0.32\textwidth, trim={2cm 1.5cm 5cm 1.5cm},clip]{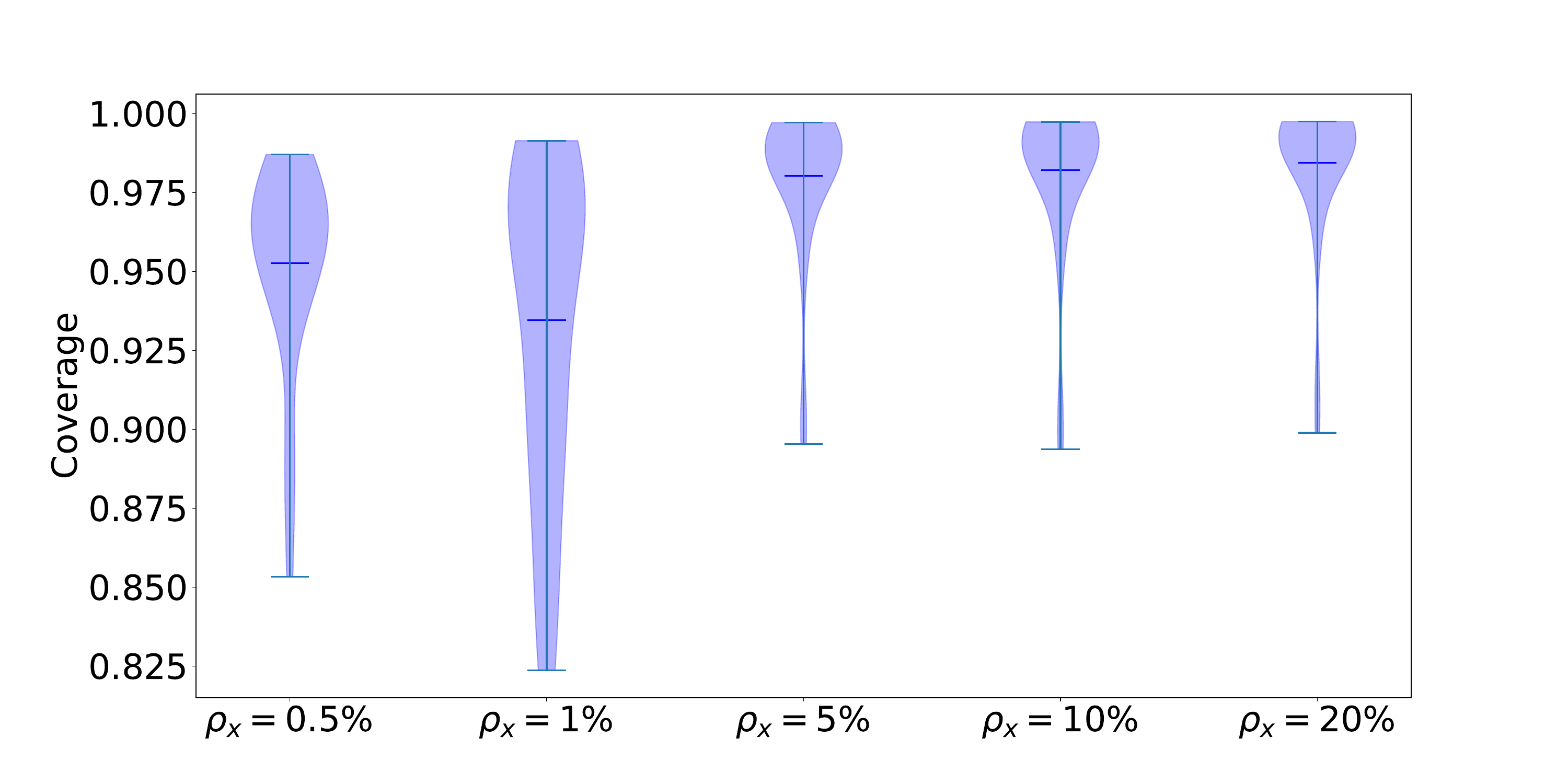}}
    \vspace{-3mm}
    \caption{2-D Euler problem: Violin plot showing the distribution of relative errors and uncertainty estimates with respect to the system parameters for ECLEIRS-UQ trained using randomly distributed $\rho_x$ spatial and $20\%$ temporal sampled data with no added noise.}
    \label{fig:Violin_Euler_sparsecompare}
\end{figure}

The results showing the effect of spatial sampling density in the training
data on the performance of ECLEIRS-UQ are shown in \figref{Violin_Euler_sparsecompare}. We observe a slight decrease in relative error with denser spatial data, while a spatial sampling of $1\%$ deviate from this pattern slightly. The results also show an increase in correlation coefficient and coverage with an increase in spatial data. Overall, while ECLEIRS-UQ is robust in maintaining low errors with the decrease in spatial data, there is a slight decrease in the quality of uncertainty estimates with sparser data. Similar behavior is also observed for IRS-UQ and PI-IRS-UQ and the results are not discussed for brevity. These results indicate that the proposed approach for quantifying uncertainty in latent dynamics models works well in providing accurate estimates.

\subsection{2-D shallow water equation} 

In the third numerical experiment, we consider 2-D shallow-water equations, which govern the propagation of water depth $h$. These equations can be written in the form 
\begin{equation}
    \frac{\partial \pmb{q}}{\partial t} + \nabla_x \cdot \pmb{f}(\pmb{q}) = \pmb{S}(\pmb{q})
\end{equation}
where $\pmb{q} = [h,\; h\pmb{u}]^T$ is the solution vector, $\pmb{f}(\pmb{q}) = [h\pmb{u}, \; h\pmb{u}\otimes\pmb{u}]^T$ is the flux vector and $\pmb{S}(\pmb{q})$ is the vector of source terms. The source terms include Coriolis and bathymetry terms which results in  loss of space-time conservation of momentum. Despite the presence of these terms, mass conservation equation still exhibits a nonlinear conservation form
\begin{equation}
    \frac{\partial h}{\partial t} + \nabla_x \cdot (h \pmb{u}) = 0.
\end{equation}
We solve these equations on a square domain $\Omega = [0, 1] \times [0, 1]$ with reflective boundary conditions at the four sides. The system is parametrized with initial conditions
\begin{equation}
    h(\pmb{x}, t = 0) = \max\!\left(
h_0 + A_h \exp\!\left(
-\frac{(\pmb{x}-\pmb{x}_h)^2}{2\sigma_h^2}
\right) - b(x,y),
\, h_{\min}
\right), \quad \pmb{u}(\pmb{x}, t = 0) = 0,
\end{equation}
where 
\begin{equation}
    b(\pmb{x})=A_b \exp\!\left(
-\frac{(\pmb{x}-\pmb{x}_b)^2}{2\sigma_b^2}
\right)
\end{equation}
describes the bathymetry conditions. For the simulations, we use fixed bathymetry parameters $A_b = 1.0$, $\sigma_b = 0.1$ and $\pmb{x}_b = 0$. The other fixed parameters are background water depth $h_0 = 1$, minimum depth $h_{\text{min}} = 10^{-6}$ and the Coriolis parameter is $10$. The rest three initial condition parameters are used to parameterize this system $\pmb{\nu} = [\pmb{x}_h, \;A_h]^T \in \mathbb{R}^3$. We run $200$ simulations for different Halton-randomly sampled parametric combination with the parametric domain $\pmb{\nu} \in \mathcal{V} := [0.3, \; 0.7] \times [0.3, \; 0.7] \times [2.0, \; 2,4]$. From this parameteric distribution, we use $75\%$ randomly selected parameters to train the model and the rest $25\%$ of parameters to test the model performance.

For generating training and validation data, we solve these equations using finite-volume method on a uniform Cartesian grid with MUSCL-type piecewise-linear reconstruction and a minmod slope limiter in space, Rusanov numerical fluxes at cell interfaces, and a second-order SSP-RK2 scheme for time integration. We use $100$ grid points in both $x_1$ and $x_2$ direction and a CFL of $0.5$ for time-marching the solution. Through this numerical experiment, we demonstrate the benefits of imposing nonlinear conservation law structure on the water-depth height prediction. This numerical experiment provides a simplified example demonstrating the role of conservation-structure embedded latent dynamics models with uncertainty estimation for natural disaster modeling such as Tsunamis and floods. 

\begin{figure}
    \centering
    \subfigure[Relative error ($\epsilon_{r}$)]{\includegraphics[width=0.32\textwidth, trim={0.5cm 1.0cm 5cm 2cm},clip]{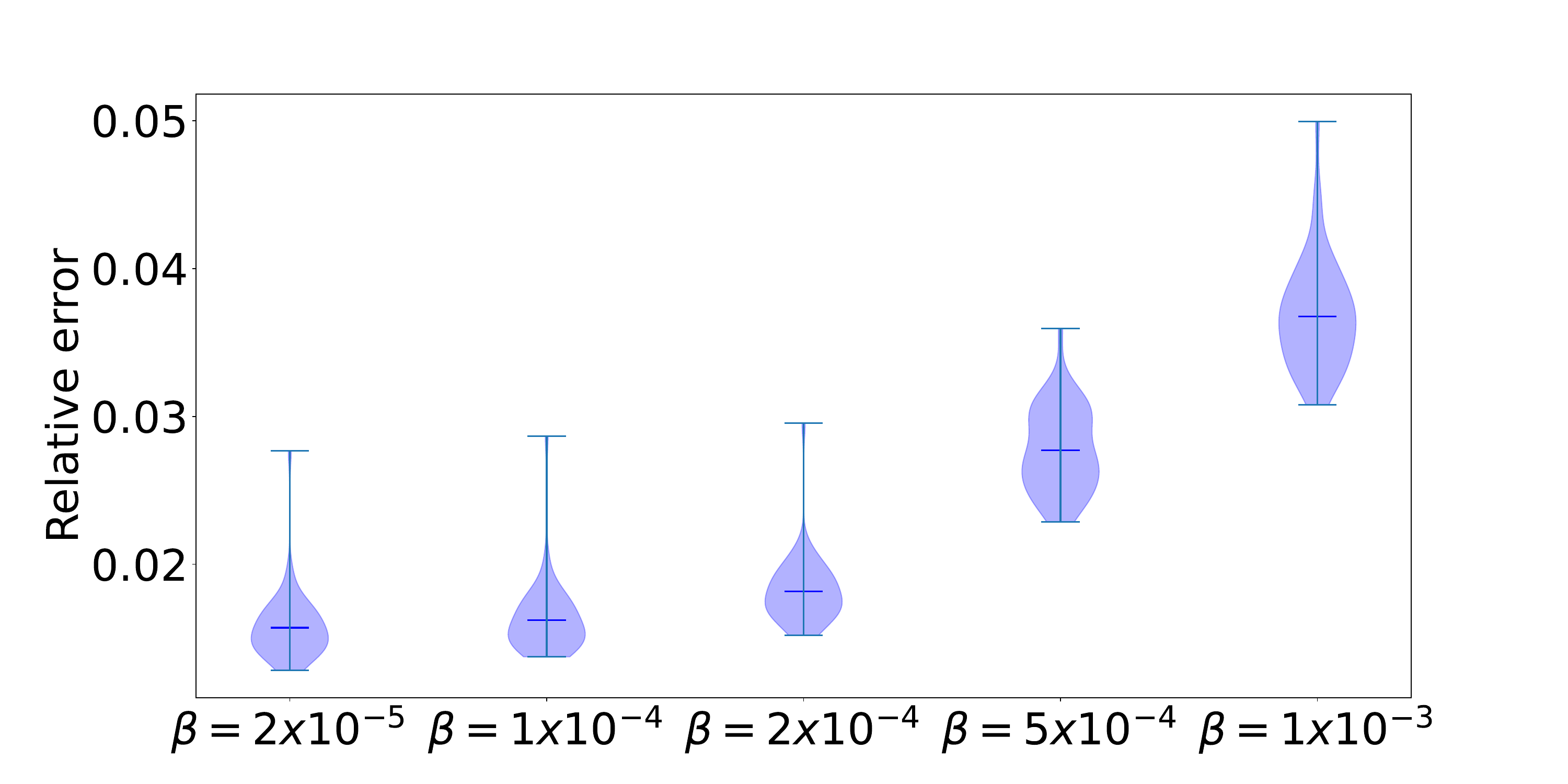}}
    \subfigure[Correlation Coefficient ($r$)]{\includegraphics[width=0.32\textwidth, trim={0.5cm 1.0cm 5cm 2cm},clip]{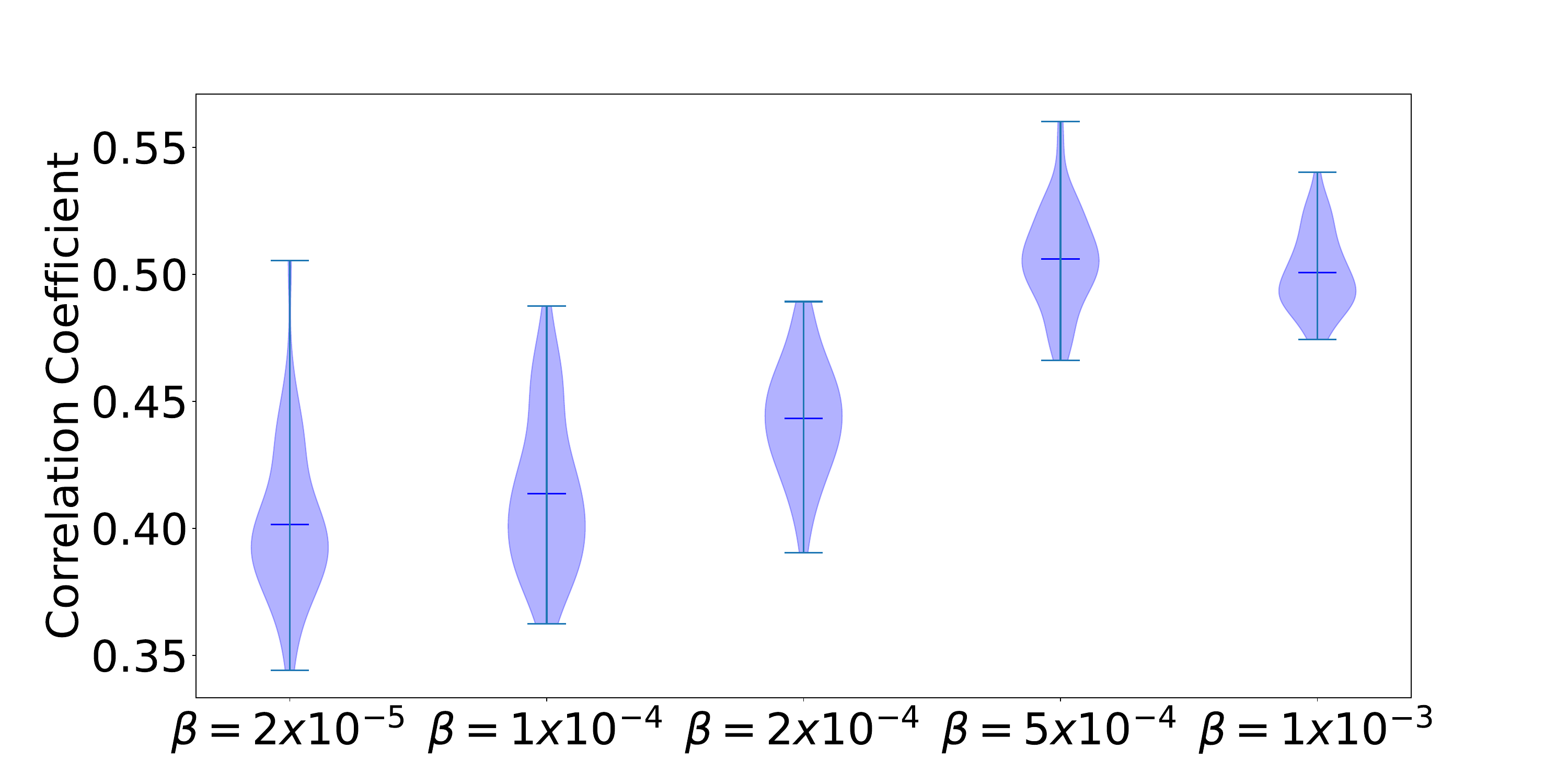}}
    \subfigure[Coverage ($\text{cov}_{2\sigma}$)]{\includegraphics[width=0.32\textwidth, trim={0.5cm 1.0cm 5cm 2cm},clip]{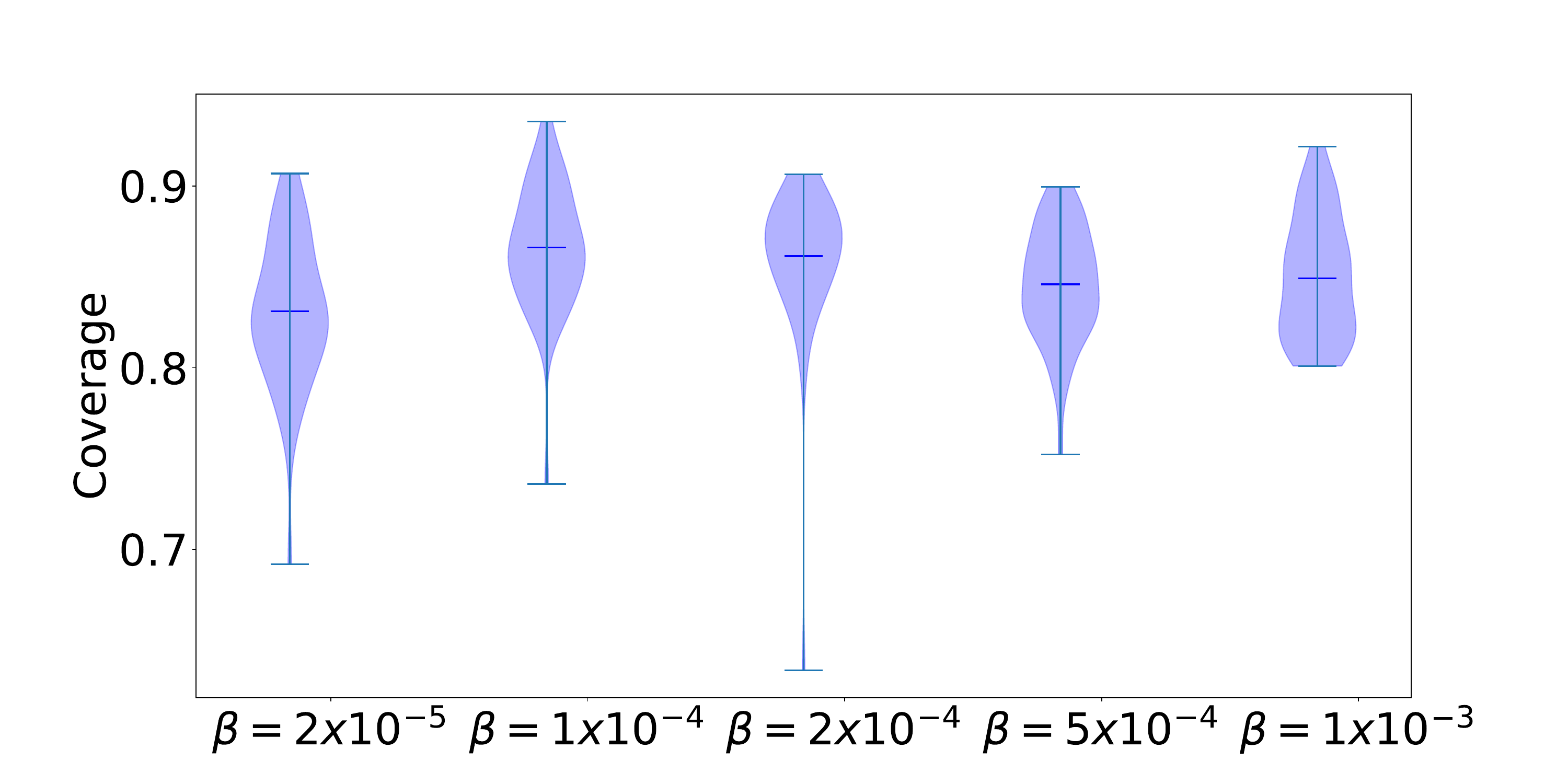}}
    \vspace{-3mm}
    \caption{2-D shallow water problem: Violin plot showing the distribution with respect to the system parameters for ECLEIRS-UQ trained on $10\%$ spatial and $100\%$ temporal sampled clean data with different values of KL-divergence parameter ($\beta$).}
    \label{fig:Violin_Shallow_ecleirs_compbeta}
\end{figure}

Similar to the other two numerical experiments, we first assess the influence of the KL-divergence parameter on prediction accuracy and quality of uncertainty estimates, as shown in \figref{Violin_Shallow_ecleirs_compbeta}. We observe an increase in prediction error with an increase in $\beta$. Conversely, there is a significant improvement in uncertainty metric such as mean correlation coefficient until $\beta = 5\times 10^{-4}$, while the mean coverage remains relatively similar with an increase in $\beta$. These results reveal the same accuracy-uncertainty trade-off observed in the previous numerical experiments. The consistency of this trend across all benchmark problems suggests that the influence of $\beta$ is primarily governed by the variational latent-space formulation rather than the specific governing equations. For the rest of the study, we choose $\beta = 10^{-4}$ as it provides lower error and remains consistent with the value we chose for other numerical experiments, although a cost of weaker correlation in the testing set. 

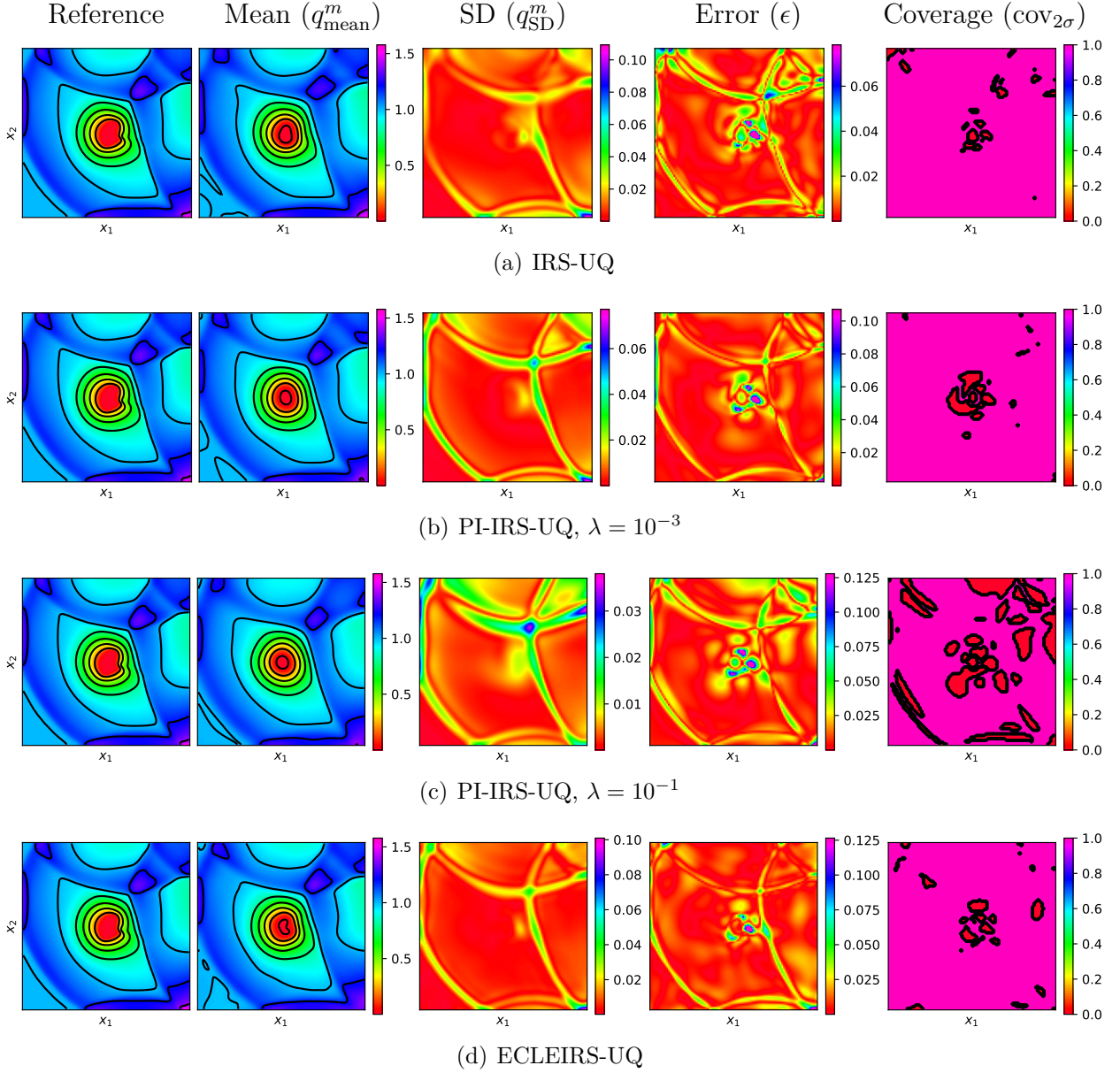
\begin{figure}[t!]
    \centering
    \begin{picture}(0,20)    
    \put(-215,0){Reference}
    \put(-140,0){Mean ($q^m_{\text{mean}}$)}    
    \put(-40,0){SD ($q^m_{\text{SD}}$)}    
    \put(60,0){Error ($\epsilon$)}    
    \put(140,0){Coverage ($\text{cov}_{2\sigma}$)}    
    \end{picture}    
    \subfigure[\label{fig:STnRatio_0p6} IRS-UQ]{\includegraphics[width=\textwidth,trim={0cm 37.8cm 0cm 38cm},clip]{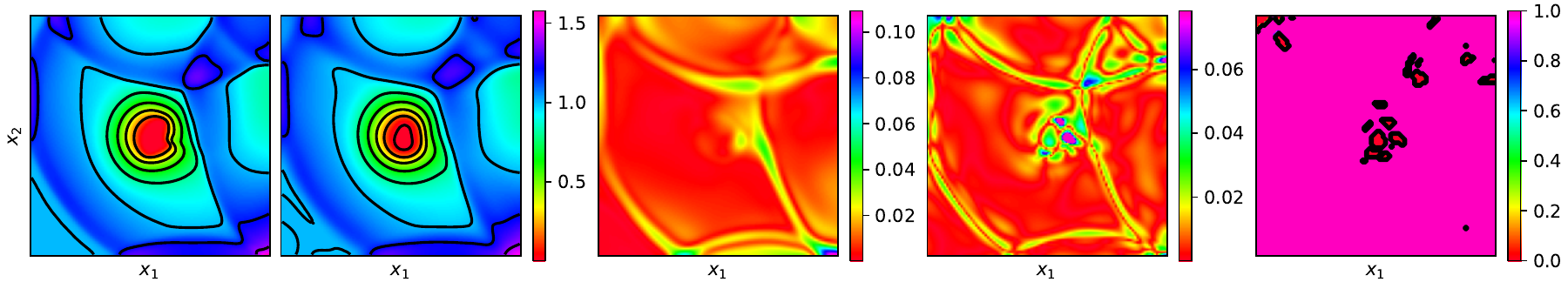}}
    \subfigure[\label{fig:STnRatio_0p6} PI-IRS-UQ, {$\lambda = 10^{-3}$} ]{\includegraphics[width=\textwidth,trim={0cm 37.8cm 0cm 38cm},clip]{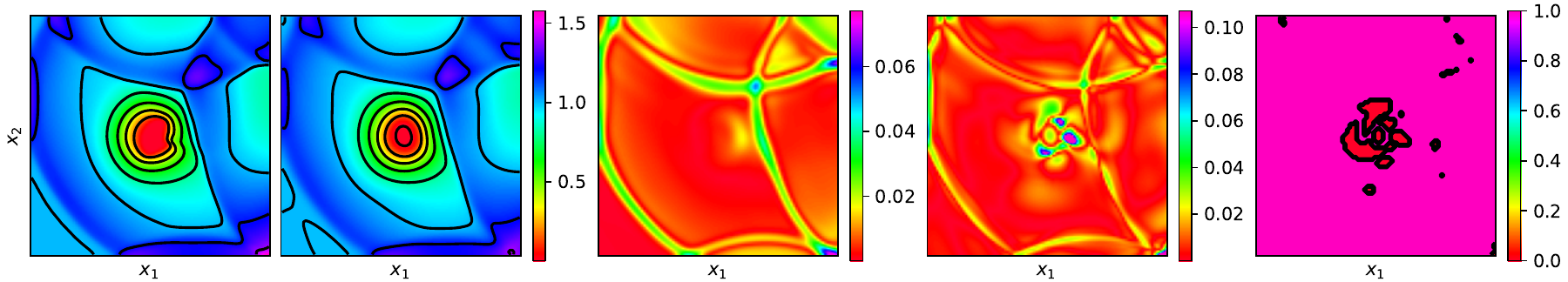}}
    \subfigure[\label{fig:STnRatio_0p6} PI-IRS-UQ, {$\lambda = 10^{-1}$} ]{\includegraphics[width=\textwidth,trim={0cm 37.8cm 0cm 38cm},clip]{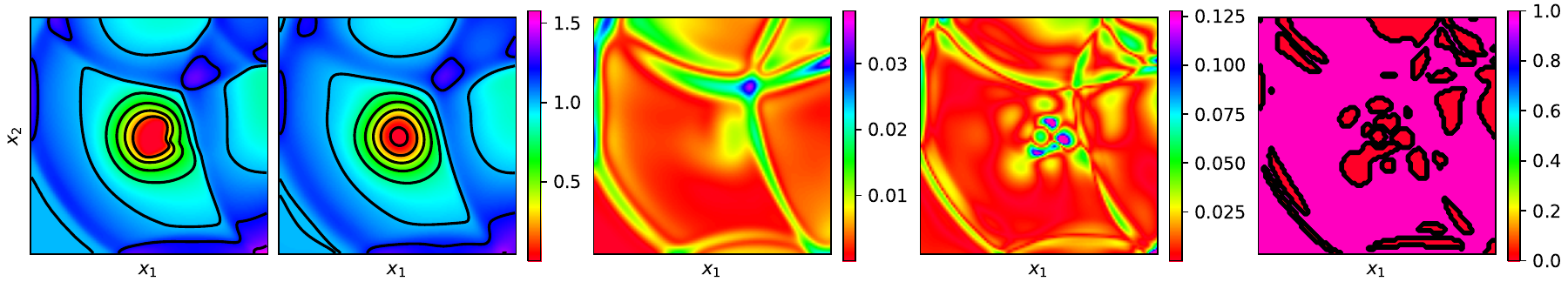}}
    \subfigure[\label{fig:STnRatio_0p6} ECLEIRS-UQ ]{\includegraphics[width=\textwidth,trim={0cm 37.8cm 0cm 38cm},clip]{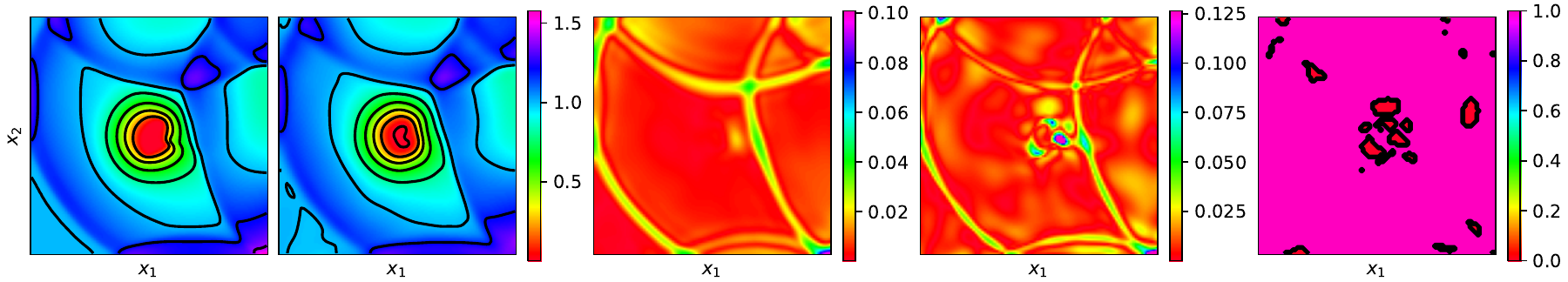}}    
    \caption{2-D shallow water-depth prediction: Contour of water depth prediction at $\pmb{\nu}=[0.60, 0.65, 2.1]$ for different models trained on $20\%$ spatial and $100\%$ temporal sampled data without added noise.}
    \label{fig:Contour_2DShallow}
\end{figure}

The contour plots showing the spatial distribution of the prediction, error and uncertainty metrics for all the modeling approaches are shown in \figref{Contour_2DShallow}. These results indicate that the standard deviation of the solution prediction correlates well with error, thereby serving as a good metric for estimating uncertainty in the solution prediction.  While the correlation appears good, there are high error regions near the central bathymetry region which is not well captured by the estimated uncertainty for all the modeling approaches. This is also observed by looking at coverage contours which show lower coverage towards the central bathymetry region for all the models. For other spatial regions, the predicted uncertainty bounds cover the reference solution well for IRS-UQ, PI-IRS-UQ with $\lambda = 10^{-3}$ and ECLEIRS-UQ. Despite localized deficiencies near the bathymetry feature, the uncertainty estimates remain informative over the majority of the computational domain, indicating that the uncertainty representation captures the dominant sources of predictive error.

\begin{figure}[t]
    \centering
    \subfigure[Relative error ($\epsilon_r$) - clean]{\includegraphics[width=0.49\textwidth, trim={2cm 1.5cm 4cm 2cm},clip]{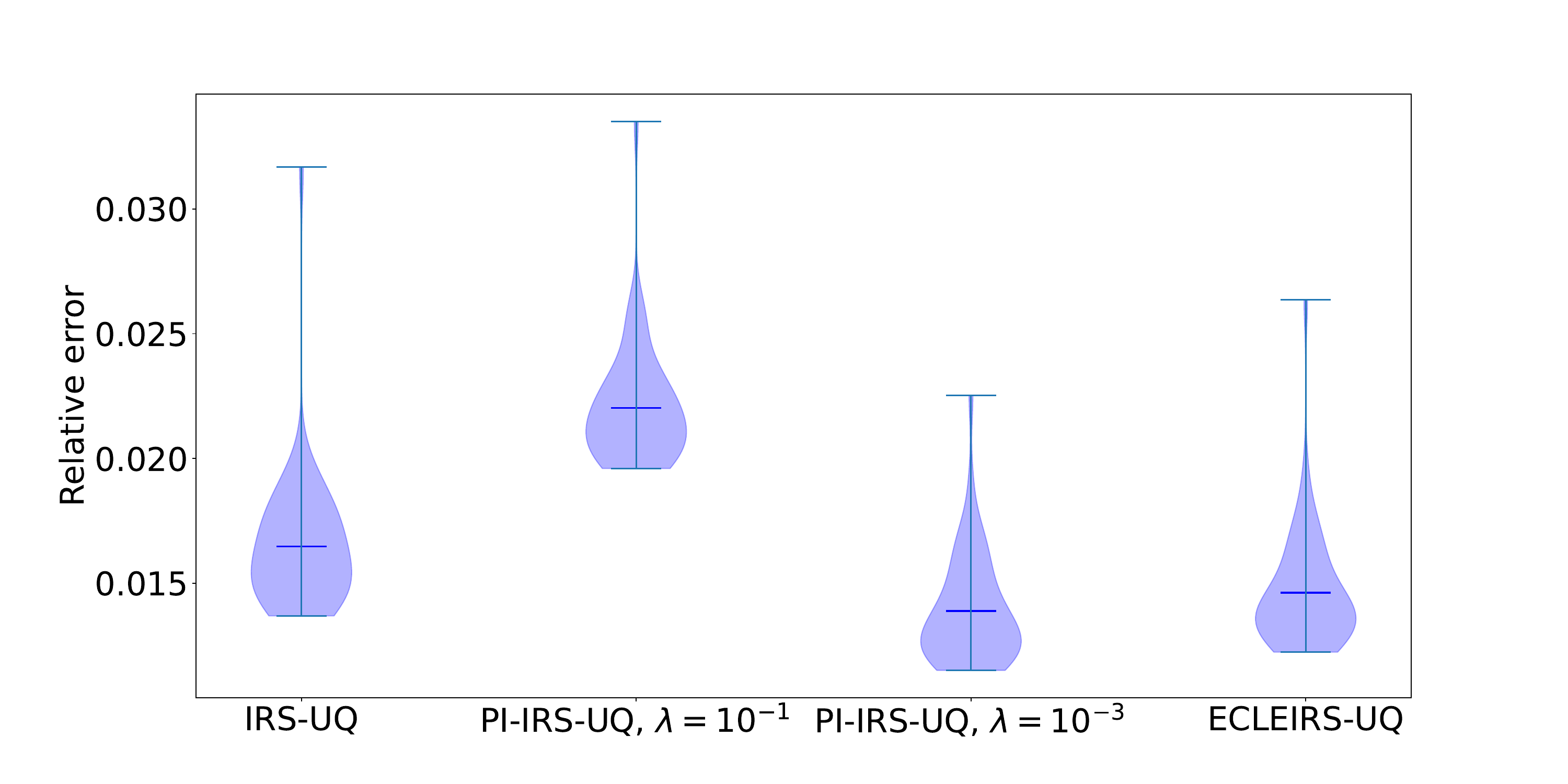}}
    \subfigure[Relative error ($\epsilon_r$) - noisy]{\includegraphics[width=0.49\textwidth, trim={2cm 1.5cm 4cm 2cm},clip]{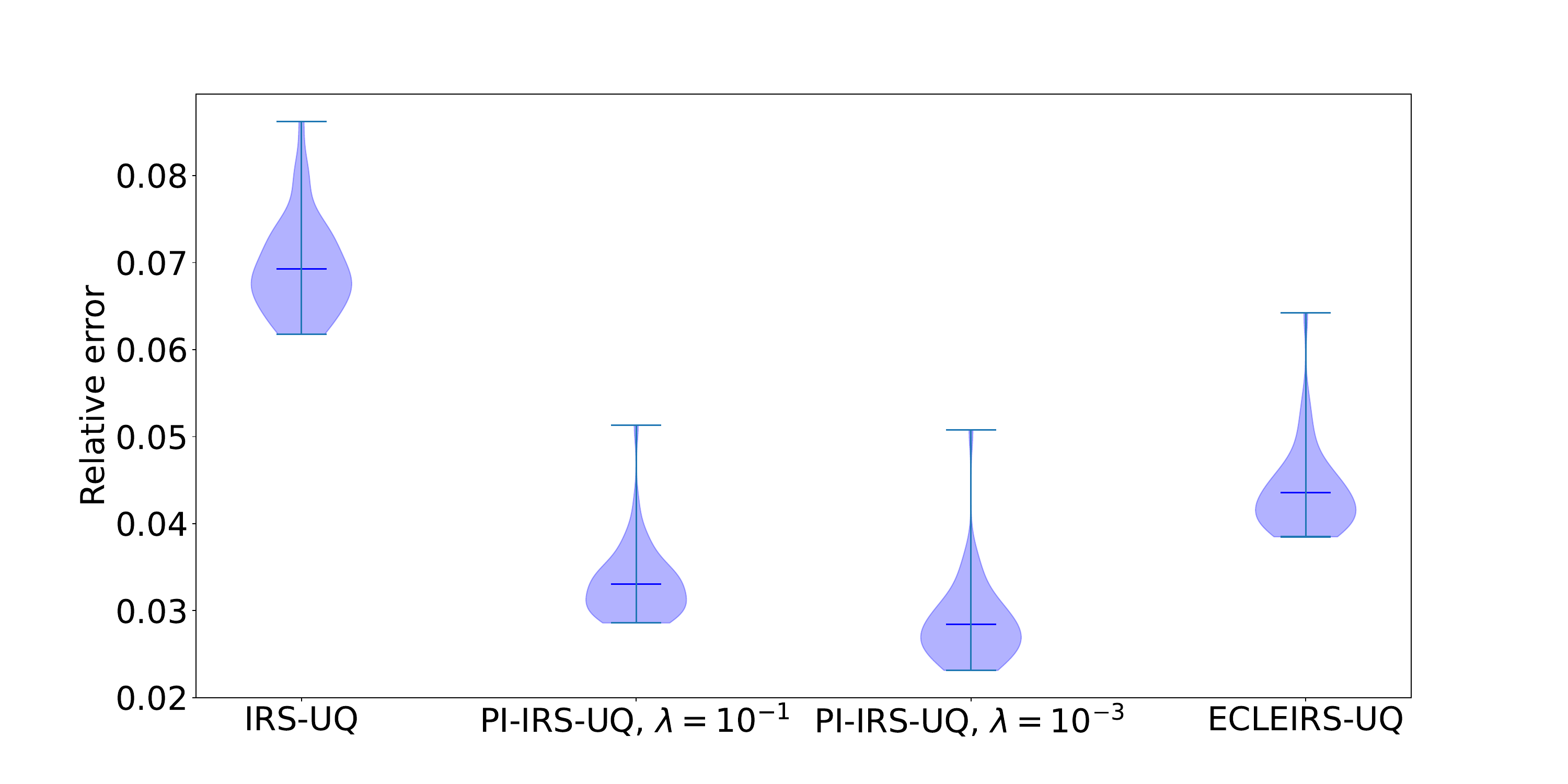}}\\
    
    \subfigure[Correlation Coefficient ($r$) - clean]{\includegraphics[width=0.49\textwidth, trim={2cm 1.5cm 4cm 2cm},clip]{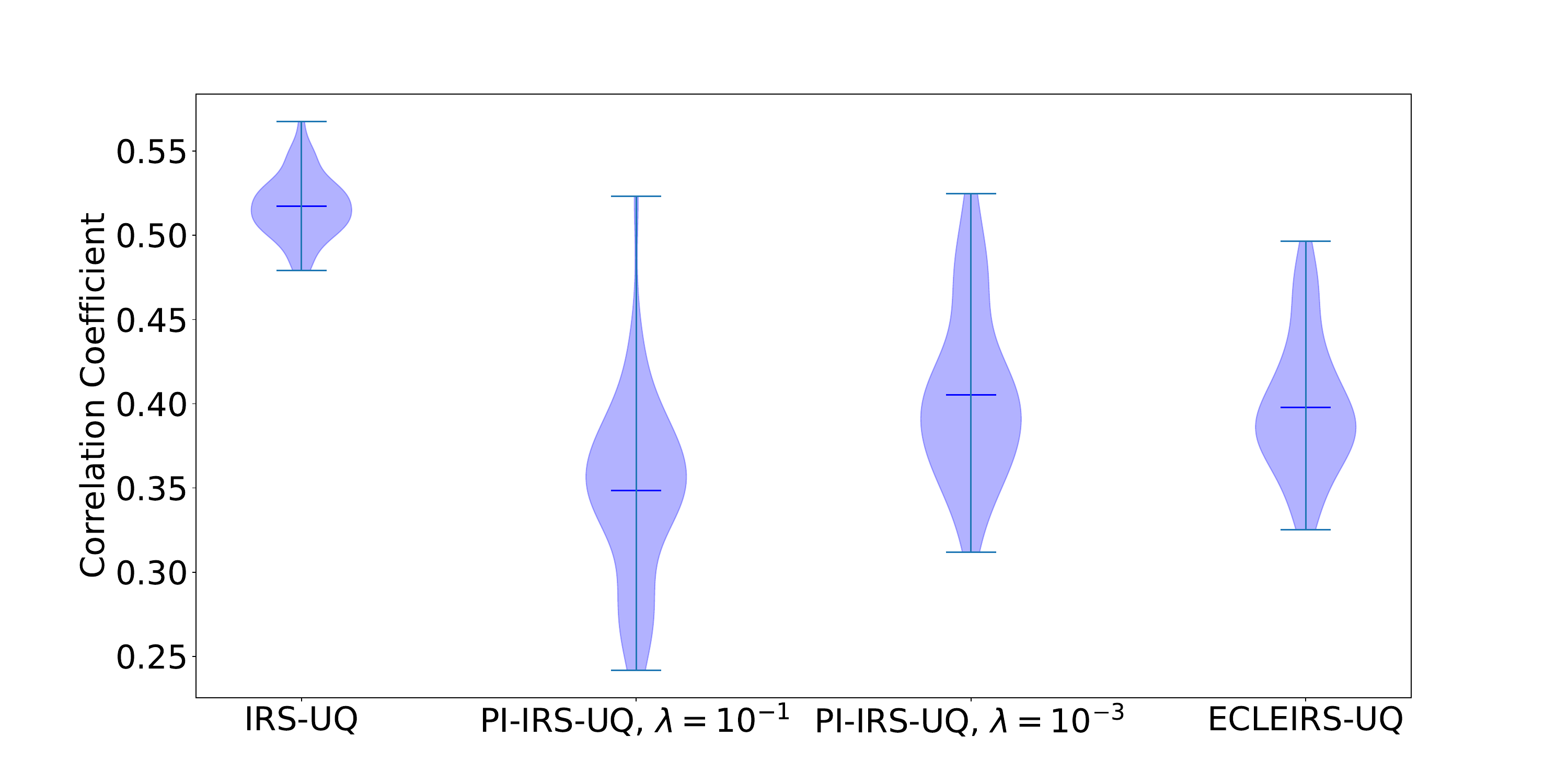}}
    \subfigure[Coverage ($\text{cov}_{2\sigma}$) - clean]{\includegraphics[width=0.49\textwidth, trim={2cm 1.5cm 4cm 2cm},clip]{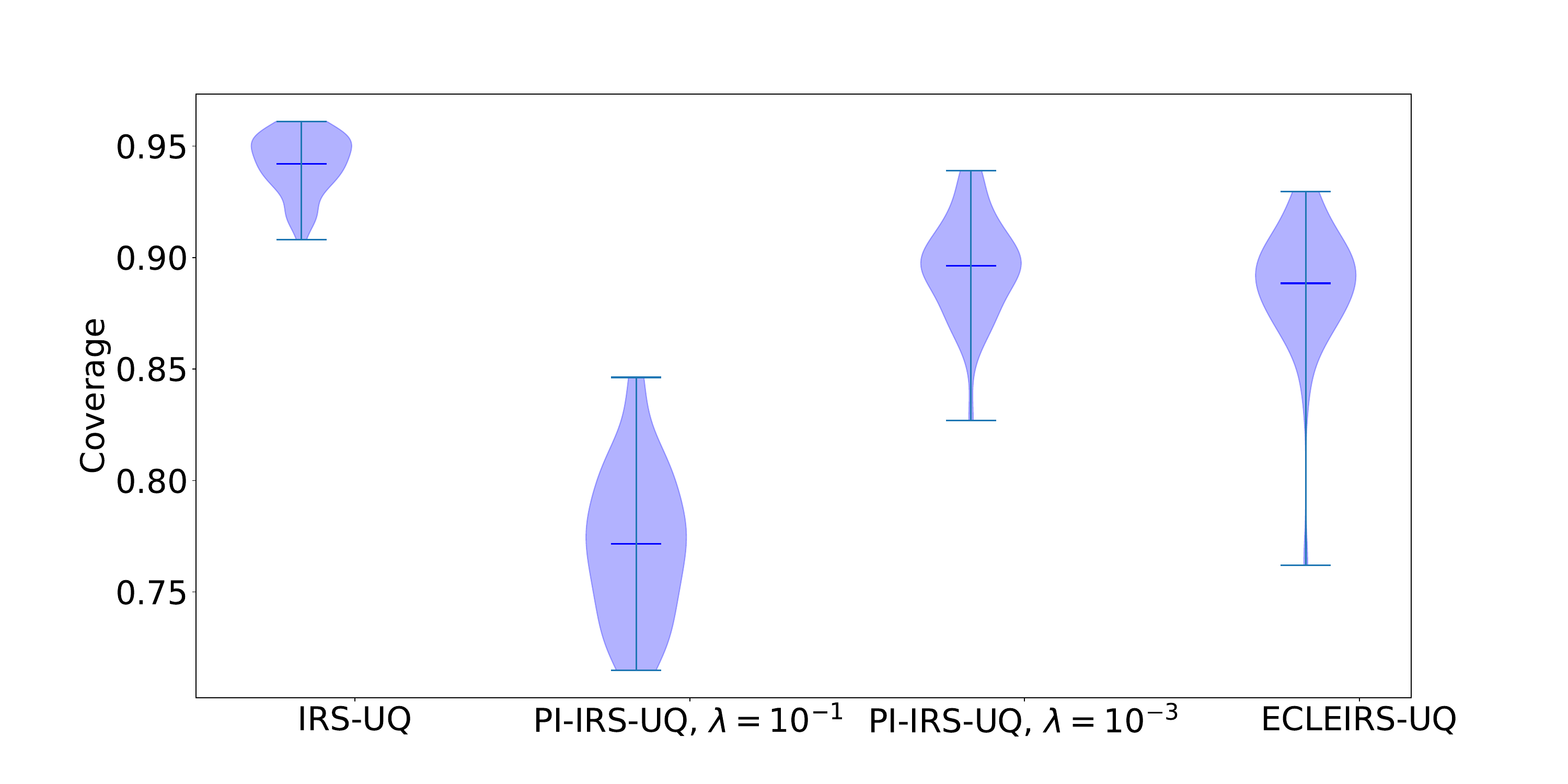}}
    \vspace{-3mm}
    \caption{2-D shallow water problem: Violin plot showing the distribution of relative errors and uncertainty estimates with respect to the system parameters for different modeling approaches. The clean training dataset implies randomly distributed $20\%$ spatial and $100\%$ temporal sparse data with no added noise. The noisy dataset implies randomly distributed $1\%$ spatial and $20\%$ temporal sampled data with added Gaussian noise with standard deviation of 0.2.}
    \label{fig:Violin_Shallow_clean}
\end{figure}

\begin{figure}[t]
    \centering
    \subfigure[Relative error ($\epsilon_r$)]{\includegraphics[width=0.32\textwidth, trim={1cm 1.0cm 3cm 1.5cm},clip]{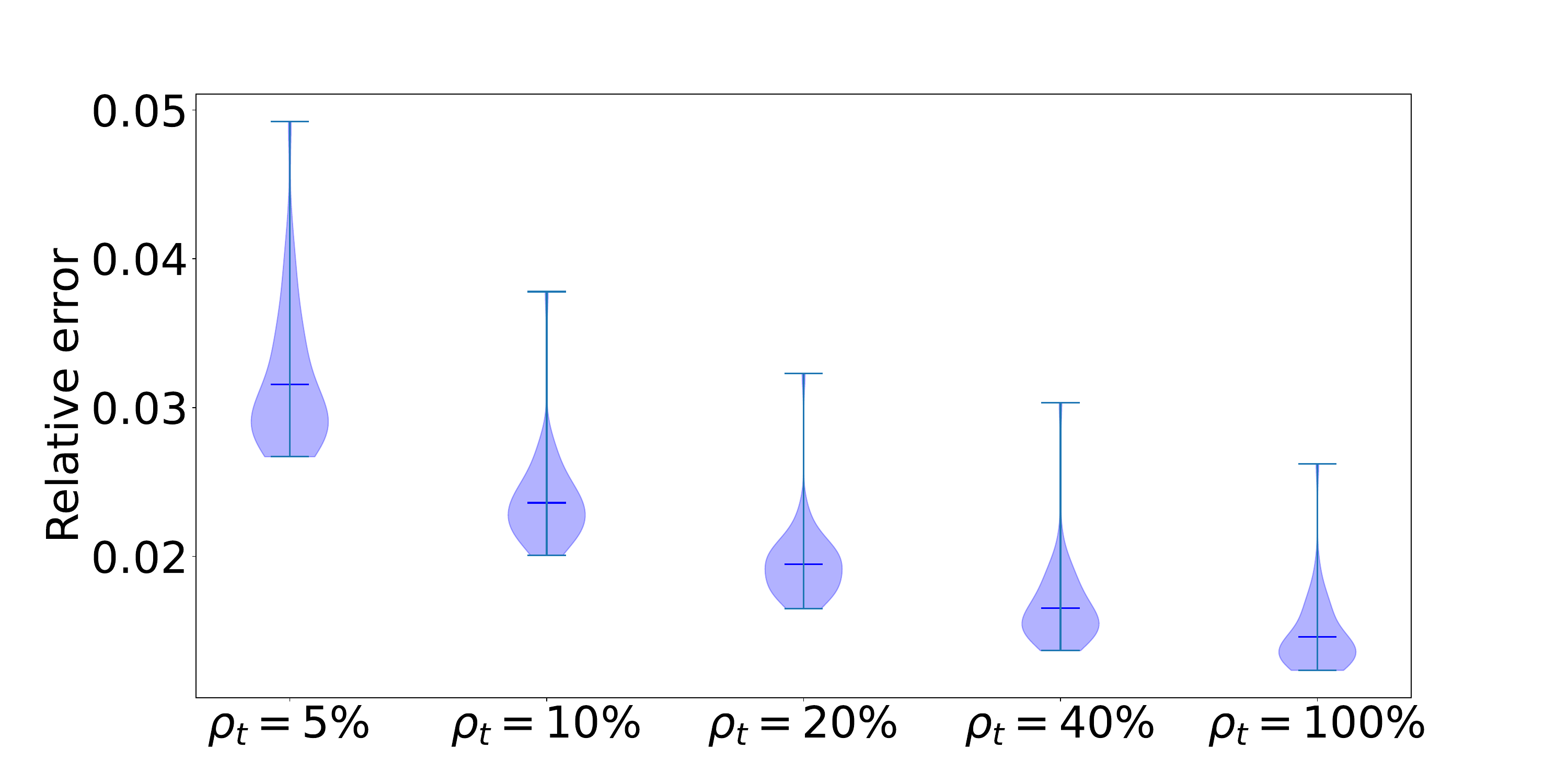}}
    \subfigure[Correlation Coefficient ($r$)]{\includegraphics[width=0.32\textwidth, trim={1cm 1.0cm 3cm 1.5cm},clip]{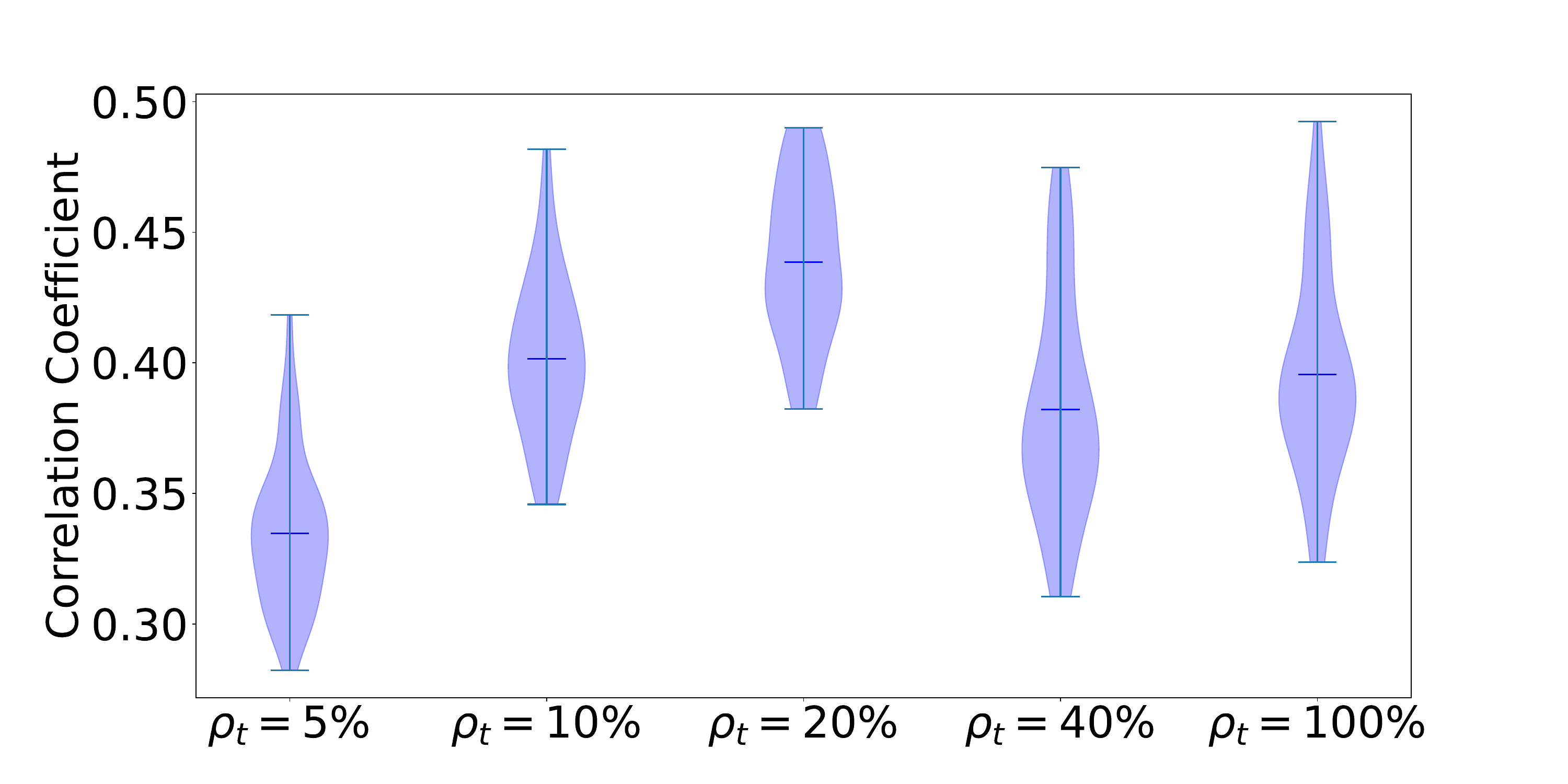}}
    \subfigure[Coverage ($\text{cov}_{2\sigma}$)]{\includegraphics[width=0.32\textwidth, trim={1cm 1.0cm 3cm 1.5cm},clip]{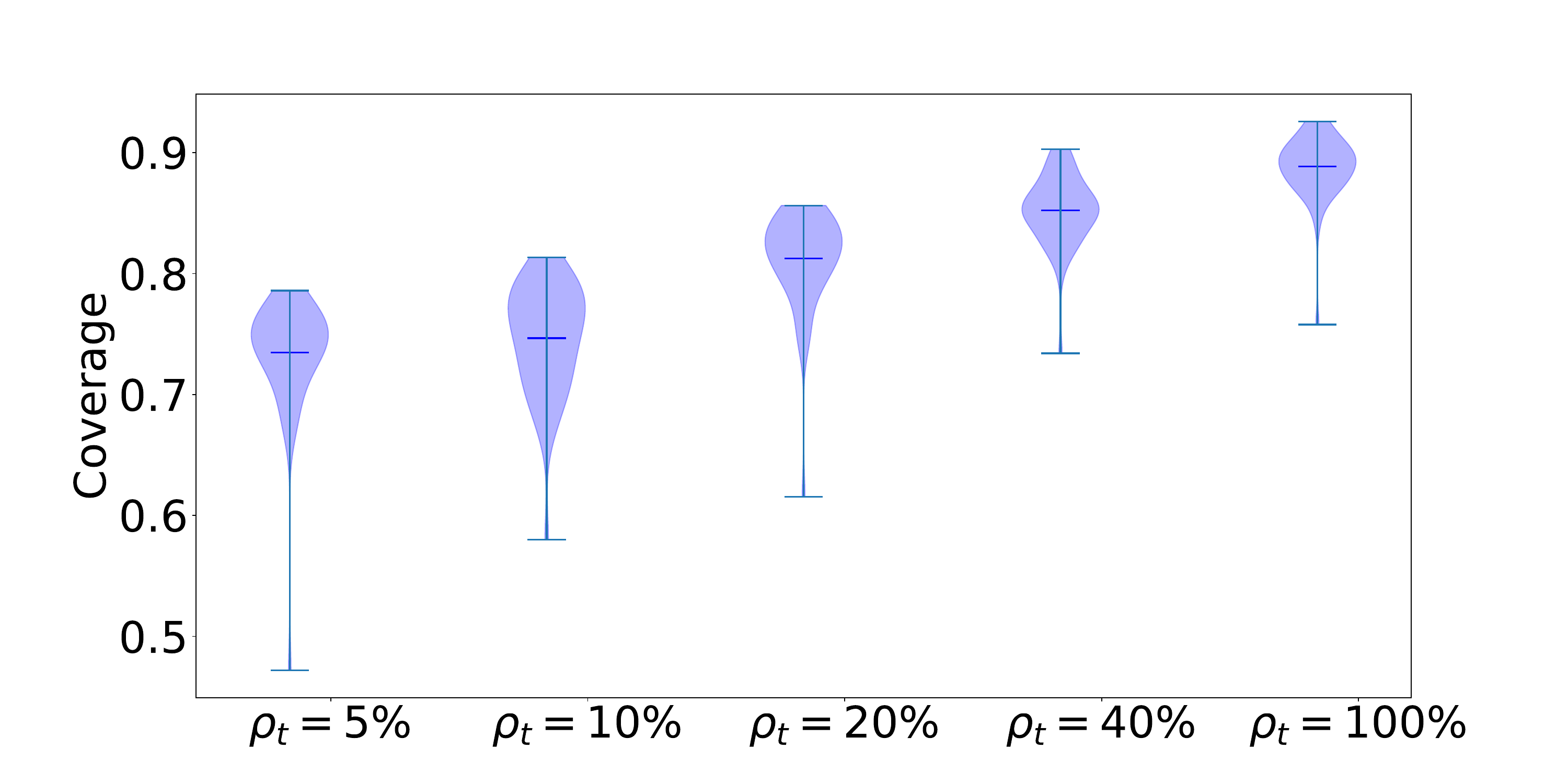}}
    \vspace{-3mm}
    \caption{2-D Shallow water problem: Violin plot showing the distribution of relative errors and uncertainty estimates with respect to the system parameters for ECLEIRS-UQ trained using randomly distributed $\rho_t$ temporal and $20\%$ spatial sampled data with no added noise.}
    \label{fig:Violin_Shallow_sparsecompare}
\end{figure}

The contour plots depicted the uncertainty estimates for a single temporal instance at a fixed parameter. To assess the distribution of errors and uncertainty estimates across the full space-time domains for testing parameters, we analyze the violin plots. The violin plots comparing the distribution of errors and uncertainty estimates for different modeling approaches trained on higher-resolution and clean data are shown in \figref{Violin_Shallow_clean}. The results indicate that IRS-UQ, PI-IRS-UQ with $\lambda = 10^{-1}$ and ECLEIRS-UQ perform similarly with low relative errors, with the latter two being slightly better. These models also perform similarly for uncertainty metrics, with IRS yielding the best uncertainty metrics with higher correlation and coverage. This observation highlights an important distinction between prediction accuracy and uncertainty quality. A model may provide highly accurate predictions while simultaneously yielding uncertainty estimates that are less correlated with the true prediction errors. For this training data scenario, PI-IRS-UQ with $\lambda = 10^{-1}$ appears to exhibit higher errors and worse uncertainty metrics than the other learned models. The relative error for models trained with noisy dataset shows IRS-UQ performance worsens as the quality of data degrades. Conversely, PI-IRS-UQ with $\lambda = 10^{-1}$ exhibits much better predictions, which is opposite to the trend observed when these models were trained with denser and clean data. Note that PI-IRS-UQ with $\lambda = 10^{-1}$ resulted in the best accuracy for the other two numerical experiments. This observation indicates that the performance of PI-URS-UQ significantly depends on the selected value of penalty parameter $\lambda$. While ECLEIRS-UQ exhibits higher errors for sparser and noisier dataset, this performance decrease is not as severe as IRS-UQ. The variation of the performance of ECLEIRS-UQ with variation in temporally sparse training data is shown in \figref{Violin_Shallow_sparsecompare}. The results show a monotonic decrease in the mean relative error with denser temporal data. We also observe better correlation-coefficient and coverage metrics with denser training data; these metrics asymptote to a high coverage and an low-medium correlation coefficient. 

The results obtained for this numerical experiment indicate that all modeling approaches trained using variational neural fields struggle to provide high correlation between estimated uncertainty and errors in prediction. From the contour plots, we observe that correlation is higher closer to the propagating waves, while it becomes lower in other regions. This trend is similar to that observed for the 2D Euler equation; however, the reflected shocks in that case were sharper. This may contribute to the higher correlation coefficient observed for the 2D Euler problem compared with the 2D shallow water problem. These observations suggest that uncertainty estimation is most effective in regions containing sharp gradients, propagating wave fronts, and localized nonlinear features. Conversely, uncertainty estimates become less informative in smooth regions where prediction errors are inherently small and spatially diffuse. From a physical perspective, this behavior is acceptable because uncertainty is primarily concentrated in dynamically important regions rather than being distributed uniformly throughout the domain. Taken together, the three benchmark problems demonstrate that variational latent neural field-based latent dynamics models can provide accurate reduced-order predictions together with meaningful uncertainty estimates. Among the approaches considered, ECLEIRS-UQ consistently provides the greatest conservation properties and strong robustness to sparse and noisy training data while maintaining predictive accuracy.

\section{Conclusions}
\label{sec:Section8}

While reduced-order modeling approaches such as latent dynamics models have accelerated computational simulations in a parameterized setting, these methods estimate solutions without providing an indication of predictive uncertainty. As a result, these methods can provide overconfident predictions, which can negatively affect applications where these models are deployed. Therefore, it is important to develop methods that do not merely give accurate predictions but also provide an estimate of confidence or uncertainty in these predictions. 

In this article, we propose a latent dynamics modeling framework for parameterized physical systems that is applicable to arbitrarily sampled space-time solution data, while also embedding conservation structure and providing uncertainty estimates for model predictions. The uncertainty estimation capability is introduced in the framework by developing variational latent neural fields that integrate an efficient Gaussian process-inspired surrogate into the auto-decoder block of neural fields. This framework models PDE solutions as a random field that is sampled through this proposed variational framework. The uncertainty estimates obtained using this framework are defined with respect to the system parameters, thereby enabling confidence estimation when the model is evaluated for in-distribution or out-of-distribution parameters. Additionally, we incorporate exact nonlinear conservation structure in the framework by writing the random solution-flux vector in a space-time divergence-free form, and learning a manifold representation of the high-dimensional vector potential of this form as a variational latent neural field.

The proposed framework for structure-preserving latent dynamics models with uncertainty estimates (ECLEIRS-UQ) is compared to other variants where variational latent neural fields are directly used to represent parameterized PDE dynamics (IRS-UQ) or when these approaches are augmented with a physics-informed training loss (PI-IRS-UQ). These methods are compared for three problems: 1) 1-D advection problem, 2) 2-D Euler problem and 3) 2-D shallow-water problem, and the ability of different models to accurately representing the solution while also providing robust uncertainty estimates is comprehensively assessed. The numerical experiments indicated that PI-IRS-UQ and ECLEIRS-UQ provide more accurate prediction for both interpolation and extrapolation scenarios compared to IRS-UQ, especially for scenarios where training data quality is low due to sparsity and noise. Furthermore, all three approaches provide useful predictive uncertainty
estimates, as observed from high coverage and correlation metrics for both interpolation and extrapolation parameters. Additionally, we proved that mean of the predicted solution-flux vector obtained using the ECLEIRS-UQ framework respects the governing conservation structure by construction, thereby allowing physically admissible results even for out-of-distribution scenarios. Taken together, these results demonstrate that variational latent neural fields provide a robust framework for latent dynamics modeling with uncertainty estimates. 

While the proposed method is demonstrated to be robust and efficient for the given problems, we plan to scale the method to more complex multi-scale physics and larger problems using domain-decomposition integrated approaches. Such approaches are integral in providing accurate results while reducing the spectral bias of neural fields in capturing high-frequency multiscale structures. Future work will also involve integrating recent generative models, such as score-based diffusion models \cite{Song2020} and flow matching \cite{Lipman2023}, for estimating uncertainty in latent dynamics models and comparing it to the proposed variational latent neural field framework. These generative approaches may provide richer representations of multi-modal uncertainty and could potentially overcome some of the Gaussian assumptions employed in the present work. Finally, future work will investigate the use of uncertainty estimates as active learning indicators for adaptive data acquisition and online model refinement. Such strategies may enable reduced-order models that autonomously improve their predictive capability in regions where uncertainty is high. Overall, the proposed framework represents a step towards physically consistent, uncertainty-aware, reduced-order models capable of supporting reliable scientific predictions in parameterized nonlinear dynamical systems.

\section{Acknowledgements}

We would like to thank Ben S. Southworth for data generation for the 2-D Euler problem and also Ryosuke Park and Marc Charest as the primary developers of the \emph{flecsim} package we used for these calculations. We would also like to thank Dan Serino and Khoa Nguyen for insightful discussions on this topic. This work was supported by Laboratory Directed Research and Development program of Los Alamos National Laboratory under project number 20230068DR. Los Alamos National Laboratory Report LA-UR-26-25104.

\bibliographystyle{unsrt}
\bibliography{ref_bbl}

\end{document}